\documentclass[apj]{openjournal}
\usepackage{graphicx}
\usepackage{amsmath,amsfonts,amssymb}
\setcitestyle{numbers,square,comma}
\usepackage[breaklinks,colorlinks,urlcolor=blue,citecolor=blue,linkcolor=black]{hyperref}
\usepackage{color}
\usepackage{multirow}

\newcommand{\be}{\begin{equation}}
\newcommand{\ee}{\end{equation}}
\newcommand{\LCDM}{$\Lambda$CDM}

\begin{document}

\begin{flushright}
FERMILAB-PUB-20-242-A, KCL-PH-TH/2020-33 \\ KEK-Cosmo-257, KEK-TH-2231 \\ IPMU20-0070, PI/UAN-2020-674FT \\ RUP-20-22
\end{flushright}

\title{The first three seconds: \\
A Review of Possible Expansion Histories of the early Universe  \hspace{5cm}}

\begin{abstract}
It is commonly assumed that the energy density of the Universe was dominated by radiation between reheating after inflation and the onset of matter domination 54,000 years later. While the abundance of light elements indicates that the Universe was radiation dominated during Big Bang Nucleosynthesis (BBN), there is scant evidence that the Universe was radiation dominated prior to BBN. It is therefore possible that the cosmological history was more complicated, with deviations from the standard radiation domination during the earliest epochs. Indeed, several interesting proposals regarding various topics such as the generation of dark matter, matter-antimatter asymmetry, gravitational waves, primordial black holes, or microhalos during a nonstandard expansion phase have been recently made. In this paper, we review various possible causes and consequences of deviations from radiation domination in the early Universe -- taking place either before or after BBN -- and the constraints on them, as they have been discussed in the literature during the recent years. 
\\
\vspace{1.2cm}
\\
\end{abstract}

\author{Rouzbeh Allahverdi$^1$}
\author{Mustafa A. Amin$^2$}
\author{Asher Berlin$^3$}
\author{Nicol\'as Bernal$^4$}
\author{Christian T.~Byrnes$^5$}
\author{M. Sten Delos$^6$}
\author{Adrienne L. Erickcek$^6$}
\author{Miguel Escudero$^7$}
\author{Daniel G. Figueroa$^{8}$}
\author{Katherine Freese$^{9,10}$}
\author{Tomohiro Harada$^{11}$}
\author{Dan Hooper$^{12,13,14}$}
\author{David I. Kaiser$^{15}$}
\author{Tanvi Karwal$^{16}$}
\author{Kazunori Kohri$^{17,18}$}
\author{Gordan Krnjaic$^{12}$}
\author{Marek Lewicki$^{7,19}$}
\author{Kaloian D. Lozanov$^{20}$}
\author{Vivian Poulin$^{21}$}
\author{Kuver Sinha$^{22}$}
\author{Tristan L. Smith$^{23}$}
\author{Tomo Takahashi$^{24}$}
\author{Tommi Tenkanen$^{25,}$\footnote{Corresponding author}}
\author{James Unwin$^{26}$}
\author{Ville Vaskonen$^{7,27,\,{\rm a}}$}
\author{Scott Watson$^{28}$}

\affiliation{$^1$Department of Physics and Astronomy, University of New Mexico, Albuquerque, NM 87131, USA}
\affiliation{$^2$Department of Physics \& Astronomy, Rice University, Houston, Texas 77005, USA}
\affiliation{$^3$Center for Cosmology and Particle Physics, Department of Physics, New York University, New York, NY 10003, USA}
\affiliation{$^4$Centro de Investigaciones, Universidad Antonio Nari\~no, Carrera 3 Este \# 47A-15, Bogot\'a, Colombia}
\affiliation{$^5$Department of Physics and Astronomy, Pevensey II Building, University of Sussex, BN1 9RH, UK}
\affiliation{$^6$Department of Physics and Astronomy, University of North Carolina at Chapel Hill, \\ Phillips Hall CB3255, Chapel Hill, North Carolina 27599, USA}
\affiliation{$^7$Physics Department, King's College London, London WC2R 2LS, UK}
\affiliation{$^8$Instituto de F\'isica Corpuscular (IFIC),  University of Valencia-CSIC, E-46980, Valencia, Spain}
\affiliation{$^{9}$Physics Department,
University of Texas, 2515 Speedway, STOP C1600, Austin, TX 78712, USA}
\affiliation{$^{10}$Oskar Klein Centre for Cosmoparticle Physics, Stockholm University, 10691 Stockholm, Sweden}
\affiliation{$^{11}$Department of Physics, Rikkyo University, Toshima, 
Tokyo 171-8501, Japan}
\affiliation{$^{12}$Fermi National Accelerator Laboratory, Theoretical Astrophysics Group, Batavia, IL 60510, USA}
\affiliation{$^{13}$University of Chicago, Kavli Institute for Cosmological Physics, Chicago, IL 60637, USA}
\affiliation{$^{14}$University of Chicago, Department of Astronomy and Astrophysics, Chicago, IL 60637, USA}
\affiliation{$^{15}$Department of Physics, Massachusetts Institute of Technology, Cambridge, MA 02139, USA}
\affiliation{$^{16}$Center for Particle Cosmology, Department of Physics \& Astronomy, University of Pennsylvania, Philadelphia, PA 19104, USA}
\affiliation{$^{17}$Theory Center, IPNS, KEK, and Sokendai, Tsukuba, Ibaraki 305-0801, Japan}
\affiliation{$^{18}$Kavli IPMU (WPI), UTIAS, The University of Tokyo, Kashiwa, Chiba 277-8583, Japan}
\affiliation{$^{19}$Faculty of Physics, University of Warsaw ul.\ Pasteura 5, 02-093 Warsaw, Poland}
\affiliation{$^{20}$Max Planck Institute for Astrophysics, Karl-Schwarzschild Str. 1, Garching, 85748, Germany}
\affiliation{$^{21}$Laboratoire Univers \& Particules de Montpellier (LUPM), CNRS \& Universit\'e de Montpellier (UMR-5299), \\ Place Eug\`ene Bataillon, F-34095 Montpellier Cedex 05, France}
\affiliation{$^{22}$Department of Physics and Astronomy, University of Oklahoma, Norman, OK, 73019, USA}
\affiliation{$^{23}$Department of Physics and Astronomy, Swarthmore College, 500 College Ave., Swarthmore, PA 19081, USA}
\affiliation{$^{24}$Department of Physics, Saga University, Saga 840-8502, Japan}
\affiliation{$^{25}$Department of Physics and Astronomy, Johns Hopkins University, 3400 N. Charles Street, Baltimore, MD 21218, USA}
\email{t.tenkanen@gmail.com}
\affiliation{$^{26}$Department of Physics, University of Illinois at Chicago, 845 W Taylor Street, Chicago, IL 60607, USA}
\affiliation{$^{27}$NICPB, R\"avala 10, 10143 Tallinn, Estonia}
\email{ville.vaskonen@gmail.com}
\affiliation{$^{28}$Department of Physics, Syracuse University, Syracuse, NY 13214, USA}

\maketitle

\pagebreak
\tableofcontents

\pagebreak
\setlength{\parskip}{5pt}

\section{Introduction}
\label{sec:introduction}

More than forty years ago, Steven Weinberg famously summarized the state of early-universe cosmology in his book, {\it The First Three Minutes} \cite{Weinberg1977}. Weinberg began his account by describing an early time when the Universe was filled with radiation. Since Weinberg's book first appeared, cosmologists have made enormous progress in understanding the physics of the very early Universe. Yet there remains a gap in our understanding of cosmic history -- a gap that spans the first few seconds. How, during that brief time interval, did the energy density of the universe come to be dominated by relativistic particles? In other words, what set the conditions at the start of Weinberg's analysis?

\begin{figure}[b]
\centering
\includegraphics[width=4in]{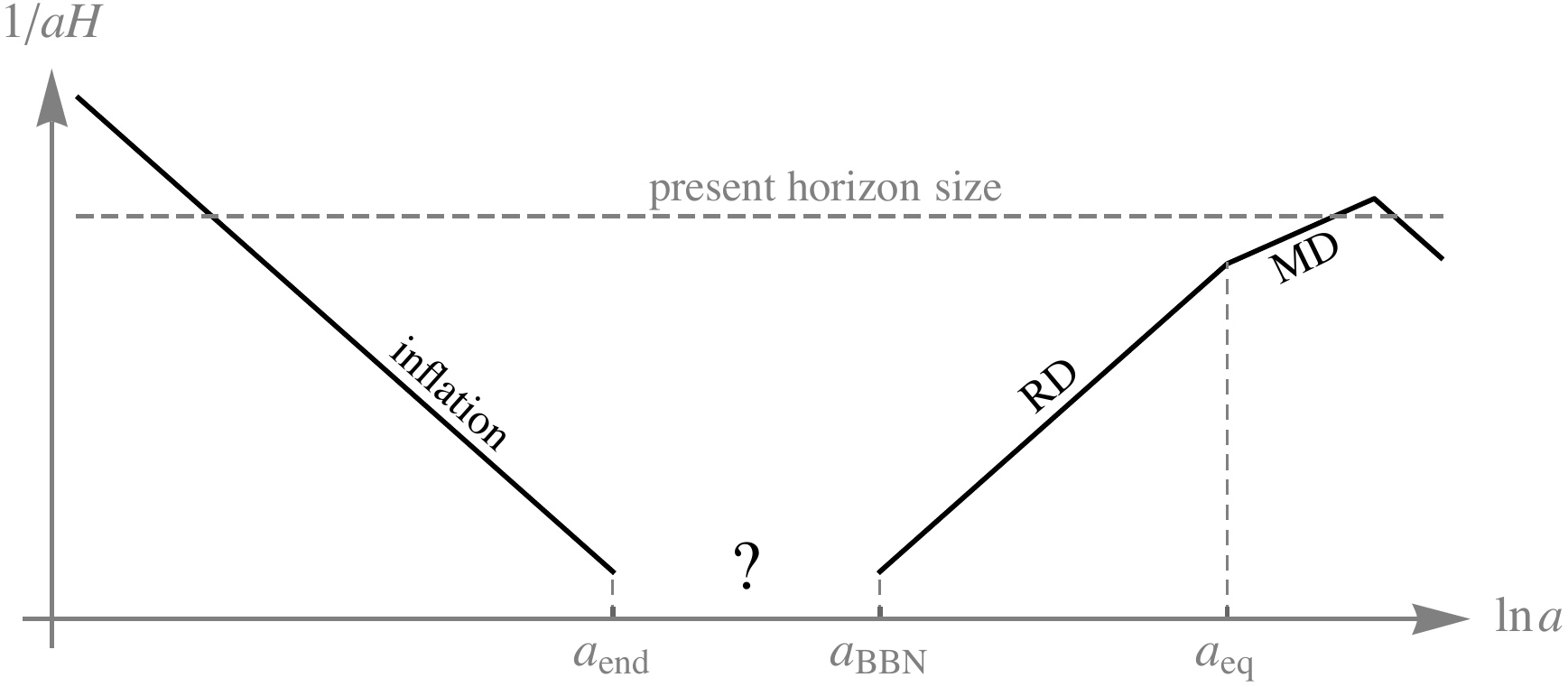} 
\caption{The evolution of the comoving Hubble scale $1/(aH)$ as a function of the scale factor $a$. The expansion history between inflation and the subsequent, post-BBN era remains unknown. Here $a_{\rm end}$ marks the end of inflation, $a_{\rm BBN}$ denotes the commencement of BBN, and $a_{\rm eq}$ is the time of matter-radiation equality at redshift $z\simeq 3400$.}
\label{fig:Fig_expansion}
\end{figure}

The earliest epoch we can confidently gain information about is inflation, an early period of accelerated expansion in the very early Universe~\cite{Starobinsky:1980te,Sato:1980yn,Guth:1980zm,Linde:1981mu,Albrecht:1982wi,Linde:1983gd}, which makes predictions consistent with high-precision measurements of the temperature anisotropies in the Cosmic Microwave Background (CMB)\footnote{For alternatives to cosmic inflation, see e.g. Refs. \cite{Battefeld:2014uga,Brandenberger:2016vhg}.}~\cite{Hinshaw:2012aka,Bennett:2012zja,Martin:2015dha,Akrami:2018odb,Akrami:2019izv}
. These observations limit the energy scale during inflation to be smaller than $\mathcal{O}(10^{16})\,{\rm GeV}$, well below the Planck scale $M_{\rm P} = 1.22\times 10^{19}\,{\rm GeV}$. Meanwhile, constraints on the effects of neutrinos on the CMB, as well as Big Bang Nucleosynthesis (BBN) constraints, reveal that the Universe had to be radiation dominated (RD) by the time neutrinos decoupled from the thermal plasma, at the energy scale $\mathcal{O}(1)\, {\rm MeV}$~\cite{Kolb:1990vq,Peebles:1994xt,Dodelson:2003ft,Weinberg:2008zzc,Gorbunov:2011zz}. Between these two epochs stretches a period of cosmic history that is not well constrained by observations (see Fig.~\ref{fig:Fig_expansion}). Although the time interval between the end of inflation and neutrino decoupling is small ($\sim 1\, {\rm sec}$), the energy scale can drop during this period by almost 20 orders of magnitude and the Universe may have expanded by up to 60 e-folds~\cite{Liddle:2003as,Dodelson:2003vq,Tanin:2020qjw}.

The rate at which the Universe expands depends on the (spatially averaged) total energy density $\langle \rho \rangle$ and pressure $\langle p \rangle$ of the combination of fluids that fill the Universe. In particular, in a spatially flat Friedmann-Lema\^{i}tre-Robertson-Walker (FLRW) Universe filled with components that can be modelled as perfect fluids, the Einstein field equations imply that the scale factor $a(t)$ evolves as
\be
a(t) \propto t^{\frac{2}{3(1+w)}} ,
\label{aw}
\ee
where $t$ is cosmic time and the equation of state parameter, $w$, is given by
\be
w=\frac{\langle p\rangle}{\langle \rho \rangle} \,.
\ee
During inflation the equation of state must be $w \simeq - 1$, whereas RD expansion corresponds to $w \simeq 1/3$. Somehow the Universe must have transitioned from $w \simeq -1$ to $w \simeq 1/3$ between the end of inflation and neutrino decoupling. 

According to standard cosmology, after an initial inflationary phase the Universe quickly entered a period of radiation domination, followed by an era of matter domination, and finally transitioned to the dark energy dominated period which we reside in today. However, this simple four stage evolution of the Universe need not necessarily be the case. Indeed, at the moment we do not have much observational information about the state of the Universe prior to BBN that occurred during radiation dominance at temperature $T \gtrsim 5\,{\rm MeV}$~\cite{Kawasaki:1999na,Kawasaki:2000en,Hannestad:2004px,Ichikawa:2005vw,IKT07,deSalas:2015glj,Hasegawa:2019jsa}. It is therefore possible that the cosmological history featured, for example, an additional early matter-dominated (EMD) epoch due to e.g. slow post-inflationary reheating or massive metastable particles that dominated the total energy density. Potentially more exotic scenarios that change the expansion history of the Universe compared to the standard RD case can also be realized, and they can have important consequences for the physics of the early Universe.

BBN is a cornerstone of the standard cosmological model. At present, the primordial abundances of helium and deuterium are measured with 1\% precision. This makes BBN a powerful arena to test nonstandard cosmological expansion histories. A common lore is that the Universe must be radiation dominated prior to BBN. This is due to the fact that the abundances of light elements are extremely sensitive to the expansion rate, thermalization of neutrino background, and the time-evolution of the neutron-to-proton ratio and their measured values agree very well with the predictions obtained assuming standard RD expansion. Likewise, measurements of the CMB can be used to achieve information about the expansion history of the early Universe. However, the standard flat \LCDM\ model, while successful at explaining numerous cosmological data sets, currently faces several potential challenges. There are theoretical motivations to consider physics beyond \LCDM, such as the unknown nature of inflation, dark matter (DM), dark energy, and mechanisms to explain neutrino masses. As observations have become increasingly precise, several anomalies have appeared, including the Hubble tension~\cite{Verde:2019ivm}, the $\sigma_8$ tension~\cite{Battye:2014qga}, the $A_{\rm lens}$ anomaly~\cite{Calabrese:2008rt,Aghanim:2018eyx}, the coincidence problem~\cite{Fitch:1997cf,Zlatev:1998tr}, the small-scale structure problem~\cite{Bullock:2017xww}, and the anomalous measurement by EDGES~\cite{Bowman:2018yin}. Theorists are hence motivated in multiple ways to postulate new physics beyond \LCDM.

\begin{table}
\begin{center}
\normalsize
\begin{tabular}{| c | c |}
\hline
Quantity / abbreviation & Explanation \\
\hline
\hline
$a$ & Scale factor \\
BAO & Baryonic acoustic oscillations \\
BBN & Big Bang Nucleosynthesis \\
CL & Confidence level \\
CMB & Cosmic Microwave Background \\
(C)DM & (Cold) Dark matter \\
EMD & Early matter-dominated \\
EoS & Equation of state \\
GW & Gravitational wave \\
$H \equiv \dot{a}/a$ & Hubble parameter \\
$H_0$ & Hubble parameter at present \\
$g_*$ & Number of energy degrees of freedom \\
$g_{*,s}$ & Number of entropy degrees of freedom \\
LSS & Large scale structure \\
MD & Matter-dominated \\
$M_{\rm P} \equiv 1 / \sqrt{ G} = 1.22 \times 10^{19} \, {\rm GeV}$ & The Planck mass \\ 
$N \equiv \ln(a_2/a_1)$ & Number of e-folds between the time $a_1$ and $a_2$ \\
$N_{\rm eff}$ & Effective number of neutrinos \\
NFW & Navarro-Frenk-White profile \\
$\Omega_{\rm i} \equiv \rho_{\rm i}/\rho_{\rm crit}$ & Energy density fraction of component $i$ \\
PBH & Primordial black hole \\
PT & Phase transition \\
RD & Radiation-dominated \\
$\rho_{\rm i}$ & Energy density of component $i$ \\
$\rho_{\rm crit} \equiv 3H^2 M_{\rm P}^2/(8\pi)$ & Critical energy density \\
SGWB & Stochastic gravitational wave background\\
SM & Standard Model \\
$T_{\rm reh}$  & Reheating temperature \\
UCMH & Ultracompact minihalo \\
vev & Vacuum expectation value \\
$w\equiv \langle p\rangle/\langle \rho \rangle$ & Equation of state parameter \\
$z$ & Redshift \\
\hline
\end{tabular}
\caption{A list of the most common quantities and abbreviations used in the paper.}
\end{center}
\label{table:abbreviations}
\end{table}

Notably, early periods of nonstandard cosmology can alter the cosmological abundances of particle species. In particular, they can have a significant impact on our expectations for the DM relic abundance, since nonstandard eras can change the expansion rate of the Universe~\cite{Scherrer:1984fd,Kamionkowski:1990ni}, lead to entropy injections that may dilute the DM relic abundance~\cite{Steinhardt:1983ia}, or provide a nonthermal production mechanism for DM~\cite{Chung:1998ua,Moroi:1999zb}.  Furthermore, a period of nonstandard cosmology may also impact small scale structure formation, potentially enhancing or impeding the formation of dense ultracompact DM microhalos~\cite{Erickcek:2011us,Barenboim:2013gya,Fan:2014zua,Delos:2018ueo,Redmond:2018xty}, axion miniclusters~\cite{Nelson:2018via,Visinelli:2018wza}, and primordial black holes (PBH)s~\cite{Khlopov:1980mg,Polnarev:1986bi,Georg:2016yxa,Harada:2016mhb,Cotner:2016cvr,Harada:2017fjm,Kokubu:2018fxy}. The survival of nongravitational relics (including, for example and in addition to DM, matter-antimatter asymmetry \cite{Adshead:2017znw,Hertzberg:2013jba,Hertzberg:2013mba,Takeda:2014eoa,Cline:2019fxx}, 
primordial magnetic fields \cite{Enqvist:1998,Bassett:2000aw,DiazGil:2008tf,Adshead:2016iae,Patel:2019isj}, or effective number of relativistic species \cite{Baumann:2015rya,Green:2019glg}) will depend on the details of the expansion history and post-inflationary thermalization of radiation. Along with large-scale gravitational effects associated with the expansion history, smaller scale gravitational effects -- including horizon-scaled variations of expansion history and associated non-Gaussianities \cite{Chambers:2007se,Chambers:2008gu,Chambers:2009ki,Bond:2009xx,Kohri:2009ac,Leung:2013rza,Suyama:2013dqa,Imrith:2018uyk,Imrith:2019njf,Garcia:2020mwi}, imprints from small-scale clustering of matter \cite{Erickcek:2011us,Amin:2019ums,Musoke:2019ima}, PBHs \cite{GarciaBellido:1996qt,Green:2000he,Bassett:2000ha,Hidalgo:2011fj,Torres-Lomas:2014bua,Suyama:2004mz,Garcia-Bellido:2017mdw,Garcia-Bellido:2017aan,Carr:2017edp,Carr:2018nkm, Cotner:2019ykd,Ezquiaga:2019ftu,Carr:2020gox}, and gravitational waves resulting from violent nonlinear dynamics  \cite{Easther:2006vd,Easther:2006gt,Dufaux:2007pt,GarciaBellido:2007af,Dufaux:2008dn,Dufaux:2010cf,Zhou:2013tsa,Figueroa:2016ojl,Figueroa:2016wxr,Tranberg:2017lrx,Adshead:2018doq,Lozanov:2019ylm} -- provide hope for future observational probes of the time before BBN.

All of the above aspects are topics of this paper. In the following we review various possible causes and consequences of deviations from radiation domination in the early Universe -- taking place either before or after BBN -- and the constraints on them, as they have been discussed in the literature during the recent years. We emphasize that this paper is not an introduction to the topic and the underlying physics but a review of recent advances related to nonstandard cosmological epochs. Several interesting proposals, regarding for example the generation of DM or matter-antimatter asymmetry during a nonstandard expansion phase, warrant further research. We hope that this review helps in guiding this endeavour.

The paper is organized into three sections: Sec. \ref{sec:causes} concentrates on the causes, Sec. \ref{sec:consequences} on the consequences, and Sec. \ref{sec:constraints} on the constraints on nonstandard expansion eras. Each section is further divided into subsections, which discuss the general topic of the section in more detail from both the model-building as well as observational point of view. Finally, in Sec.~\ref{sec:summary}, we summarize our discussion. The paper contains various quantities and abbreviations that are used throughout the paper. They are laid out in Table~\ref{table:abbreviations}.

\section{Causes of nonstandard expansion phases}
\label{sec:causes}

In this section, we will discuss several causes of nonstandard expansion eras, taking place either before or after BBN. Such causes of nonstandard expansion can broadly be grouped as (i) modifications to the properties of the standard, \LCDM, components, and (ii) the introduction of new components altogether. These two categories can be further refined into modifications to the background and/or perturbative evolution. These categories are not exclusive, and some models may fall within multiple descriptions, as we will discuss.

\subsection{Post-inflation reheating (Authors: M. A. Amin, D. I. Kaiser  \& K. D. Lozanov)}
\label{sec:reheating}

In this section, we focus on the period right after inflation. For reviews, see Refs.~\cite{Bassett:2005xm,Frolov:2010sz,Allahverdi:2010xz,Amin:2014eta,Lozanov:2019jxc}. In some of the simplest models of inflation consistent with observations \cite{Martin:2015dha,Akrami:2018odb,Akrami:2019izv}, inflation is driven by a slowly rolling scalar field with a potential $V(\phi)\propto \phi^{\alpha}$. The shape of the potential affects the characteristics of the primordial perturbations which ultimately seed the CMB anisotropies; for values of the inflaton field relevant to scales probed by the CMB, observations constrain power potential to be more shallow than quadratic ($\alpha < 2$). The dynamics at the end of inflation, on the other hand, are determined by the shape of the potential near its minimum, as well as by the inflaton's couplings to other fields, including SM fields and/or intermediaries (see Fig.~\ref{fig:Fig_Reheating}).\footnote{We consider a power-law behavior of the potential during inflation for highlighting the relative flatness (compared to quadratic) of the potential for large field values. In some models, such as those with ``Hill-top" potentials \cite{Boubekeur:2005zm} or ``Natural Inflation"  models \cite{Freese:1990rb} this might not be the best parametrization.}

\subsubsection{Single-field phenomena}

\begin{figure}[b]
\centering
\includegraphics[width=6.5in]{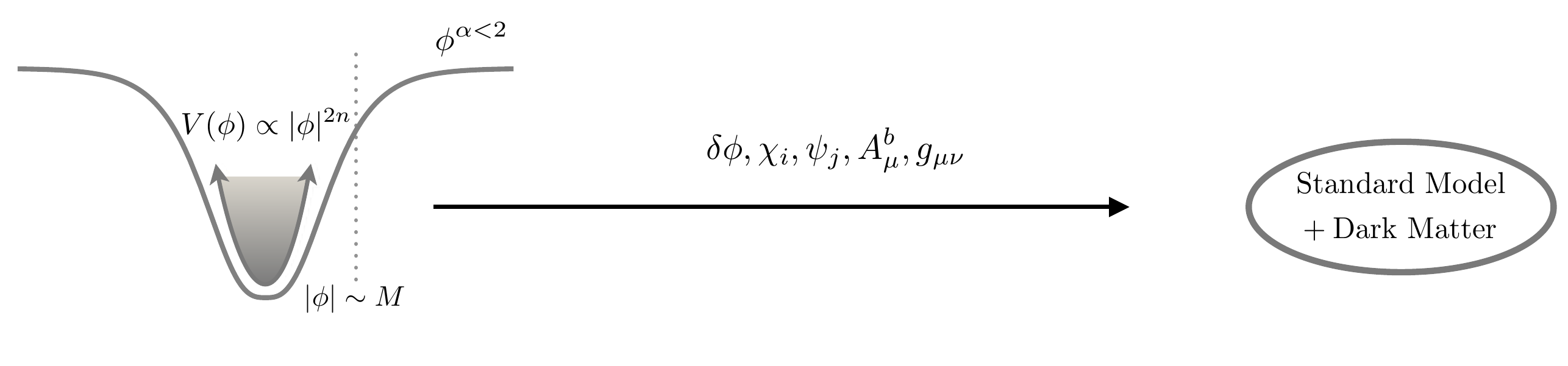} 
\caption{The energy stored in the almost homogeneous inflaton is tranferred to its own perturbations, as well as other fields, and eventually to the SM fields and DM.}
\label{fig:Fig_Reheating}
\end{figure}

We first consider typical effects in single-field models that have negligible couplings to other fields. As shown in Fig.~\ref{fig:Fig_Reheating}, at large field values the potential takes the form $V (\phi) \propto \phi^\alpha$ with $\alpha < 2$ (to be consistent with observations), near the minimum we assume a power-law of the form $V (\phi) \propto \vert \phi \vert^{2n}$, and we assume that the potential interpolates smoothly between these two forms at an intermediate scale $\vert \phi \vert \sim M$. 

During inflation, the inflaton field $\phi (t, {\bf x}) = \varphi (t) + \delta \phi (t, {\bf x})$ will be dominated by the spatially homogeneous component, with $\vert \delta \phi / \varphi \vert \ll 1$. After slow-roll evolution has ended, the homogeneous condensate $\varphi (t)$ will oscillate around the mininum of its potential. The (time-averaged) equation of state will be determined by the power $n$ \cite{Turner:1983he},
\begin{equation}
w=\frac{n-1}{n+1}\,,
\label{wn}
\end{equation}
which for $n = 2$ gives $w = 1/3$, and for $n=1$, yields $w=0$. 

(Almost) spatially homogeneous, oscillating scalar fields are unstable. They become spatially inhomogeneous either due to self-interactions, or gravitational clustering. In particular, the effective mass of the inflaton fluctuations, $m_{\rm eff}^2 = V_{, \phi \phi} (t)$, depends on $\varphi (t)$. Here $V_{, \phi \phi} \equiv {\rm d}^2 V (\phi) / {\rm d} \phi^2 \vert_{\phi = \varphi}$. The quasi-periodic oscillations of $\varphi (t)$ can drive a resonant transfer of energy from the inflaton condensate into shorter-wavelength fluctuations \cite{Kofman:1994rk,Shtanov:1994ce,Kofman:1997yn}. For reviews, see Refs.~\cite{Bassett:2005xm,Frolov:2010sz,Allahverdi:2010xz,Amin:2014eta,Lozanov:2019jxc}. For a single inflaton field minimally coupled to gravity (and neglecting coupled metric perturbations), the equation of motion for Fourier modes of the inflaton fluctuations takes the form
\begin{equation}
    \delta \ddot{\phi}_k + 3 H \delta \dot{\phi}_k + \left[ \frac{ k^2}{a^2} + V_{, \phi \phi} (t) \right] \delta \phi_k = 0 ,
    \label{deltaphieom}
\end{equation}
where overdots denote derivates with respect to cosmic time $t$, $H \equiv \dot{a} / a$, and $k$ is comoving wavenumber. As $\varphi (t)$ oscillates quasi-periodically at the end of inflation, the modes $\delta \phi_k (t)$ behave like oscillators with time-varying frequencies. In the limit $\vert V_{, \phi \phi} \vert \gg H^2$, solutions to Eq.~(\ref{deltaphieom}) generically take the form
\begin{equation}
    \delta \phi_k (t) = {\cal P}_k^+ (t) \, e^{\mu_k t} + {\cal P}_k^- (t) \, e^{-\mu_k t} ,
    \label{deltaphisolution}
\end{equation}
where ${\cal P}_k^{\pm} (t) = {\cal P}_k^{\pm} (t + \tau)$, with $\tau$ the period of oscillation of $V_{, \phi \phi} (t)$. The quantity $\mu_k$ is known as the Floquet exponent. It is generally a complex function of $k$, $\tau$, and the amplitude of $V_{, \phi \phi} (t)$. Modes $\delta \phi_k (t)$ whose wavenumbers $k$ lie within so-called ``resonance bands" $\Delta k$, for which ${\rm Re} (\mu_k ) \neq 0$, will become exponentially amplified due to parametric resonance. Such exponential growth corresponds to a rapid, nonadiabatic transfer of energy from the homogeneous condensate into higher-momentum inflaton particles. Moreover, the resonance process can involve the collective decay of multiple inflatons from the condensate into higher-momentum particles, making the process intrinsically nonperturbative. 

\begin{figure}[t]
\centering
\includegraphics[width=6.5in]{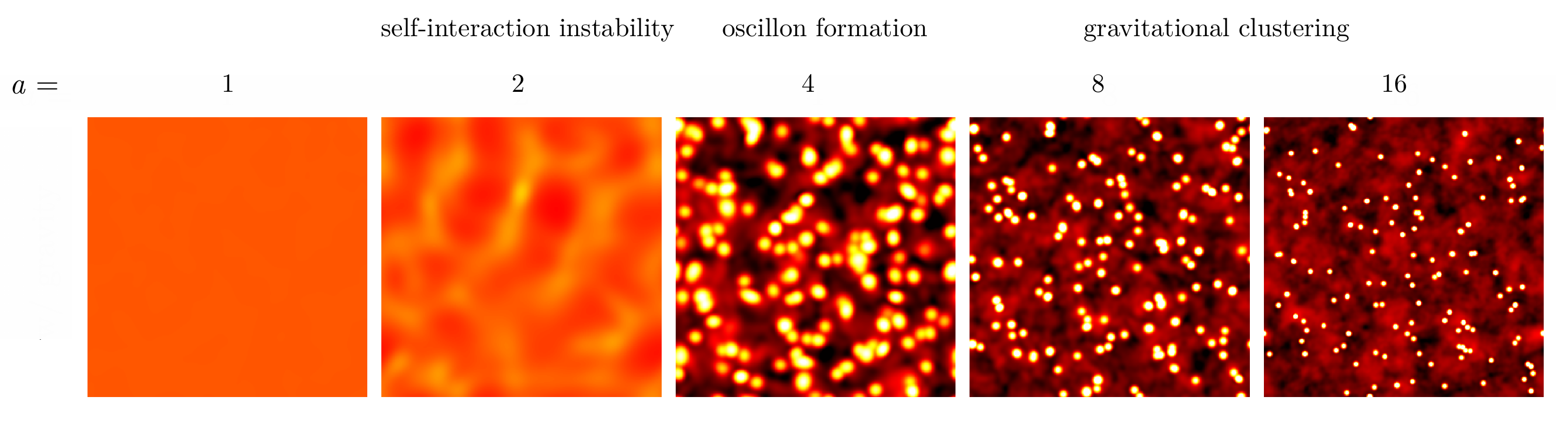} 
\caption{For single-field inflaton potentials with quadratic minima which flatten as we move away from the minimum ($n=1$, $M\ll M_{\rm P}$ in Fig.~\ref{fig:Fig_Reheating}), the almost homogeneous inflation field fragments rapidly into dense solitons (called oscillons) after inflation due to an attractive self-interaction. The oscillons then cluster gravitationally. The equation of state $w$ throughout this phase is approximately zero. Each oscillon has fixed physical size and amplitude, neither of which redshift with expansion. The above boxes are comoving and roughly horizon-sized initially. Lighter colors represent higher densities. Figure based on Ref.~\cite{Amin:2019ums}.}
\label{fig:Fig_Oscillons}
\end{figure}

For single-field models with $n = 1$, equation of state remains $w=0$ with or without fragmentation of the homogeneous condensate. Even in this simple case, the post-inflation epoch can host interesting nonlinear physics. In particular, for $M \ll M_{\rm P}$, as the field explores the flatter-than-quadratic region of the potential ($\phi\gtrsim M$ with $\alpha<2$), broad self-resonance drives copious production of oscillons \cite{Amin:2010xe,Amin:2010dc,Amin:2011hj,Gleiser:2011xj,Lozanov:2017hjm,Hong:2017ooe,Fukunaga:2019unq}. Oscillons are highly overdense, spatially localized, exceptionally long-lived field configurations \cite{Bogolyubsky:1976yu,Gleiser:1993pt,Copeland:1995fq,Kasuya:2002zs,Hindmarsh:2006ur,Amin:2010jq,Zhang:2020bec} where the field amplitude in their cores does not redshift as the universe expands. The rapid formation of oscillons from an almost homogeneous condensate at the end of inflation, and their slower gravitational clustering \cite{Amin:2019ums} is shown in Fig.~\ref{fig:Fig_Oscillons}.\footnote{Note that oscillons are long-lived (with lifetimes $\sim 10^8$ oscillations of the field \cite{Zhang:2020bec} in some models) but not perfectly stable and will evolve by slowly emitting scalar radiation.} In regions of parameter space that do not yield such resonant phenomena ($M\gtrsim M_{\rm P}$), the inflaton condensate will fragment and cluster due to gravitational interactions on a typical time-scale $\Delta N_{\rm frag} \simeq 5$ \cite{Jedamzik:2010dq,Easther:2010mr,Musoke:2019ima}, although the equation of state remains $w \simeq 0$. Either of these single-field scenarios with $n = 1$ would yield a long duration during which the Universe remained MD rather than transitioning to RD expansion, and hence could lead to conflicts with observational constraints similar to moduli fields discussed in Sec.~\ref{sec:moduli}. 

For $n\ne 1$, fragmentation of the condensate is also inevitable (independent of the value of $M$). Resonant amplification of small inhomogeneities will again fragment the homogeneous condensate. Unlike the $n=1$ case, here the equation of state will approach that of radiation, $w = 1/3$ after sufficient time, even in the absence of couplings to other fields \cite{Lozanov:2016hid}. This counter-intuitive result arises because for the $n>1$ case, the resonant band structure is such that the transfer of energy from the homogeneous condensate to relativistic waves of the inflaton field continues to be efficient as the homogeneous, oscillating field approaches the minimum of the potential due to expansion. This leads to a universe filled with relativistic waves of the inflaton field, with an equation of state $1/3$ \cite{Lozanov:2016hid,Lozanov:2017hjm}. Note that the produced waves are relativistic, and remain so, because the effective mass of the field at the minimum of the potential is zero for $n>1$. The number of e-folds between the end of inflation and the beginning of RD epoch in this scenario is given by \cite{Lozanov:2016hid,Lozanov:2017hjm}
\begin{equation}
\Delta N_{\rm{rad}}\gtrsim \frac{n+1}{3}\ln\left[\frac{k}{\Delta k}\frac{\sqrt{ 8 \pi} \, M}{M_{\rm P}}\frac{|4-2n|}{n+1}\right]\qquad n\ne 0,\, 2\,,
\label{Nbr}
\end{equation}
where $\Delta k/ k$ is the relative width of the narrow resonance band. Eq.~(\ref{Nbr}) holds for $n\ne 0, \, 2$ and $M\lesssim M_{\rm P}$, and yields $\mathcal{O}(1) \lesssim \Delta N_{\rm rad} \lesssim {\cal O} (10)$. For $n=2$, Eq.~(\ref{wn}) indicates that the homogeneous evolution already has $w=1/3$. In that case, if $M\ll M_{\rm P}$, then $\Delta N_{\rm rad}\lesssim 1$ due to broad self-resonance. If the inflaton field couples to other light fields, the duration to RD epoch will be shortened. In this sense, Eq.~(\ref{Nbr}) provides an upper bound on the duration between the end of inflation and onset of RD epoch. This upper bound can significantly reduce the uncertainty in inflationary observables such as the tensor-to-scalar ratio $r$ and spectral tilt $n_{\rm s}$ \cite{Lozanov:2016hid}. 

Our discussion of single field models assumes that the coupling to other fields is sufficiently weak to not play a major role in the single field dynamics for a significant number of e-folds after inflation. Of course, the inflaton must couple to other fields (even if very weakly) to eventually yield a radiation-dominated, thermal universe filled with Standard Model particles before BBN. For example, to prevent an indefinite MD period for the $n=1$ case, couplings to other relativistic daughter fields must be invoked to drain energy away from the inflaton. This is true whether the universe is filled with oscillons, or includes gravitationally bound inflaton structures, or even if the inflaton remains homogeneous. For the $n>1$ case, we can get to radiation domination without couplings to other fields, however, couplings to other fields must be again invoked to eventually lead to the requisite particle content, and thermalization before BBN. The detailed nature of the decay and thermalization will depend on the nature of the coupling to daughter fields (which may be Standard Model fields or intermediaries), the homogeneous/inhomogeneous state of the field, and could be perturbative or non-perturbative.\footnote{For an example of a non-perturbative, explosive gauge field production from mergers of oscillons see Ref. \cite{Amin:2020vja}.}

\subsubsection{Multifield phenomena}

In models in which the inflaton field $\phi$ couples directly to other fields (which we may collectively denote as $\chi$), the $\chi$ fields typically acquire $\phi$-dependent effective masses, $m_\chi^2 = m_\chi^2 (\phi)$. After inflation, these couplings can drive even more efficient transfer of energy out of the inflaton condensate than in single-field models. Such fast processes can hasten the transition to $w = 1/3$, though in some cases an incomplete decay of the inflaton can yield a later phase of matter-dominated expansion, with $w \sim 0$, prior to neutrino decoupling.

For scalar fields $\chi$ that are minimally coupled to gravity, modes $\chi_k$ obey an equation of motion of the same form as Eq.~(\ref{deltaphieom}), with $V_{, \phi \phi} \rightarrow m_{\chi}^2 (\varphi (t))$. Couplings between $\phi$ and $\chi$ are typically less constrained by observations than are self-couplings of $\phi$, and hence there can be broad resonance bands (with $\Delta k \gtrsim m_\chi$) within which the Floquet exponents for modes $\chi_k$ are large (${\rm Re} (\mu_k) \sim {\cal O} (m_\phi)$, where $m_\phi$ is the mass of the oscillating inflaton).  Given the nonperturbative nature of such resonances, these decays can produce daughter particles much more massive than the inflaton \cite{Kofman:1997yn,Bassett:2005xm,Allahverdi:2010xz,Frolov:2010sz,Amin:2014eta,Lozanov:2019jxc}. In some models (such as those involving spontaneous symmetry breaking), a tachyonic instability may occur, for which $( [k^2 / a^2 ] + m_{\chi}^2 ) < 0$, which likewise yields a rapid, nonperturbative amplification of certain modes $\chi_k$ \cite{Greene:1997ge,Felder:2000hj,DeMelo:2001nr,Arrizabalaga:2004iw,Dufaux:2006ee,Amin:2019qrx}. 

Models with multiple scalar fields $\phi^I$ that have nonminimal couplings to gravity can have even richer resonance structure, since (upon performing a conformal transformation to the Einstein frame) the field-space manifold acquires nontrivial curvature \cite{Kaiser:2010ps}. In such models, couplings can arise between the fields from noncanonical kinetic terms as well as from direct couplings in the interaction potential. For large dimensionless nonminimal couplings $\xi_I \geq {\cal O} (10^2)$, as one finds in popular models like Higgs inflation \cite{Bezrukov:2007ep}, these couplings can yield especially efficient resonances at the end of inflation \cite{Bassett:1997az,Tsujikawa:1999iv,Tsujikawa:1999jh,Tsujikawa:1999me,Bezrukov:2008ut,GarciaBellido:2008ab,Child:2013ria,Ema:2016dny,DeCross:2015uza,DeCross:2016cbs,DeCross:2016fdz,Sfakianakis:2018lzf}. The rapid production of particles with wavenumber $k \geq H$ can drive $w \rightarrow 1/3$ within $\Delta N_{\rm rad} \sim {\cal O} (1)$ after the end of inflation \cite{Nguyen:2019kbm,vandeVis:2020qcp}. 

The exponential amplification of modes driven by parametric resonance and/or tachyonic instability does not persist indefinitely. Rather, the phase of rapid growth ends due to backreaction on the inflaton oscillations from the produced particles, as well as from rescattering among modes $\chi_k$ of different $k$. Such backreaction effects typically grow quickly, within $\Delta t_{\rm br} \sim {\cal O} (1) \times \mu_k^{-1} \ll H^{-1}$ \cite{Bassett:2005xm,Allahverdi:2010xz,Frolov:2010sz,Amin:2014eta,Lozanov:2019jxc}. In many models, backreaction grows to be significant before resonance can completely transfer energy from the inflaton condensate into decay products, which can leave a significant fraction of the total system energy, $\gtrsim {\cal O} (10\%)$, in the inflaton field.

The subsequent evolution of the equation of state is largely model dependent. For example, in models with minimally coupled scalar fields and potentials of the form $V (\phi, \chi) = m_\phi^2 \phi^2 / 2 + \kappa \phi \chi^2 + g^2 \phi^2 \chi^2$, the couplings between $\phi$ and $\chi$ can give rise to a long-lived, nontrivial period of expansion with $0 < w \lesssim 1/3$, depending on the values of the coupling constants $\kappa$ and $g$ \cite{Podolsky:2005bw,Dufaux:2006ee,Amin:2019qrx} (see Ref.~\cite{Antusch:2020iyq} for recent long-duration simulations). At late times in such models, the expansion of the Universe will dilute the energy of relativistic decay products so much that the energy remaining in the inflaton condensate can dominate the total energy density again, giving rise to a matter-dominated phase with $w \simeq 0$. If the inflaton potential is non-quadratic at the minimum, then similar to the single-field case (see discussion near Eq.~\eqref{Nbr}), the equation of state always approaches $1/3$, albeit the time-scales to approach radiation domination can be shortened based on strength of the coupling to other fields \cite{Antusch:2020iyq}.

If the inflaton couples to vector bosons, the Universe can transition rapidly to $w = 1/3$ after inflation \cite{Davis:2000zp,GarciaBellido:2003wd,Braden:2010wd,Allahverdi:2011aj,Deskins:2013lfx,Adshead:2015pva,Adshead:2017xll,Cuissa:2018oiw}. For example, if the inflaton is an axion-like field, then a coupling of the form $\lambda \phi F \Tilde{F}$ (where $F_{\mu\nu}$ is the gauge-boson field-strength tensor) can lead to a virtually instantaneous transition to RD phase at the end of inflation, for sufficiently large coupling $\lambda$ \cite{Adshead:2019igv,Adshead:2019lbr}. Moreover, nearly all of the energy of the inflaton condensate can be transferred to gauge fields in such processes, potentially avoiding a primordial period of expansion with $w \simeq 0$. 

In other regions of parameter space, for sufficiently small amplitudes of the inflaton oscillations $\varphi$ and for weak couplings $\kappa$ and $g$, resonances will remain subdominant, and the decay of the inflaton condensate will occur mostly due to perturbative decays. The decay rates may be estimated from Feynman diagrams for the tree-level couplings. Even for models in which strong resonances occur at early times after inflation, such perturbative decays can play an important role at late times, providing a mechanism to complete the transfer of energy from the inflaton condensate into light decay products. The rate of such perturbative decays can be enhanced in the case of bosonic fields $\chi$. In those cases, Bose enhancement yields correction factors to the tree-level decay rates that one would estimate in vacuum, $\Gamma_\chi^{\rm vac}$, of the form $\Gamma_\chi \propto \Gamma_\chi^{\rm vac} \, n_\chi$, where $n_\chi$ is the occupation number. Then the change of the occupation number, $\dot{n}_\chi = \Gamma_\chi$, can yield exponential growth rates for $n_\chi$ rather than typical linear rates of growth in vacuum \cite{Kofman:1997yn}. The perturbative decay of the $\phi$ condensate will be complete when $\Gamma^{\rm vac}_\chi \sim H$, after which the energy will reside in relativistic decay products and the Universe will become RD. 

If the inflaton couples to fermions $\psi$, the effective mass $m_\psi (\varphi (t))$ can become time-dependent at the end of inflation, much as for bosonic fields $\chi$. This can yield a very different distribution of fermionic decay products than would result from perturbative decays, although the Pauli exclusion principle prevents the occupation number for any fermionic mode to grow larger than 1 \cite{Greene:1998nh,Greene:2000ew,Peloso:2000hy,Tsujikawa:2000ik,Adshead:2015kza,Adshead:2017znw}. In models in which the inflaton couples to both bosonic and fermionic fields, perturbative decays of the inflaton condensate into fermions can help complete the transfer of energy out of the inflaton and avoid a prolonged phase of $w \simeq 0$.

Finally, we note that if significant dissipative effects due to particle production are present during inflation (for example, during warm inflation \cite{Berera:1995ie,Berera:2008ar}), and a significant thermal bath is already present when inflation ends, the duration of transition to radiation domination from the end of inflation might be shortened.

\subsubsection{Future directions}

The period immediately after the end of inflation can feature rich, nonlinear and nonperturbative dynamics. Parametric resonance and tachyonic instabilities can transfer energy efficiently from the inflaton condensate into higher-momentum decay products. In both single-field and multifield models, the efficient production of relativistic decay products can drive a transition in the equation of state from $w \simeq -1$ during inflation to $w = 1/3$ within a few e-folds after the end of inflation. If only slow, perturbative processes are involved, the transition can take considerably longer. Details of the energy transfer depend on the model; in some cases, the rapid transition to RD expansion can slide into a later phase of matter-dominated expansion (with $w \simeq 0$) unless some mechanism exists (such as perturbative decays) to completely drain the energy from the inflaton condensate. Although the details of nonperturbative post-inflationary dynamics remain model dependent, recent work has made progress towards more model-independent means of characterizing this phase \cite{Amin:2015ftc,Ozsoy:2015rna,Figueroa:2016wxr,Amin:2017wvc}. 

For a given model, the full nonlinear dynamics of fields after inflation can now be explored numerically using a number of codes which evolve scalar and vector fields on a lattice \cite{Felder:2000hq,Frolov:2008hy,Easther:2010qz,Huang:2011gf,Sainio:2012mw,Child:2013ria,Amin:2018xfe,Giblin:2019nuv,Lozanov:2019jff,Figueroa:2020rrl}. After such nonlinear evolution, however, the spectrum of decay products is typically far from an equilibrium distribution. The processes and time-scales over which such distributions of particles relax into thermal equilibrium remain much less understood, and an important area for further research; progress may be made by exploiting methods developed for the study of other systems, such as quark-gluon plasma in the context of heavy ion collisions \cite{McDonough:2020tqq}. Another interesting area of research concerns incorporating gravitational effects more fully into our understanding of post-inflation reheating \cite{Nambu:1996gf,Taruya:1997iv,Bassett:1998wg,Bassett:1999mt,Bassett:1999ta,Tsujikawa:2002nf,BasteroGil:2007mm,BasteroGil:2010nm,Jedamzik:2010dq,Martin:2019nuw,Giblin:2019nuv}.

\subsection{Extra stages of inflation (Authors: T. Tenkanen \& V. Vaskonen)} 
\label{sec:extrainflation}

\begin{figure}[t]
\centering
\includegraphics[width=0.5\textwidth]{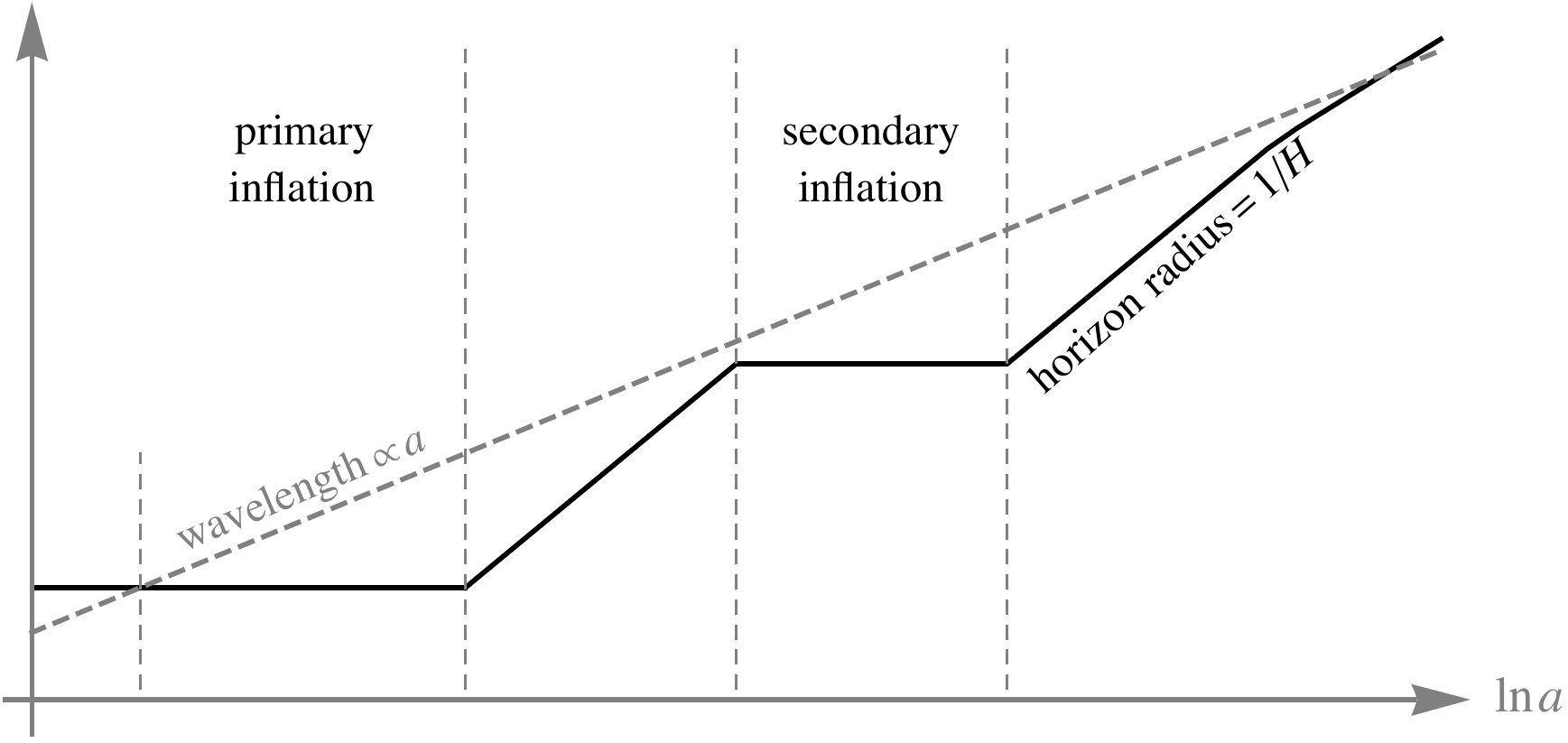}
\caption{The black curve shows the evolution of the Hubble horizon $1/H$ as a function of the scale factor $a$, whereas the gray dashed line shows, as an example, a scale that re-enters horizon after the matter-radiation equality. \\}
\label{fig:secinfl}
\end{figure}

In this section we discuss scenarios in which the primary or primordial inflationary phase was followed by a secondary or even multiple extra stages of inflation. These additional inflationary stages may have been punctuated by an era of either standard RD expansion or by some type of nonstandard expansion; see Fig.~\ref{fig:secinfl} for illustration. Because the standard way of treating the dynamics of the early Universe usually contain only one period of inflation -- the primordial one --, any additional stages of inflation can be seen as a nonstandard addition regardless of how the Universe expanded between the inflationary stages. The energy scales at which the additional inflationary stages may have occurred are not particularly limited, except for the usual requirements that the upper limit for the energy scale of inflation is $\rho_{\rm tot}\simeq (10^{16}\,{\rm GeV})^4$ (see Sec.~\ref{sec:inflation}) and any nonstandard phase must have ended prior to BBN at $T\sim 1$\,MeV (see Sec.~\ref{sec:CMB_BBN_constr}).

\subsubsection{Multistep inflation}

We begin by discussing scenarios of multistep inflation by which we refer to models where (nonthermal) inflation occurs in at least two phenomenologically different stages with distinct characteristics, as originally suggested in Refs.~\cite{Starobinsky:1986fxa,Silk:1986vc}. This type of scenarios can be generically divided into three categories: those in which the same scalar field(s) that were responsible for the primary inflation caused also the secondary stage of inflation (for instance because the inflaton potential has two separate flat regions), so-called hybrid inflation models in which inflation at its last stages was driven not by the energy density of the inflaton field but by the vacuum energy density induced by an interaction to another field, and those in which another scalar field that was an energetically subdominant ``spectator" field during the primary phase of inflation either acquired a large vacuum expectation value (vev) during the first inflationary stage or was kept frozen in its initial value with a finite energy density and later on became the dominant component, thus starting a new stage of inflation. Examples of the first category above include scenarios studied recently in e.g. Refs.~\cite{Artymowski:2016ikw,Kannike:2017bxn,Pi:2019ihn} (see also Refs.~\cite{Raveendran:2016wjz,Maeda:2018sje} for recent studies on double inflation in multifield models with a rapid turn in field-space), while scenarios belonging to the second category were originally introduced in Refs.~\cite{Linde:1991km,Linde:1993cn} and which have been studied in a large number of works since. However, generically hybrid models predict a blue-tilted spectrum for primordial perturbations\footnote{See Eq. \eqref{Pzeta} for a parameterization of the curvature power spectrum.} and are therefore now ruled out by observations~\cite{Akrami:2018odb}.

A secondary stage of inflation started by a spectator field was originally studied in Refs.~\cite{Starobinsky:1986fxa,Silk:1986vc} and has been studied more recently in e.g. Refs.~\cite{McDonald:2003xi,Moroi:2005kz,Moroi:2005np,Dimopoulos:2011gb,Kawasaki:2012gg,Enomoto:2012uy,Kohri:2012yw,Byrnes:2014xua,Inomata:2017vxo,Hardwick:2017fjo,Torrado:2017qtr,Bae:2017tll,Enqvist:2019jkb,Carr:2019hud,Tada:2019amh}. Typically in scenarios where the secondary stage of inflation starts due to a spectator field $\sigma$ acquiring a vev during the first stage of inflation, the vev has to be very large for the field to become energetically dominant while still in slow-roll to trigger a new period of inflation: for example, for a quadratic spectator potential $V_\sigma \propto \sigma^2$ the criterion is $|\sigma| \gtrsim M_{\rm P}/\sqrt{4\pi}$. A natural mechanism for this, based on the stochastic evolution of light scalar fields in (quasi-) de Sitter space was discussed in Ref.~\cite{Hardwick:2017fjo}. Another possibility is based on the type of scenarios studied in e.g. Ref.~\cite{Silk:1986vc}, where the authors considered the potential 
\be
    V = V_0 + \lambda_\phi\phi^4\left[\ln\left(\frac{\phi^2}{f^2}\right)-\frac12\right] + \lambda_\sigma \sigma^4\,,
\ee
where $f$ is a mass scale related to the ultraviolet completion of the theory. Here both fields are assumed to have a large vev initially, $\phi_i\,,\sigma_i > M_{\rm P}$ and the field $\phi$ is assumed to drive the first stage of inflation. If $\lambda_\sigma$ is not very small, the stochastic quantum fluctuations of $\sigma$ acquired during the first stage of inflation were smaller than its initial value, and
the field remained essentially frozen at $\sigma_i$. The first stage of inflation then ended due to $\phi$ rolling down to $\phi \lesssim M_{\rm P}$, and the second stage began shortly after when the field $\sigma$ became the energetically dominant species. In practice, however, also a much simpler chaotic potential with a suitable choice of parameters can permit similar dynamics.

Finally, we note that in addition to the scenarios discussed above, also the recently proposed Transplanckian Censorship Conjecture \cite{Bedroya:2019tba} has motivated studies on the occurrence of multiple stages of inflation, see e.g. Refs. \cite{Mizuno:2019bxy,Li:2019ipk} (however, see also Ref. \cite{Dvali:2020cgt}).

\subsubsection{Thermal inflation}

Next, we discuss scenarios where the secondary period of inflation is caused by thermal effects. In this case the inflationary stage is known as thermal inflation~\cite{Lyth:1995hj,Lyth:1995ka}, and it is is driven by a scalar field that at a high temperature is far from its zero temperature vev. Part of the total energy density of the Universe is then in the form of vacuum energy, and as temperature drops, the vacuum energy can become the dominant energy density component, therefore causing a period of inflation. Eventually, the thermal inflation ends when the scalar field either tunnels or rolls down to a lower energy minimum of its potential. Of particular interest is the former case,  where the vacuum energy dominance ends with a first-order phase transition~\cite{Coleman:1977py,Callan:1977pt,Linde:1980tt,Linde:1981zj} (see also Sec.~\ref{sec:PTsandBaryogenesis}). This happens by formation of bubbles of the true vacuum which expand, finally turning the whole Universe into the new phase. These bubbles nucleate sparsely and are initially small. They then expand by many orders of magnitude and as they collide, most of the energy is in very thin regions around the bubble walls that move almost at the speed of light~\cite{Bodeker:2009qy,Bodeker:2017cim,Ellis:2018mja,Ellis:2019oqb}. After the bubbles have collided, the scalar field oscillates around the true vacuum, potentially causing a short period of matter-dominance before decaying and reheating the radiation bath. Similarly, a short MD period can be caused in the case where the thermal inflation ends with the scalar field rolling down to the new minimum.

A period of thermal inflation can be realized, for example, in classically scale-invariant models or strongly-interacting theories~\cite{Randall:2006py,Konstandin:2011dr,Marzola:2017jzl,Marzo:2018nov,Baratella:2018pxi,vonHarling:2017yew,Bruggisser:2018mrt,Ellis:2019oqb}~\footnote{We note that the early thermal inflation models considered flat directions in supersymmetric and grand unified theories (see e.g. Refs.~\cite{Lyth:1995hj,Lyth:1995ka,Barreiro:1996dx,Adams:1997de,Hui:1998dc,Asaka:1999xd}).}. In this case, the scalar potential is of the form
\be \label{eq:Vcsi}
    V \simeq V_0 + \frac{3 g^4 \phi^4}{4\pi^2}\left[\ln\left(\frac{\phi^2}{v_\phi^2}\right) - \frac{1}{2} \right]\, +  g ^2 T^{\,2} \phi^2 \,,
\ee
where $g$ is a coupling constant and $v_\phi$ the vev of $\phi$. The scalar field is stuck in a metastable phase that is separated from the true vacuum by a potential energy barrier generated by purely thermal effects (the last term in Eq.~\eqref{eq:Vcsi}). Therefore, a very long period of supercooling is possible, because the potential energy barrier shrinks as the temperature decreases, unlike in models where the potential energy barrier is dominated by nonthermal terms. The potential energy barrier can also disappear at a nonzero temperature in models that are not classically scale-invariant. In this case the supercooling period is typically not very long.

An interesting consequence of thermal inflation ending to a first-order phase transition is the formation of a stochastic gravitational wave background in bubble collisions (see Sec.~\ref{sec:GWBGs}). In addition, the collisions of very energetic bubble walls may lead to copious formation of PBHs~\cite{Hawking:1982ga,Kodama:1982sf,Khlopov:1998nm} (see also Sec. \ref{sec:formationPBHs}). Alternatively, in the case that the thermal inflation ends by the field rolling down its potential PBHs can form from large density perturbations generated when the thermal inflation potential turns from convex to concave at the high-temperature minimum~\cite{Dimopoulos:2019wew}.

Finally, a secondary nonthermal stage of inflation can be caused by symmetry restoration by nonthermal effects as the particles produced in preheating after inflation can for a long time remain out of thermal equilibrium~\cite{Kofman:1995fi,Tkachev:1995md,Felder:2000sf}. Also this inflationary stage ends with a symmetry breaking phase transition that can be of first order~\cite{Khlebnikov:1998sz,Rajantie:2000fd}.

A common feature of all these scenarios is that if the secondary stage of inflation lasted for a sufficiently long time, this becomes the ``primordial" inflation whose imprints on the structure of the Universe at different scales are the only ones we can observe. On the other hand, if the secondary phase only lasted for a short time, 
the initial conditions for large and small structure formation decouple so that the first stage determines the spectrum on large scales and the second the spectrum on small scales~\cite{Silk:1986vc}. In addition to providing initial conditions for large scale structure formation, the primary stage of inflation provides initial conditions also for all the stages that follow it. This has been argued to alleviate the problems some inflationary models face with initial conditions~\cite{Artymowski:2016ikw}.

\subsection{Heavy particles and dark sectors (Authors: N. Bernal \& J. Unwin)}
\label{sec:dark_sectors}

In this section we discuss how nonstandard cosmological histories may arise from different particle physics scenarios including heavy particles or dark sectors. 

\subsubsection{Nonstandard cosmology from heavy particles}

We start with the intuitive idea that a long lived heavy particle $\phi$ could source an EMD epoch. The state $\phi$ could initially be in thermal equilibrium with the SM states, or could be out-of-equilibrium at all times. Once $\phi$ becomes nonrelativistic its contribution $\rho_\phi$ to the energy density of the Universe will grow relative to the contribution due to SM radiation $\rho_R$. Their energy densities evolve with the scale factor according to  $\rho_\phi\propto a^{-3}$ while  $\rho_R\propto a^{-4}$. Thus, even if $\phi$ initially contributes a small component to the total energy density, then provided it is sufficiently long-lived it will grow  to dominate the energy density of the Universe and lead to an EMD epoch. These ideas first arose in the context supersymmetry and string theory, see e.g. Refs.~\cite{Moroi:1999zb,Vilenkin:1982wt,Coughlan:1983ci,Starobinsky:1994bd, Dine:1995uk}, and the particular case of string  moduli fields will be discussed in Sec. \ref{sec:moduli}.

More generally, any massive meta-stable state could potentially come to dominate the energy density of the early Universe. Specifically, in the early Universe the evolution of $\rho_\phi$ and the SM entropy density $s$ are governed by the following coupled Boltzmann equations~\cite{Chung:1998rq} 
\begin{align}
	&\dot\rho_\phi+3H\rho_\phi=-\Gamma_\phi\,\rho_\phi\,,
		\label{eq:cosmo2} \\
	&\dot{s}+3Hs=+\frac{\Gamma_\phi\,\rho_\phi}{T}\,,
	\label{eq:cosmo3}
\end{align}
where $\Gamma_\phi$ is the $\phi$ decay width, $H$ is the Hubble parameter, and $T$ is the temperature of the SM photons. Notably,  the evolution of $T$ can be tracked via Eq.~\eqref{eq:cosmo3}, since the entropy density $s(T)\equiv(\rho_R+p_R)/T=2\pi^2\,g_{*,s}(T)\,T^3/45$.  Additionally, under the assumption that the SM bath maintains internal equilibrium at all times in the early Universe, its temperature dependence follows from its energy density $\rho_R(T)=\pi^2\,g_*(T)\,T^4/30$. Here $g_*(T)$ and $g_{*,s}(T)$ correspond to the effective number of relativistic degrees of freedom for  SM energy and entropy densities, respectively.  The Hubble expansion rate $H$ is defined by $H^2=8\pi(\rho_\phi+\rho_R)/(3M_P^2)$. While $\Gamma_\phi\ll H$ (such that the radiation bath is mainly composed of particles that are not produced by decays) the entropy is separately conserved in each sector and the scale factor and temperature scale as $T\propto a^{-1}$ (up to variations of $g_{*,s}$). The expansion rate will be different depending on the fractional distribution of the total energy density. Specifically, during the era of MD one has $H\propto  T^{3/2}$, whereas for the RD case $H\propto  T^2$. Also note that once $\phi$ decays, the entropy conservation in the bath no longer holds (which corresponds to when the energy density in the relativistic decay products of $\phi$ becomes significant), and the evolution switches to $T\propto a^{-3/8}$ and thus $H\propto T^4$~\cite{Scherrer:1984fd,Giudice:2000ex} until the Universe transitions to the RD era.

Going beyond an EMD epoch, one can envisage instead that the Universe may have been dominated at one stage by a particle species with a general equation of state $w$, such that $\rho_\phi\propto a^{-3(1+w)}$~\cite{Turner:1983he}. In this case the Boltzmann equation~\eqref{eq:cosmo2} depends on $w$ and (assuming the scalar field is decaying at zero-momentum) is instead given by%
\footnote{If $\phi$ is coherently oscillating in a potential $V(\phi)\propto\phi^n$, the right hand side should read $-\frac{2n}{2+n}\Gamma_\phi\,\rho_\phi$~\cite{Turner:1983he}.}
\begin{equation} \label{eq:rhophigamma}
	\dot\rho_\phi+3\,(1+w)H\rho_\phi=-\Gamma_\phi\,\rho_\phi\,.
\end{equation}
 Thus, while the $\phi$ field dominates the energy density of the Universe the Hubble rate scales as  $H\propto a^{-\frac32(1+w)}\propto T^{\frac32(1+w)}$. 

As in the MD case, once the energy density in the relativistic decay products of $\phi$ becomes significant this evolution switches to $H\propto T^4$ (regardless of the value of $w$~\cite{Maldonado:2019qmp}).
Simple cases corresponding to MD and RD expansion are found for $w=0$ and $w=1/3$, respectively. Examples with more general $w$ can occur for oscillating scalar fields with certain potentials (e.g. periodic~\cite{Choi:1999xn,Gardner:2004in} or special constructions~\cite{DEramo:2017gpl,DiMarco:2018bnw}), in brane world cosmologies~\cite{Okada:2004nc,Meehan:2014bya}, and for scalar-tensor theories~\cite{Catena:2004ba, Dutta:2016htz}. An especially well motivated case is that of kination~\cite{Barrow:1982ei, Ford:1986sy, Spokoiny:1993kt} in which $\phi$ is a ``fast-rolling" field whose kinetic term $\dot\phi$ dictates the expansion rate of the post-inflation Universe, implying an equation of state $w\approx 1$.  In most models of interest $|w|\leq1$, with $w=-1$ corresponding to dark energy or quintessence domination~\cite{Ratra:1987rm, Caldwell:1997ii}. For models with $w>1$ there are concerns about  superluminal propagation, although it has been claimed this may be resolved in certain cases~\cite{DEramo:2017gpl}. Importantly, for $w>1/3$ the energy density $\rho_\phi$ gets diluted faster than radiation, and if $\rho_\phi\ll\rho_R$ at $T\sim 1$ MeV, the late decay of $\phi$ typically has a negligible impact.

\subsubsection{The transition to radiation domination} 

There are two main possibilities through which a period of nonstandard cosmology can come to an end in the case that it is sourced by a population of heavy state $\phi$ or similar physics. Firstly, if the energy density $\rho_\phi$ redshifts faster than radiation ($w>1/3$) then after some duration the SM radiation will dominate the energy density, leading to a smooth transition to a RD Universe. Kination ($w=1$) is the prime example of this scenario. Alternatively, the period of nonstandard cosmology could end due to the decays of $\phi$ to lighter degrees of freedom which are produced relativistically. This leads to a rather more abrupt exit from the era of nonstandard expansion which coincides with $3H\sim\Gamma_\phi$. For scenarios with $w<1/3$  (such as an EMD epoch) decays of $\phi$ provide the principle route to a RD Universe prior to BBN.

A common model assumption is that the $\phi$ states can decay to SM particles (and potentially also states beyond the SM). Note that if the decay products of $\phi$ are produced with relativistic momenta, then when the population of $\phi$ decay at $H\sim\Gamma_\phi$ this causes the Universe to transition from a $\phi$-sourced era of nonstandard cosmology to a RD Universe. Importantly, for successful BBN the temperature at the end of the $\rho_\phi$ dominated phase must satisfy $T_{\rm reh}\gtrsim \mathcal{O}(1)$~MeV
(see Sec.~\ref{sec:CMB_BBN_constr}), where $T_{\rm reh}$ is determined by the decay rate, $\Gamma_\phi=3H(T=T_{\rm reh})$, giving
\begin{equation}\label{eq:Tend}
    T_{\rm reh}^4\equiv\frac{5}{4\pi^3\,g_*(T_{\rm reh})}\,M_P^2\,\Gamma_\phi^2\,.
\end{equation}
Note that if $T_{\rm reh}$ is below the electroweak phase transition temperature, one must address questions regarding the origin of the baryon asymmetry; for discussion and potential resolutions see~e.g. Refs.~\cite{Davidson:2000dw, Allahverdi:2010im,Randall:2015xza,Bernal:2017zvx} and Sec.~\ref{sec:PTsandBaryogenesis}.

This transition to a RD Universe can have a number of important consequences, most prominently the dilution of frozen-out cosmological abundances and the nonthermal production of states, which we discuss next. For a particle species $X$ which is decoupled from the thermal bath its comoving number density $Y\equiv n_X/s$ is typically expected to be constant. However, the decays of $\phi$ can lead to changes in the comoving number density of decoupled species. The first possibility is that $n_X$ is increased due to (nonthermal) production of $X$ via direct $\phi$ decays to these $X$ states. In particular, it has been observed that by adjusting the $\phi$ branching fraction to DM one can reproduce the observed relic density in a large range of scenarios~\cite{Moroi:1999zb,Jeannerot:1999yn,Fujii:2002kr,Gelmini:2006pw,Acharya:2008bk,Acharya:2009zt,Chu:2013jja,Moroi:2013sla,Baer:2014eja,Kane:2015qea,Drees:2017iod,Bernal:2018hjm,Kaneta:2019zgw,Allahverdi:2019jsc}. 

Alternatively, if $\phi$ decays dominantly to SM states then this leads to heating of the thermal bath relative to the decoupled species, breaking the common scaling of $n$ and $s$~\cite{Scherrer:1984fd,Chung:1998rq,Giudice:2000ex,McDonald:1989jd}.  This scenario is often referred to as an ``entropy injection'' (or ``entropy dump'') since in the instantaneous $\phi$ decay approximation it appears to imply a jump in the entropy density, although in actuality the temperature of the bath simply cools more slowly due to $\phi$ decays and at no point does the temperature increase~\cite{Scherrer:1984fd}. Notably, similar to nonthermal production one can adjust the magnitude of dilution due to entropy injections to obtain the observed DM relic abundance in many scenarios. Indeed, this is the premise behind Flooded DM~\cite{Randall:2015xza} and is also used in related scenarios which use entropy production to adjust the DM abundance after thermal freeze-out \cite{Scherrer:1984fd, Asaka:2006ek, Patwardhan:2015kga, Berlin:2016vnh, Tenkanen:2016jic, Berlin:2016gtr, Hamdan:2017psw,Hardy:2018bph, Bernal:2018ins,Chanda:2019xyl}. A dedicated discussion on the consequences of nonstandard expansion phases on DM models is presented in Sec.~\ref{sec:dark_matter}

The magnitude of the entropy injection during the transition to RD epoch due to $\phi$ decays depends on the  contribution to the total energy density from the $\phi$ states relative to the SM radiation.  Assuming the instantaneous decay approximation, in which all $\phi$ states decay simultaneously at $T=T_{\rm reh}$, the entropy prior and after decays are related via
\begin{equation}
\frac{s_{\rm before}}{s_{\rm after}}=\left[\frac{\rho_{R}(T_{\rm reh})}{\rho_\phi(T_{\rm reh})}\right]^{3/4},
\label{zeta}
\end{equation}
up to variations of the relativistic degrees of freedom.
Instantaneous $\phi$ decays is a reasonable approximation provided no physical processes of interest (e.g.~freeze-out) occur near the point of $\phi$ decays~\cite{Scherrer:1984fd}. Since it is assumed that below some critical temperature $T_*$ the  $\phi$ energy density starts evolving as $a^{-3(1+w)}$, whereas the radiation continues to evolve as $a^{-4}$, over time the $\rho_\phi$ grows relative to $\rho_R$. Thus to calculate the entropy injection one needs to track both the initial contributions to the energy density at $T=T_*$ and the duration from this point to $T=T_{\rm reh}$. Specifically, these energy densities can be followed from inspection of the Friedmann equation (for $H(T_*)>H(T)>\Gamma_{\phi}$)
\begin{equation}
	H(a)^2= H(a_*)^2\left[\mathcal{R}_\phi\left(\frac{1}{\Delta a}\right)^{3(1+w)}+\mathcal{R}_{R}\left(\frac{1}{\Delta a}\right)^{4}\right],
\label{eq:FE}
\end{equation}
where $\Delta a(T)\equiv a(T)/ a(T_*)$ and $\mathcal{R}_i\equiv \rho_i/(\rho_\phi+\rho_{R})\big|_{T=T_*}$.

Provided that $\phi$ is sufficiently long lived that it becomes the dominant energy density component, then between $T=T_*$ and $T=T_{\rm reh}$ the scale factor grows by the following factor
\begin{equation}
\Delta a_\Gamma\equiv \Delta a(T_{\rm reh}) \simeq \left[\frac{H(a_*)^2\,\mathcal{R}_\phi}{\Gamma_\phi^2}\right]^{\frac{1}{3(1+w)}}.
\end{equation}
If one then assumes that prior to $T=T_*$ the $\phi$ are relativistic then $\rho_{\phi}(T_*)  =\rho_R(T_*)\,\mathcal{R}_\phi/\mathcal{R}_R$. Further supposing all the energy in $\phi$ is transferred to radiation when it decays, then it follows that the magnitude of the entropy dump is 
\begin{equation}
	\Delta s= \left[\frac{\rho_{R}(T_{\rm reh})}{\rho_\phi(T_{\rm reh})}\right]^{3/4}
=
	\left[\frac{\rho_{R}(T_*)}{\rho_{\phi}(T_*)}\frac{\Delta a_{\Gamma}^{-4}}{\Delta a_{\Gamma}^{-3(1+w)}}\right]^{3/4}
=
	\left[\frac{\mathcal{R}_{R}}{\mathcal{R}_\phi}\,\frac{1}{\Delta a_{\Gamma}^{1-3w}}\right]^{3/4}
\simeq
	\frac{\mathcal{R}_{R}^{3/4}}{\mathcal{R}_\phi^\frac{1}{1+w}}\,\left[\frac{T_{\rm reh}}{T_*}\right]^\frac{1-3w}{1+w}.
\label{1.3e9}
\end{equation}
This ratio is important since it implies that for decoupled particle species which are not significantly produced due to $\phi$ decays, then the associated comoving number density will be reduced by a (potentially large) factor $\Delta s$ as the states are diluted relative to the SM radiation. Note that this dilution also applies to particle  asymmetries~\cite{Bramante:2017obj}. In particular, it dilutes any pre-existing baryon asymmetry in the Universe. This is discussed in detail in Sec.~\ref{sec:PTsandBaryogenesis}.

We highlight that in determining the magnitude of the entropy injection the temperature of the visible sector bath at which $\rho_\phi$ starts evolving as $a^{-3(1+w)}$, denoted $T_*$, is a critical parameter. In the simple case where $\phi$ is initially in thermal contact with the SM prior to decoupling and then becomes nonrelativistic, it is reasonable to identify $T_*\sim m_\phi$. However, if the $\phi$ decouple very early, or are never in thermal contact, the $\phi$ could well be thermally distributed with a different temperature and $T_*$ is a free parameter. We shall discuss establishing temperature difference between sectors shortly. Another important example is the case in which $\phi$ is a bosonic field $a$ oscillating in its potential, such that the average equation of state is $w \simeq 0$ (matter-like), e.g.~an axion-like particle (ALP) or any scalar field oscillating in a quadratic potential.
The ALP case is distinct from the heavy particle scenario since $a$ becomes matter-like once it begins to oscillate around its minimum which occurs for ALP mass $m_a \sim H$~\cite{Preskill:1982cy,Abbott:1982af}, corresponding to $T_*\sim \sqrt{3 m_{a}\, M_{\rm P}}$, for a simple quadratic potential $V(a) \supset m_a^2\,a^2$.

\subsubsection{Nonstandard cosmology beyond heavy particles} 

While our discussion thus far has been in the context of heavy elementary particles, the state responsible for determining the equation of state of the Universe need not be elementary and could be composite states, nonperturbative states~\cite{Cotner:2016cvr,Kusenko:2008zm}, or other forms of matter. For instance, a natural proposal is that a period of MD could arise if the energy density of the Universe is mainly composed of PBHs which subsequently evaporate.  Requiring that the Universe transitions to RD epoch prior to BBN implies an upper mass bound on the PBHs, around $6\times 10^5$~kg~\cite{Hooper:2019gtx}. Additionally, it was recently pointed out that PBHs could be very efficiently produced from the preheating instability so that they dominate the energy density of the Universe, and therefore the reheating no longer occurs because of inflaton decays, but rather due to evaporation of the PBHs~\cite{Martin:2019nuw, Martin:2020fgl}. PBHs are discussed further in Sec. \ref{sec:formationPBHs}.

Dark sectors\footnote{It is common to refer the set of states which are charged under the SM gauge groups, or otherwise with significant couplings to SM states, as the {\em visible sector}.  In contrast to this a {\em dark sector} (or hidden sector)  typically refers to a set of states which are neutral singlets under the SM gauge groups and with weak direct coupling to SM states.} 
offer another interesting alternative through which to introduce nonstandard elements to cosmology. Although the dark and visible sectors could be in equilibrium in the early Universe, this need not necessarily be the case and in the converse scenario there can potentially be significant temperature differences between the two sectors.  Such temperature differences can arise due to asymmetric inflationary reheating of sectors~\cite{Adshead:2016xxj,Hardy:2017wkr,Adshead:2019uwj}. Moreover, even in the case that the visible and dark sector are initially  decoupled with similar temperatures, subsequently evolution can lead to  temperature differences.  Examples in which decoupled dark sectors dynamically develop different temperatures include  self-heating via semi-annihilation processes~\cite{Kamada:2017gfc} or self-interactions~\cite{Carlson:1992fn,Kamada:2018hte,Bernal:2020gzm}, also via number-depleting annihilations via 3-to-2~\cite{Carlson:1992fn,Dolgov:1980uu,Pappadopulo:2016pkp} or 4-to-2~\cite{Bernal:2015xba, Heikinheimo:2016yds,Bernal:2017mqb,Heikinheimo:2017ofk} processes, and in co-decaying DM models~\cite{Dror:2016rxc}.
 
To appreciate (one reason) why temperature differences between sectors can be  important, consider the case in which the temperature of $\phi$ states $T_{\phi}$ is significantly lower than that of the SM photon temperature $T$. Then the $\phi$ become nonrelativistic at $T_\phi\sim m_\phi$ which could be much earlier than the point at which $T\sim m_\phi$.  Thus the states $\phi$ need not necessarily be extremely heavy, relative to SM states, in order to give rise to an EMD era. Furthermore, the temperature of the visible sector at which $\phi$ becomes nonrelativistic corresponds to $T_*$ in Eq.~\eqref{1.3e9} and, notably, if $T_*$ can be treated as a (largely) free parameter this permits a large range in the magnitude of the entropy injection due to $\phi$ decays, and increases the potential impact of dilution on decoupled particle species.

\subsection{Moduli fields (Authors: K. Sinha \& S. Watson)}
\label{sec:moduli}

String theories and M-theory compactifications yield a large number of scalar fields, or moduli.  A well-studied example is  Calabi-Yau compactification where these moduli correspond to the complex structure and K\"ahler, also below deformations of the internal manifold \cite{Greene:1996cy, Aspinwall:2004jr}. Generating masses for moduli has been a central program in string phenomenology and cosmology \cite{Giddings:2001yu,Silverstein:2004id,Grana:2005jc, Kachru:2003aw, Douglas:2006es}. This is necessitated by several considerations: massless scalars lead to long range interactions which are severely constrained by ``fifth force" experiments. Moreover, if the couplings of moduli to different sectors of matter fields are non-universal, the corresponding long-range forces would violate the equivalence principle. An additional question is that of predictability: the vacuum expectation values of moduli determine the parameters of the low-energy theory and thus should be fixed. Moduli stabilization has been a central program in string theory over several decades and the construction of the effective potential draws upon many ingredients that are central to  string theory: branes, fluxes, and strong dynamics in the hidden sector. Explicit constructions are still being actively discussed by the community \cite{Dasgupta:2019gcd, Randall:2019ent, Cribiori:2019hrb,Garg:2018reu,Ooguri:2018wrx}. 

From the perspective of low-energy phenomenology, an important question about moduli stabilization is the resulting mass $m_\sigma$ of the modulus $\sigma$. The physics of moduli  should decouple  at scales $E \ll m_\sigma$. Nevertheless, decoupled moduli can still impact the pattern of the  low-energy particle spectrum in many situations, for example when they are responsible for  supersymmetry breaking \cite{Conlon:2007xv,Acharya:2008zi,Heckman:2008qt,Dutta:2009uf}. On the other hand, the impact of light moduli on the evolution of the early Universe can also lead to many different observational predictions \cite{Acharya:2008bk,Heckman:2008jy,Easther:2013nga,Kane:2015jia}. In the early Universe, moduli are typically displaced from the minimum of their effective potential and undergo coherent oscillations that can mimic a MD era \cite{Kane:2015jia, Acharya:2009zt}.  Since moduli couple to other fields gravitationally, their decay width is given parametrically as $ \Gamma_\sigma \sim m_\sigma^3 / M^2_{\rm P}$. Depending on the mass $m_\sigma$, moduli can decay late enough to affect the successes of BBN. This is the so-called cosmological moduli problem, which first appeared as the Polonyi problem in early versions of spontaneously broken supergravity. Stated most generally, the problem is that supersymmetry breaking in the hidden sector with gravity mediation often contains scalar fields that have weak scale masses and couple gravitationally. After inflation, these fields  behave like nonrelativistic matter and dominate the energy density of the Universe until after BBN, spoiling its successful predictions \cite{Banks:1995dp,Banks:1995dt,Coughlan:1983ci,Banks:1993en}.

The cosmological moduli problem may actually be more of an opportunity if the masses of the moduli result in decays prior to BBN. Indeed, this situation leads to new cosmological predictions and an alternative to a standard thermal history. That is, if a modulus is massive enough to decay before BBN, but not so massive as to decouple entirely from low energy physics (a situation that commonly occurs in UV complete examples), its effects on DM, baryogenesis, and structure formation can lead to novel and interesting predictions~\cite{Kane:2015jia}. The late decay of moduli can be viewed as a second reheating of the Universe (after the first, inflationary one), with the production of relativistic SM and other light (beyond the SM) particles. Depending on the masses and couplings of the modulus to other fields, such as DM and baryons, this framework can lead to new expectations for post-inflationary cosmology prior to BBN.

The critical question then is: what is the mass $m_\sigma$  of a modulus in string theory, or, framed another way, what is the behaviour of the curvature of the effective potential near its minimum? Given that the scales relevant in string theory are the Planck scale, the string tension, and the typical size of the extra dimensions (the Kaluza-Klein scale), one would expect a priori that $m_\sigma$ is determined by a combination of these scales and that there is no obvious role of a low-energy scale in the stabilization mechanism. If that were the case, moduli would be too massive and decay too early to leave discernible imprints on pre-BBN cosmology. 

In explicitly constructed examples of moduli stablization, a common result is that some moduli remain parametrically lighter than the string scale. This can be understood at the level of effective supergravity, since the key is the connection between moduli stabilization and supersymmetry breaking. The effective potential determines the moduli F-terms dynamically. This, combined with the K\"ahler metric of the matter sector, fixes the strength of gravity mediation. More concretely, the scalar potential  in the supergravity limit is given by
\begin{equation} \label{GSWKS2}
V=e^{K/M^2_{\rm P}}  \left( \sum_\sigma \left| D_\sigma W  \right\vert^2 - {\frac{3}{M^2_{\rm P}} |W|^2} \right) \,\,,
\end{equation}
where $W(\sigma)$ and $K(\sigma)$ are the  superpotential and K\"ahler potential, respectively. The gravitino mass is defined as $m^2_{3/2} = \langle e^{K/M^2_{\rm P}} |W|^2/M^4_{\rm P} \rangle$. At the extremum, $\partial_{\sigma} V = 0$ and $\partial_{\sigma \sigma} V \, > \, 0$. Moreover, an almost vanishing cosmological constant requires  $V \simeq 0$. Then, both terms in Eq.~\eqref{GSWKS2} are  forced to be $\sim m^2_{3/2}$. Since the modulus mass matrix is positive definite, the smallest eigenvalue should then be $\mathcal{O}(m_{3/2})$. For a model with a single modulus with nonvanishing F-term, the trivial conclusion is  $\partial_{\sigma \sigma} V \, \sim \, m^2_{3/2}$. 

This set of conditions is restrictive enough to yield $m_{\sigma} \sim m_{3/2}$. Indeed, taking into account a redefinition to obtain canonical kinetic terms, one obtains \cite{Kane:2015jia} 
\begin{equation} \label{GSWKS3}
m_{\sigma} \, \sim \, \langle \sigma \rangle \, m_{3/2} \,\,\sim \,\, \mathcal{O}(1-100) \, \times \, m_{3/2}.
\end{equation}
We note that to remain in the supergravity limit, stabilization schemes typically require $ \langle \sigma \rangle \, \gtrsim \mathcal{O}(1)$ in Planck units. The exact hierarchy between the mass of the modulus and the gravitino mass, however, depends on the specific stabilization scheme. For example, in M-theory compactifications the pre-factor is $\mathcal{O}(1)$, while for type IIB Kachru-Kallosh-Linde-Trivedi (KKLT) compactifications the pre-factor is $\mathcal{O}(10-100)$.

The corresponding reheat temperature is given by $T_{\rm reh} \sim g_*^{-1/4} \sqrt{\Gamma_\sigma \, M_{\rm P}} $ or
\begin{equation}
 \label{GSWKS4}
T_{\rm reh} = c^{1/2} \left(\frac{10.75}{g_*}\right)^{1/4} \left( \frac{m_{\sigma}}{50\, {\rm TeV}}\right)^{3/2}\, {\rm MeV} \,, 
\end{equation}
where $g_*$ is the number of relativistic degrees of freedom at $T_{\rm reh}$, and $c$ is a model-dependent $\mathcal{O}(0.1-1)$ number. It is intriguing that the lower bound on the modulus mass required to avoid the cosmological moduli problem corresponds to the gravitino mass scale required to address the electroweak hierarchy problem.

Although the appeal to supergravity shown above is suggestive, the expectation of light moduli at the TeV scale does not necessarily depend on having low-scale supersymmetry as a solution to the hierarchy problem. Statistical arguments in the type IIB string landscape indicate that the distribution of moduli masses is logarithmic in vacua with stabilized K\"ahler moduli  \cite{Broeckel:2020fdz}. Since explicit compactifications feature moduli in the mass range we have discussed, one can expect a preponderance of nearby vacua with similar moduli masses. To our knowledge such statistical arguments have not been advanced for M-theory compactifications, but similar results can be expected to hold in those cases as well. Indeed, there are indications from recent work that very general features of string compactifications lead to the existence of moduli with masses exponentially suppressed compared to the Planck scale. In explicit constructions, the moduli masses arise from exponentially suppressed  contributions to the superpotential coming from non-perturbative effects (in KKLT, such contributions are essential to obtain a small value of the Gukov-Vafa-Witten flux superpotential $W_0$ \cite{Demirtas:2019sip}; in Large Volume Scenarios, they occur in the superpotential term for the K\"ahler moduli).

Another interesting result from these models is that the reheat temperature is deeply connected to fundamental physics and is not a free parameter. Since moduli are ubiquitous in string theory, a nonstandard cosmological history thus appears to be a robust possibility and perhaps an inevitable one. Of course, connecting the physics of moduli to a consistent nonstandard cosmological history presents many challenges. We now turn to a discussion of some of these issues.

One class of challenges lies in consistent embeddings of the matter sector into a given string compactification and the  interactions of various particles with moduli. For example, the moduli-induced gravitino problem \cite{Endo:2006zj, Allahverdi:2013noa, Acharya:2008bk} arises because gravitinos produced from moduli themselves subsequently decay and produce DM in supersymmetric models. This tends to overclose the Universe unless the decay of the modulus to the gravitino is suppressed due to branching or kinematics. Whether or not such blocking is possible is a model-dependent question. Similarly, the moduli-induced dark radiation problem arises from the fact that moduli are either accompanied by light pseudoscalars in typical string compactifications, or themselves decay to such axion-like-particles \cite{Allahverdi:2014ppa, Acharya:2018deu, Acharya:2015zfk}. The existence of such relativistic species in the dark sector is subject to $\Delta N_{\rm eff}$ constraints from cosmological observations, in turn imposing constraints on the moduli physics. Another class of challenges lies in various features of the low-energy effective potential of the modulus. An example of this is the ``overshoot problem": since moduli are displaced from their minima during inflation, this displacement should not be such that a modulus overcomes the barrier protecting the local low-energy minimum. This problem is again model-dependent and depends on the stabilization mechanism, and may suggest a correlation of the scales of supersymmetry breaking (the barrier height) and inflation \cite{Kallosh:2004yh, He:2010uk}. Many of these issues require further theoretical investigation.  Regardless, we have seen that a nonstandard cosmological history is a robust prediction of string/M-theoretic approaches to beyond-SM physics.

There can be a number of effects of a decaying modulus on early Universe cosmology: nonthermal production of DM \cite{Kane:2009if, Allahverdi:2013noa}, late-time baryogenesis \cite{Allahverdi:2010rh, Allahverdi:2010im, Kane:2011ih}, effects on the CMB~\cite{Easther:2013nga} and LSS \cite{Iliesiu:2013rqa}, and implications for the matter power spectrum \cite{Fan:2014zua}, effective number of neutrino species \cite{Allahverdi:2014ppa}, and isocurvature constraints. We briefly discuss some of these topics, leaving a more detailed discussion for other authors in this review.

\textbf{Gravitational effects: matter power spectrum and CMB:}  We first discuss the effect of moduli on the DM halo. In a RD Universe, DM perturbations $\delta \rho / \rho$ grow logarithmically with the scale factor. In a matter-dominated Universe, on the other hand, they grow linearly: $\delta \rho / \rho \sim a(t) \sim t^{2/3}$. This implies that while in standard cosmology  the growth of structure only becomes significant after matter-radiation equality, an early MD era with cosmological moduli results in an early period of growth and a new scale at smaller wavelengths in the matter power spectrum. For $T_{\rm reh}$ near the BBN limit, the result may be the formation of Earth-sized, ultracompact microhalos \cite{Fan:2014zua,Erickcek:2011us,Erickcek:2015bda}. If the DM decouples both thermally and kinetically prior to the subsequent onset of RD epoch, this enhancement of the small-scale matter power spectrum may survive, with profound consequences for WIMP detection \cite{Erickcek:2015bda,Erickcek:2015jza, Blanco:2019eij} and axion cosmology \cite{Nelson:2018via}. For a more detailed discussion on effects on the matter power spectrum, see Sec. \ref{sec:microhalos} and Sec. \ref{sec:microhalo_constraints}.

We next turn to possible effects of moduli on the CMB \cite{Easther:2013nga}. The main point here is that the equation of state (e.g. thermal $w = 1/3$ or nonthermal $w = 0$) in the aftermath of inflation determines the expansion rate of the Universe. Thus, the rate at which primordial perturbations re-enter the Hubble horizon is impacted by an era when moduli dominate the energy density. The total number of e-folds is given by $N_k = {\rm ln} (a_{\rm end}/a_k)$,  where $a_{\rm end}$ and $a_k$ are the scale factors at the end of inflation and when the pivot scale exited the horizon, respectively. It is dependent on, apart from the inflationary potential, the value of the energy density at the end of inflation and the energy density at which the Universe is assumed to become thermalized. The energy densities in turn depend on the mass and reheating temperature of the modulus. Thus, the properties of the modulus sector impact inflationary observables through $N_k$ and conversely, CMB observations can be used to constrain the modulus sector. For a more detailed discussion on inflationary observables, see Sec. \ref{sec:inflation}.

\textbf{Nonthermal DM:} Moduli-dominated cosmological histories have a major impact on the standard thermal freeze-out paradigm of DM. The reason is that the  decay of the modulus dilutes any previously existing population of DM by a factor $\Omega_{\rm CDM} \rightarrow \Omega_{\rm CDM} (T_{\rm reh}/T_f)^3$, where $T_f$ is the DM freeze-out temperature. DM is produced during the modulus reheating process at low temperatures $\sim T_{\rm reh}$. If the number density of DM particles produced exceeds the critical value $n^c_x =(H/ \langle \sigma_xv \rangle )_{T=T_{\rm reh}}$, then the particles quickly annihilate down to this value, which is an attractor. The result is a parametric enhancement of the relic density $\Omega^{NT}_{\rm DM} \, = \, \Omega^{T}_{\rm DM} \left(T_f/T_{\rm reh} \right)$. If, on the other hand, the number density of DM particles produced from modulus decay is lower than the attractor value, no further annihilation takes place and the final relic density is simply given by $n_{\rm DM} \sim n_\sigma \times {\rm Br}_{\sigma \rightarrow {\rm DM}}  $, where ${\rm Br}_{\sigma \rightarrow {\rm DM}}$ is the branching ratio for modulus decay to DM and $n_\sigma $ is the number density of the scalar condensate. Due to the extra freedom in choosing $T_{\rm reh}$ in the first case and ${\rm Br}_{\sigma \rightarrow {\rm DM}}$ and $n_\sigma$ in the second, a much wider range of DM masses and annihilation cross-sections can satisfy the observed relic density constraint compared to the thermal freeze-out scenario. For a more detailed discussion on DM, see Sec. \ref{sec:dark_sectors} and Sec. \ref{sec:dark_matter}.

\textbf{Baryogenesis:} Just as in the case of DM, the decay of a modulus dilutes baryon asymmetry existing from any previous era and has to be regenerated. This turns out to be quite natural. The decay of the modulus provides the non-equilibrium condition required for baryogenesis and a typical value for the dilution factor\footnote{The dilution factor is defined as $Y \equiv n_\sigma/s = 3 T_{\rm reh}/(4 m_\sigma)$, where the last expression can be obtained directly from the Boltzmann equation for $n_\sigma$.} is $Y \sim 10^{-8}$. If the modulus decays to a field that has $CP$- and $B$- violating couplings to SM fields, baryogenesis can occur through usual interference between tree and loop-level diagrams. These interference terms are usually suppressed by factors like $(1/4\pi)$, so that a net factor of $\sim 10^{-10}$ for the baryon asymmetry can be obtained with $\mathcal{O}(1)$ Yukawa couplings \cite{Allahverdi:2010rh, Allahverdi:2010im}. For a more detailed discussion on baryogenesis, see Sec. \ref{sec:PTsandBaryogenesis}.

\subsection{Post-BBN modifications (Authors: T. Karwal \& T. L. Smith)}
\label{sec:postBBN}

In this section, we discuss causes of nonstandard, post-BBN expansion history. We outline examples of modifications to the properties of the standard \LCDM\ components, and scenarios which introduce new components altogether. 

These two categories can be further refined into modifications to the background and/or perturbative evolution. The conservation of stress-energy dictates how all minimally coupled  components evolve in both the (homogeneous and isotropic) background as well as at linear order in their inhomogeneous perturbations (see e.g. Ref.~\cite{Ma:1995ey}):
\begin{eqnarray}
    \dot{\bar{\rho}} &=& - 3 \frac{\dot a}{a} \bar \rho (1+w) \,,
    \label{eq:bkgcont} \\
    \dot \delta &=& - (1+w)
    (\theta + 3 \dot \phi) - 3 \frac{\dot a}{a}\left(c_s^2 - w\right) \delta \,,
    \label{eq:pertcont}\\
    \dot \theta &=& - \frac{\dot a}{a}(1-3w) \theta - \frac{\dot w}{1+w} \theta + \frac{c_s^2}{1+w} k^2 \delta - k^2 \Sigma + k^2 \psi \,,
    \label{eq:euler}
\end{eqnarray}
where dots indicate derivatives with respect to conformal time, $\bar \rho$ is the fluid's homogeneous density, $w \equiv \bar P/\bar \rho$ is its equation of state parameter, $\delta \equiv \delta \rho/\bar \rho$ is its density contrast, $\theta$ is the divergence of its velocity perturbation, $c_s^2 \equiv \delta P/\delta \rho$ is its sound speed, $\phi$ and $\psi$ are the temporal and spatial gravitational metrics in conformal Newtonian gauge, and $\Sigma$ is the fluid's anisotropic stress.

The simplest modifications to the \LCDM\ expansion history can be characterised as changes to $w$ in Eq.~\eqref{eq:bkgcont}. Models which do this can generically be characterized by the way in which the equation of state parameter, $w$, changes with time. For example, quintessence models which promote the cosmological ``constant'' to a dynamical quantity predict some time variation in the dark energy equation of state, $w_{\rm de}$, affecting the late-time expansion history relative to ``$\Lambda$''-CDM. Particular quintessence models have the property that $w_{\rm de}$ tracks the matter equation of state parameter, $w_{\rm m}$ \cite{Steinhardt:1999nw,Doran:2006kp,Calabrese:2011hg,Ahmed:2002mj}. These ``tracker'' models are particularly interesting as potential solutions to the coincidence problem of dark energy. Such tracker dark energies can arise from various mechanisms including scalar fields with particular potentials \cite{Doran:2006kp}, quintessence and k-essence models \cite{Scherrer:2005je,Scherrer:2004au,Linder:2008ya} and alternative models of gravity \cite{Ahmed:2002mj}. Models which propose additional forms of energy density, beyond what is contained in \LCDM, can also modify the post-BBN expansion history. A universe which contains a collection of scalar fields may undergo periodic epochs of anomalous expansion (see e.g. Refs.~\cite{Griest:2002cu,Kamionkowski:2014zda}). Some models of minimally coupled ``early'' dark energy (EDE) envision a scalar field which is fixed at some point in its potential by Hubble friction which becomes dynamical when the Hubble parameter drops below some critical value around the matter-radiation equality \cite{Karwal:2016vyq,Smith:2019ihp}. The fractional contribution of this field to the total energy density briefly increases and then decreases, as long as it dilutes faster than matter. Other models propose a nonminimal coupling to other components (such as the neutrino sector) which then produces a ``kick'' to the scalar field, again leading to a brief increase in the total energy density of the Universe \cite{Sakstein:2019fmf}. Due to its success at providing a compelling resolution to the Hubble tension \cite{Poulin:2018cxd}, numerous particle models have now been proposed to comprise of such EDEs, including axion-like \cite{Poulin:2018dzj,Alexander:2019rsc,Berghaus:2019cls,Smith:2019ihp} and power-law scalar fields \cite{Agrawal:2019lmo} with nonstandard kinetic terms \cite{Lin:2019qug}, and first-order phase transitions in the dark sector \cite{Niedermann:2019olb}. 

Several models of nonstandard expansion also arise due to interactions between the different components that fill the Universe.
Conservation of stress-energy in the presence of exchange of energy-momentum between different components ($X$ and $Y$) can be written as (see e.g. Ref.~\cite{Byrnes:2010ft})
\begin{equation}
\nabla_\mu T^\mu_{X \nu} = Q_\nu\,, \ \ \nabla_\mu T^\mu_{Y \nu} = -Q_\nu,
\end{equation}
where $Q_\nu$ is the energy-momentum flux exchanged between the two components. The form of $Q_\nu$ depends on the specific physics of the coupling, but a natural choice is to write the energy-momentum transfer as a linear combination of the four-velocities \cite{1984PThPS..78....1K}
\begin{equation}
    Q_\nu = \zeta_X u_{X\nu} + \zeta_Y u_{Y \nu} \,,
    \label{eq:int}
\end{equation}
where $u_{(X,Y)\nu}$ is the four-velocity of the components. 
Models with interactions of this form include interactions between dark matter and dark energy \cite{Farrar:2003uw}, developed in order to address the coincidence problem \cite{Cai:2004dk,He:2008tn}. A similar inter-species interaction is DM decaying into dark radiation \cite{Poulin:2016nat,Pandey:2019plg}, 
invoked to explain the $\sigma_8$ and Hubble tensions. 
While there are stringent bounds on all DM behaving this way, a fraction of DM annihilating into dark radiation is permitted by cosmological data. 
Depending on the rate $\Gamma$ of decay of the DM, when the expansion rate $H \sim \Gamma$, all the interacting DM decays into dark radiation over one cosmic time scale, which rapidly dilutes away. 
If this event occurs during MD, the expansion of the Universe is suddenly slowed by a factor roughly proportional to the square-root of the fraction of decaying DM, and then follows the usual dilution of matter density again. 
Close to matter-radiation equality or matter-dark-energy equality, such an event can shift the redshift of equality. 
At other times, when matter is subdominant, the effects of decaying DM on the background evolution of the Universe are minimised, with the principal effects being confined to the evolution of perturbations. 

Given the form of the energy-momentum exchange rate in Eq.~\eqref{eq:int}, at the background level we have
\begin{eqnarray}
\label{Qeq}
    \dot{\bar{\rho}}_X + 3 H \bar{\rho}_X(1+w_X) &=& Q_0\,,\\ \nonumber
    \dot{\bar{\rho}}_Y + 3 H \bar{\rho}_Y(1+w_Y) &=& -Q_0 \,,
\end{eqnarray}
which generalize Eq.~\eqref{eq:rhophigamma}. The impact of this interaction on the expansion rate can be more simply understood once we realize that the sum of the continuity equations for the interacting components is, itself, conserved. 
Therefore, at the background level, the two components behave as a single, effective, component with an effective equation of state parameter $w_{\rm eff} \equiv (\bar \rho_X w_X + \bar \rho_Y w_Y)/(\bar \rho_X + \bar \rho_Y)$. In the noninteracting case, $Q_0 = 0$, for constant equation of state parameters, we know that $\rho_{X,Y} \propto a^{-3(1+w_{X,Y})}$ so that, taking $w_X>w_Y$, $w_{\rm eff}$ monotonically transitions from $w_X$ to $w_Y$ as the Universe expands. 
In the interacting case, $Q_0 \neq 0$, as energy density flows from one component to the other, $w_{\rm eff}$ will deviate from a monotonic evolution, leading to a change in the expansion history. Note that it is possible to have $w_X < -1$ (for example, a scalar field with a non-canonical kinetic term, e.g., Ref.~\cite{Guo:2004fq}) in which case $w_{\rm eff} < -1$. In this case we can still model the background and perturbative dynamics as an effective fluid \cite{Hu:2004kh}.
A similar effect is seen when working out the dynamics in some modified gravity theories. 
For example, we can write any scalar-tensor gravity theory in the Einstein frame, in which the modified expansion is due to a nonminimal coupling between the scalar field and other components \cite{Damour:1992we}. 
In addition to this, an agnostic treatment of the microscopic dynamics of dark energy makes it indistinguishable from modified gravity \cite{Bertschinger:2008zb}. This effective treatment of modified gravity and nonstandard dark energy interactions is the main focus of the ``parameterized post-Friedmann'' formalism \cite{Hu:2007pj}. 

Finally, DM-baryon interactions \cite{Dvorkin:2013cea,Boddy:2018wzy,Munoz:2015bca} have been recently considered as a potential explanation of the anomalous measurements from EDGES \cite{Bowman:2018yin}. 
In these models the DM-baryon cross-section behaves as $\sigma= \sigma_0 v^n$, where $v$ is the relative velocity bewteen the scattering particles. This velocity dependence may arise in several natural interactions, such as electric- or magnetic-dipole interactions through light mediators ($n=-2$) or Coulomb-like interactions or Yukawa interactions through light mediators ($n=-4$). 

It is important to note that when the background equations are modified, 
in order to maintain energy-momentum conservation at the linear level, 
the perturbation equations must also be modified. 
However, there are interactions that preserve the background equations and just modify the evolution of the linear perturbations.
Thomson scattering between photons and baryons and any self-interactions are examples of such mechanisms.  In the remainder of this section we will consider the effects of scattering so that the number of each constituent remains unchanged. 

For a self-interacting relativistic component \cite{Park:2019ibn,Kreisch:2019yzn}, the linear conservation equations [Eqs.~\eqref{eq:pertcont} and \eqref{eq:euler}], remain unchanged. 
The modifications enter through the evolution of the component's anisotropic stress, $\Sigma$.
The evolution of $\Sigma$ is governed by the Boltzmann equation which, in turn, is sourced by a collision term. In Refs.~\cite{Oldengott:2014qra,Oldengott:2017fhy} it was shown that, in the case of neutrino self-interactions mediated by a massive scalar, the collision integral can be approximated to be of the form $\mathcal{C}[\mathcal{F}] = \mathcal{F}/\dot{\tau}$, where $\mathcal{F}$ is the component's distribution function and $\dot{\tau}$ is the rate at which the component's opacity changes. In the absence of a collision term, the equations of motion for the anisotropic stress $\Sigma$ cause a decrease over time of the density contrast and velocity perturbation (i.e. ``free-streaming''); self-interactions disrupt this and lead to enhanced clustering for that component. On the other hand, self-interacting DM \cite{Spergel:1999mh,Tulin:2017ara} has been considered to alleviate a variety of small-scale structure ``problems''. However, the typical interaction cross-sections that are considered are only active in collapsed, nonlinear, structures and hence have little effect on the linear physics probed by the CMB and large-scale structure observations.

Models which excite initial conditions beyond the standard adiabatic initial conditions also modify the late-time evolution of perturbations. 
Any model of inflation that involves more than one dynamically important scalar field, such as the curvaton scenario \cite{Lyth:2002my} or axion DM \cite{Marsh:2015xka}, produce nonadiabatic initial conditions, which will be discussed in more detail in Sec. \ref{sec:curvaton} (see also Ref.~\cite{Markkanen:2018gcw}). As a specific example, we mention compensated isocurvature perturbations which balance an overdensity of baryons with an underdensity of cold DM and vice-versa (see e.g. Refs.~\cite{Gordon:2009wx, He:2015msa}). This scenario leaves the local matter density perturbations unchanged but affects the plasma sound speed $c_s^2$, leading to a 
smoothing of primordial perturbations. This has been suggested as 
a solution to the $A_{\rm lens}$ anomaly \cite{Munoz:2015fdv,Smith:2015bln} and may partially alleviate the Hubble tension through large-scale fluctuations in $\Omega_{\rm b}$ \cite{Heinrich:2019sxl}. 

Finally, models which promote ``constants'' of nature to dynamical fields may also cause nonstandard cosmological evolution. Scalar-tensor theories of gravity, mentioned before, are an example of this, leading to a dynamical gravitational coupling, $G(\vec x,t)$. 
Other intriguing examples are models of a dynamical fine-structure constant \cite{Bekenstein:1982eu,Barrow:2002db}. 
These models predict a time and spatial variation in the strength of electromagnetic interactions (such as Thomson scattering) which has a unique impact on the structure of the observed CMB anisotropies (see e.g. Ref.~\cite{Smith:2018rnu,Hart:2019dxi}).

\section{Consequences of nonstandard expansion phases}
\label{sec:consequences}

In this section, we will discuss several consequences of nonstandard expansion eras, paying particular attention to models of dark matter, early Universe phase transitions and baryogenesis, models of inflation, and generation of the primordial curvature perturbations, as well as microhalos and PBHs.

\subsection{Consequences for dark matter (Authors: A. Berlin, D. Hooper \& G. Krnjaic)}
\label{sec:dark_matter}

If the early Universe was not consistently dominated by radiation as is typically assumed, the phenomenology of DM production and detection could be substantially altered from standard expectations. In particular, as discussed in Sec.~\ref{sec:dark_sectors}, if the visible sector (which contains the SM) is supplemented by a decoupled hidden (or dark) sector (which contains the DM), these two sectors can be produced independently during post-inflation reheating and maintain separate temperatures throughout cosmological evolution. Within the hidden sector, the DM could annihilate to lighter hidden sector states which ultimately decay into SM particles. If these unstable states are sufficiently heavy and long-lived, they will come to dominate the energy density of the Universe before decaying to SM particles, thereby diluting the abundances of all relics, including that of the DM~\cite{Berlin:2016gtr,Berlin:2016vnh,Tenkanen:2016jic}. This makes it possible to circumvent the well-known $\sim$100 TeV upper limit on the mass of the DM, based on partial wave unitarity~\cite{Griest:1989wd} (for related scenarios, see Refs.~\cite{Dev:2016xcp,Fornengo:2002db,Gelmini:2006pq,Kane:2015jia,Hooper:2013nia,Patwardhan:2015kga,Randall:2015xza,Reece:2015lch,Lyth:1995ka,Davoudiasl:2015vba,Cohen:2008nb,Yamanaka:2014pva,Kamionkowski:1990ni,Giudice:2000ex,Gelmini:2006pw,Grin:2007yg,Acharya:2009zt,Visinelli:2009kt,Kane:2015qea,DEramo:2017gpl,Redmond:2017tja,Visinelli:2017qga,Drees:2017iod,Bernal:2018ins,Hardy:2018bph,Drees:2018dsj,Arbey:2018uho,MB,Bernal:2019uqr,Blinov:2019rhb,Blinov:2019jqc,Gelmini:2019clw,Chanda:2019xyl,Gelmini:2019esj,Han:2019vxi,Arias:2019uol,Harigaya:2019tzu,Chowdhury:2018tzw,Betancur:2018xtj,Bernal:2018hjm}; for alternative ``freeze-in'' production of DM in scenarios with a nonstandard expansion phase, see Refs. \cite{Co:2015pka,Evans:2016zau,Redmond:2017tja,Bernal:2017kxu,DEramo:2017ecx,Visinelli:2017qga,Drees:2017iod,Bernal:2018ins,Bernal:2018kcw,Maldonado:2019qmp,Allahverdi:2019jsc,Bernal:2019mhf,Cosme:2020mck,Bernal:2020bfj,Bernal:2020qyu}).

Long lifetimes are straightforwardly realized if the decaying particle is the lightest hidden sector state. In fact, if the hidden and visible sectors are decoupled, the lightest hidden sector state will automatically be long-lived, since its width relies on a coupling that is too small to sustain thermal equilibrium between the two sectors. This picture is relatively universal, and can be found within any model in which the DM freezes out through annihilations in a heavy and highly decoupled hidden sector that is populated after inflation. In contrast to scenarios in which an additional out-of-equilibrium decay is invoked solely to dilute the initial cosmological abundances of various species, dilutions of this kind are an inevitable consequence of thermal decoupling.

In this framework, the DM coupling to the decaying particle is essential for generating the relic abundance. Initially the DM and lighter long-lived state are in chemical equilibrium until the DM freezes out by annihilating to the lighter state when the dark sector temperature falls below the DM mass. However, unlike in conventional freeze-out, there is no thermal contact with the SM and there is a substantial entropy transfer when the lighter unstable particle eventually dominates the energy density and then heats the SM radiation bath upon decay.

Although this class of scenarios can be realized within the context of a wide of range of hidden sector models, we will consider a simple vector portal model as an illustrative example. For our DM candidate, we introduce a stable Dirac fermion, $X$, which has unit charge under a spontaneously broken $U(1)_X$ gauge symmetry, corresponding to the massive gauge boson, $Z'$. The hidden sector Lagrangian contains
\be
{\cal L} \supset -\frac{\epsilon}{2} B^{\mu\nu}Z'_{\mu \nu} +   g_{\rm DM} Z^{\prime}_{\mu}  \overline X \gamma^\mu X \,,
\ee
where $Z'_{\mu \nu}$ and $B_{\mu\nu}$ are the $U(1)_X$ and hypercharge field strengths, respectively, and $\epsilon$ quantifies the degree of kinetic mixing between them~\cite{Holdom:1985ag,Okun:1982xi}. If $\epsilon$ is small (a choice which is technically natural), and the $Z^\prime$ is the lightest hidden sector particle, it will be long-lived, leading it to dominate the energy density of the Universe before decaying. The DM undergoes the process of thermal freeze-out entirely within the hidden sector, dictated by the cross-section for $X\bar{X}\rightarrow Z'Z'$. For $m_X \gg m_{Z^\prime}$ this cross-section is given by $\sigma v \simeq \pi \alpha^2_X/m^2_X$, where $\alpha_X \equiv g_{\rm DM}^2/4\pi$.

As an initial condition, we take the hidden and visible sectors to be described by separate thermal distributions, with temperatures of $T_{\rm h}$ and $T$, respectively. The ratio of these temperatures, $\xi_{\rm inf} \equiv (T_{\rm h}/T)_{\rm inf}$, is determined by the physics of inflation, including the sectors' respective couplings to the inflaton~\cite{Hodges:1993yb,Berezhiani:1995am}. From entropy conservation in each sector, we can calculate the time evolution of $\xi$ (prior to the decays of $Z'$)
\be
\frac{s_h}{s} = \frac{g^h_{*,s}}{g_{*,s}} \, \xi^3 = {\rm constant} \qquad \longrightarrow \qquad \xi = \xi_{\rm inf} \, \bigg(\frac{g^h_{*,s,{\rm inf}}}{g^h_{*,s}}\bigg)^{1/3} \, \bigg(\frac{g_{*,s}}{g_{*,s,{\rm inf}}}\bigg)^{1/3} \,,
\ee
where $s$ and $s_h$ are the hidden and visible sector entropy densities, $g_{*,s}$ and $g_{*,s}^h$ are the numbers of effective relativistic entropy degrees of freedom in the visible and hidden sectors. If the SM temperature is well above the electroweak scale, $g_{*,s} \simeq g_{*,s,{\rm inf}}$. As the temperature of the hidden sector falls below $m_X$, $g^h_{*,s}$ decreases from $g_{Z'}+(7/8)g_X$ to $g_{Z'}$, bringing $\xi$ from $\xi_{\rm inf}$ to $(13/6)^{1/3} \, \xi_{\rm inf} \approx 1.3 \, \xi_{\rm inf}$, for $m_{Z^\prime}\ll m_X$. 

As the Universe expands, $X$ will eventually freeze-out of chemical equilibrium, yielding a nonnegligible relic abundance. The evolution of the number density of $X$ (plus $\bar{X}$), $n_X$, is described by the Boltzmann equation
\be
\label{eq:Boltz1}
\dot{n}_X + 3 H  n_X = - \frac{1}{2}\langle \sigma v \rangle ~ \left(n_X^2 - \frac{n^2_{Z'}}{\, n^2_{Z', {\text{eq}}}}n_{X,{\text{eq}}}^2 \right) \,,
\ee
where $\langle \sigma v \rangle$ is the thermally averaged cross-section for the process $X\, \bar{X} \rightarrow Z' \, Z'$, and $H = [8 \pi \, (\rho_R + \rho_h)/3 \, M_{\rm P}^2]^{1/2}$ describes the expansion rate of the Universe in terms of the energy densities in the visible and hidden sectors.

\begin{figure}[t]
\begin{center}
\includegraphics[width=0.4\textwidth]{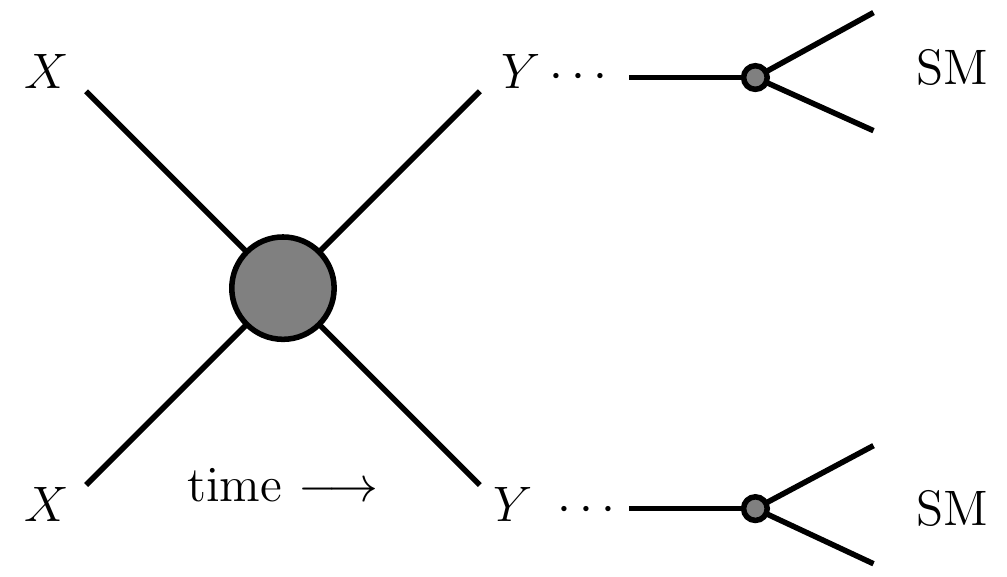}
\caption{\label{fig:schematic} A schematic diagram of the processes being considered in Sec. \ref{sec:dark_matter}. Here $X$, the DM candidate, annihilates (possibly through intermediate processes) into pairs of metastable hidden sector particles, $Y$. If the hidden sector is heavy and extremely decoupled from the visible sector (which contains the SM), then $Y$ will be long-lived, and may eventually dominate the Universe's energy density. Upon its decay into SM particles, $Y$ reheats the visible Universe and dilutes all particle abundances, including the relic density of $X$.}
\label{fig:cartoon}
\end{center}
\end{figure}

\begin{figure}[t]
\includegraphics[width=0.49\textwidth]{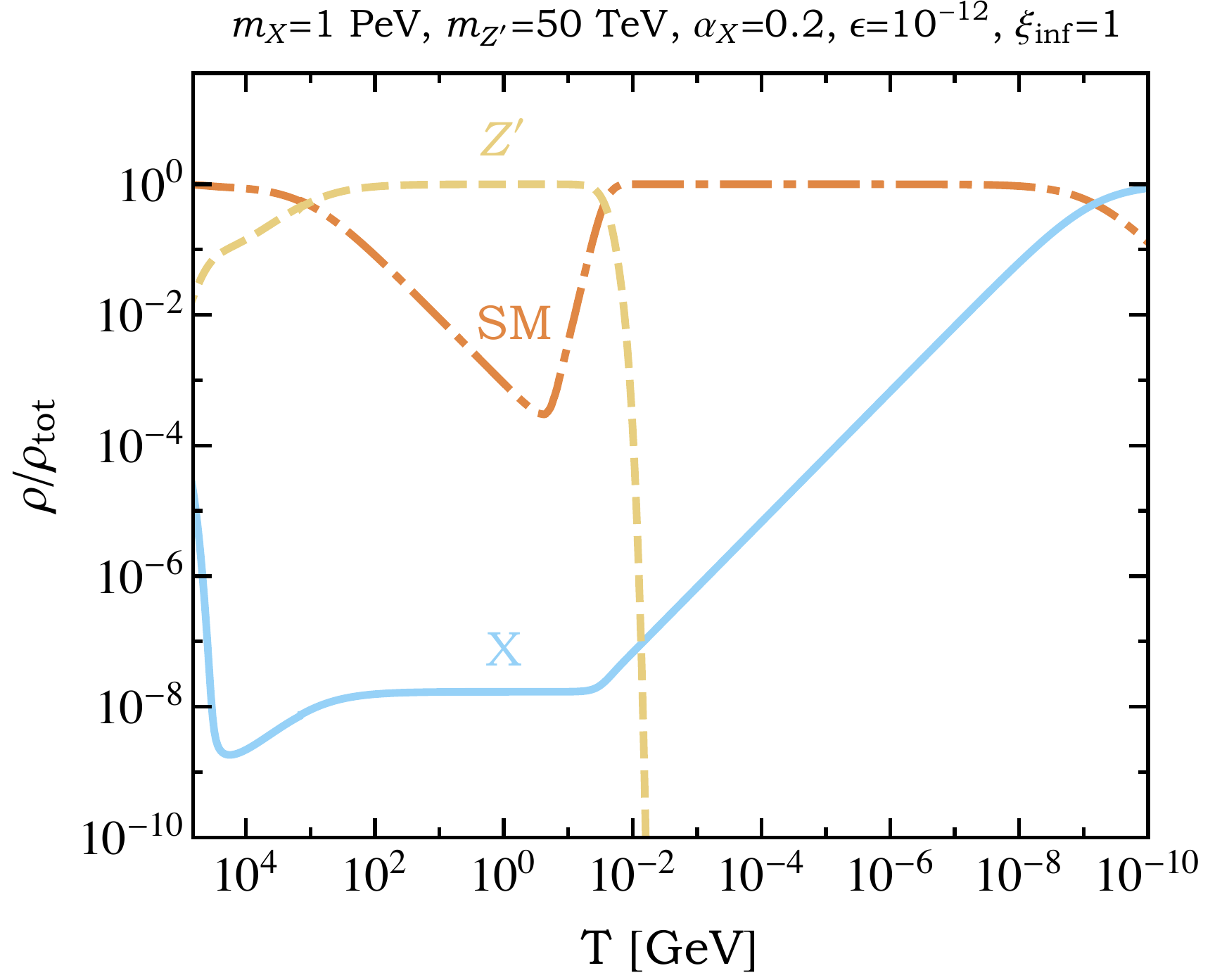}
\includegraphics[width=0.49\textwidth]{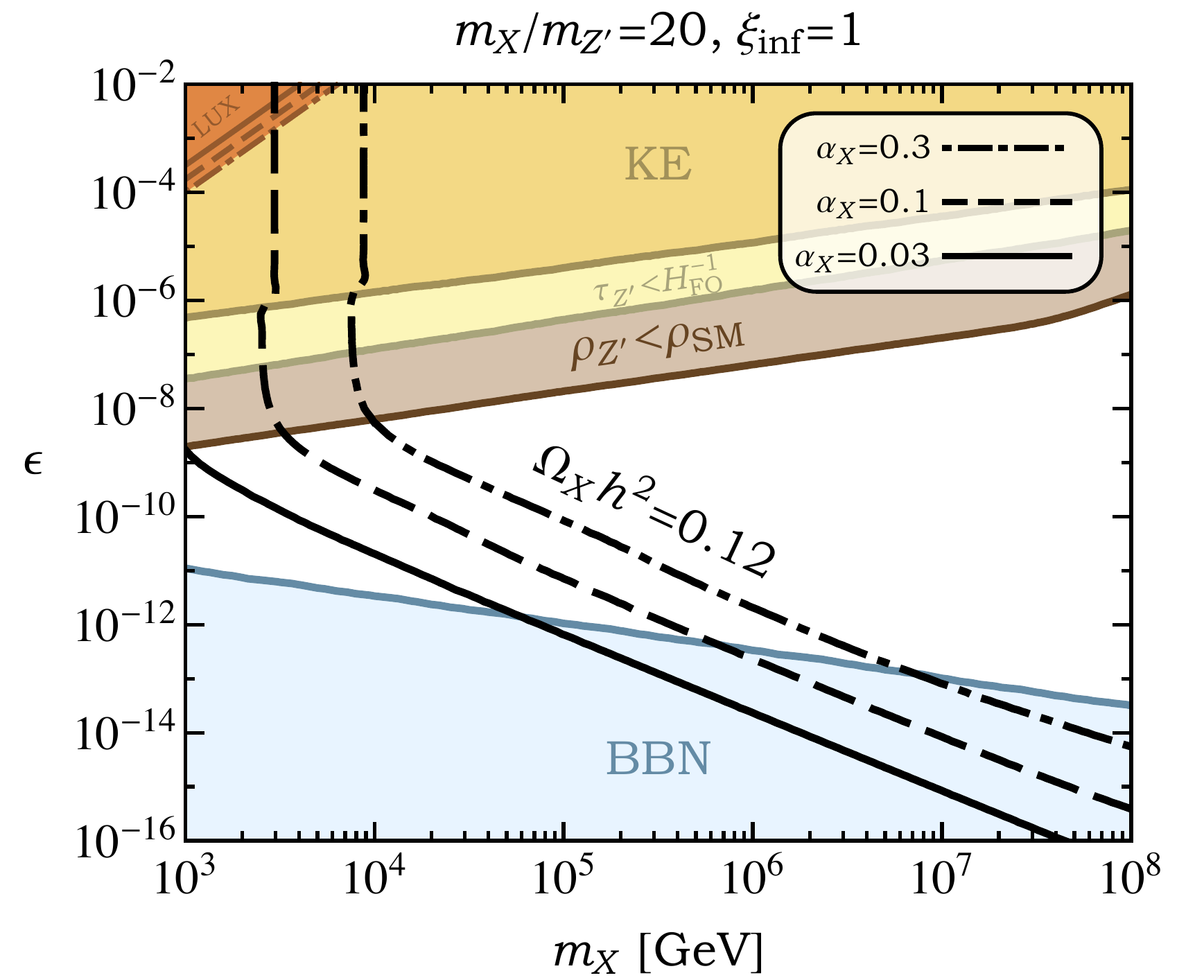}
\caption{Left: The evolution of the energy densities of DM (blue solid), of $Z'$s (yellow dashed), and in the visible sector (orange dot-dashed), as a function of the visible sector temperature. Upon becoming nonrelativistic, the $Z'$s quickly come to dominate the energy density of the Universe. When they decay, they heat the SM bath and dilute the $X$ abundance. Right: The black contours represent the regions of the $m_X-\epsilon$ plane in which the DM density is equal to the measured cosmological abundance, for three values of the hidden sector interaction strength, $\alpha_X$, and for $m_{Z'} = m_{X}/20$ and $\xi_{\rm inf} \equiv (T_{h}/T)_{\rm inf}=1$. The blue region is excluded by measurements of the light element abundances. In and above the orange and yellow regions, the hidden and visible sectors are in kinetic equilibrium during DM freeze-out, or the $Z'$ population decays before the freeze-out of $X$, respectively. In and above the brown region, the $Z'$ population never dominates the energy density of the Universe, and thus does not significantly dilute the DM relic abundance. In contrast to the case of a standard thermal relic, DM from a decoupled sector can be as heavy as $\sim$1-10 PeV without being overproduced in the early Universe.}
\label{example}
\end{figure}

In the case that $n_{Z'}$ remains close to its equilibrium value during the freeze-out of $X$, the Boltzmann equation can be solved semi-analytically. In this case, the thermal relic abundance of $X$ (plus $\bar{X}$) is given by,
\be
\label{relic1}
\Omega_X h^2 \approx 1.7 \times 10^{-10} ~ \frac{x_f \sqrt{g_\text{eff}}}{g_{*,s}} ~ \left( \frac{\sigma v_{X\bar{X}\rightarrow Z'Z'}}{\text{GeV}^{-2}} \right)^{-1}
\approx   1.6 \times 10^4 \bigg(\frac{x_f}{30}\bigg)  \bigg(\frac{0.1}{\alpha_X}\bigg)^2  \bigg(\frac{m_X}{{\rm PeV}}\bigg)^2
\bigg(\frac{\sqrt{g_{\rm eff}}/g_*}{0.1} \!  \bigg) \,, 
\ee
where $g_{\rm eff} \equiv g_{*,s} + g^h_{*,s} \, \xi^4$ at freeze-out. $x_f$, which is defined as the mass of $X$ divided by the SM temperature at freeze-out, is found to be $\sim 20 \times \xi$ over a wide range of parameters. From Eq.~\eqref{relic1}, it is clear that a PeV-scale DM candidate with perturbative couplings will initially freeze-out with an abundance that exceeds the observed DM density ($\Omega_X h^2 \gg \Omega_{\rm DM} h^2 \simeq 0.12$). It has long been appreciated, however, that this conclusion can be circumvented if the Universe departed from the standard RD picture after DM freeze-out~\cite{Fornengo:2002db,Gelmini:2006pq,Kane:2015jia,Hooper:2013nia,Patwardhan:2015kga,Randall:2015xza,Patwardhan:2015kga,Reece:2015lch,Lyth:1995ka,Davoudiasl:2015vba,Cohen:2008nb,Boeckel:2011yj,Boeckel:2009ej,Hong:2015oqa}. More specifically, as the Universe expands, the remaining $Z'$s will become nonrelativistic and quickly come to dominate the energy density of the Universe when  $\rho_{Z'} =  0.0074 \, g_{*,s} \xi_{\rm inf}^3  \,  m_{Z'} T_{\rm dom}^3 >  (\pi^2/30)g_{*} T_{\rm dom}^4$, which occurs at a visible sector temperature
\be
T_{\rm dom}   \sim 1 \, {\rm TeV} \, \times \, \xi_{\rm inf}^3 \left( \frac{m_{Z^\prime}}{50 \, \rm TeV} \right) \,.
\ee
This expression is valid so long as the $Z'$ population departs from chemical equilibrium while relativistic.  
 When the $Z'$s ultimately decay (see Fig.~\ref{fig:cartoon}), they will deposit energy and entropy into the visible sector, potentially diluting the DM abundance to acceptable levels. In the left frame of Fig.~\ref{example}, we show the evolution of the energy densities in the visible and hidden sectors, for a representative choice of parameters in this model. 

To calculate the impact of this effect, we integrate over the $Z'$ decay rate to find the factor by which the relics will be diluted~\cite{Kolb:1990vq}
\be
\frac{s_f}{s_i} \approx 680 \times \bigg( \!\frac{10^{-10}}{\epsilon} \!  \bigg) \bigg(\frac{m_{Z'}}{100\,{\rm TeV}}\bigg)^{1/2} \bigg(  \frac{\langle g^{1/3}_{*,s} \rangle^{3}}{100}   \bigg)^{1/4} \, \xi_{\rm inf}^3 \,,
\ee
where $\langle g_{*,s} \rangle$ denotes the time-averaged value over the period of decay. Combining this with Eq.~\eqref{relic1}, we find that the final DM relic abundance is given by
\be
\Omega_X h^2 \approx \frac{0.12}{\xi_{\rm inf}^3} \, \bigg(\frac{\epsilon}{10^{-13}}\bigg) \, \bigg(\frac{0.045}{\alpha_X}\bigg)^2 \, \bigg(\frac{m_X}{{\rm PeV}}\bigg)^2 \, \bigg(\frac{100 \,{\rm TeV}}{m_{Z'}}\bigg)^{1/2} \,    \bigg(\frac{x_f}{30}\bigg)    \bigg(\frac{\sqrt{g_{\rm eff}}/g_{*,s}}{0.1}\bigg) \bigg(  \frac{100}{\langle g^{1/3}_{*,s} \rangle^{3}} \bigg)^{1/4} \,.
\ee

In the right panel of Fig.~\ref{example} we plot some of the phenomenological features of this model as a function of the DM mass and the degree of kinetic mixing between the $Z'$ and SM hypercharge. The black contours denote the regions where the DM density is equal to the measured cosmological abundance, for three values of the hidden sector interaction strength, $\alpha_X$. Below the brown region, $Z'$ decays deposit significant entropy into the visible sector, reducing the final $X$ abundance. To assure consistency with BBN (see Sec.~\ref{sec:CMB_BBN_constr} for details), we require that the temperature of the Universe exceeds 10~MeV after the decays of the $Z'$ population, resulting in the following constraint: $\epsilon \gtrsim 2 \times 10^{-13} \times (100 \, {\rm TeV}/m_{Z'})^{1/2} \, (g_{*,s}/10)^{1/4}$. 

Viable phenomenology can be obtained within this class of models for a wide range of portal couplings, spanning many orders of magnitude. Depending on the degree of kinetic mixing, the hidden and visible sectors may have been entirely decoupled from one another, or kept in kinetic equilibrium through interactions of the type $\gamma f \leftrightarrow Z' f$. Quantitatively, we find that the rate for these processes exceed that of Hubble expansion if: $\epsilon \gtrsim  10^{-7} \times (T/10\,{\rm GeV})^{1/2}$ (shown as the orange region in the right panel of Fig.~\ref{example}).  Thus for smaller values of $\epsilon$, the hidden sector will not reach equilibrium with the visible sector and will remain decoupled. Furthermore, in the yellow regions of the right panel of Fig.~\ref{example}, the $Z'$ population decays prior to the freeze-out of $X$.

Many hidden sector DM models are difficult to constraint or otherwise test. In particular, the feeble couplings between the hidden and SM sectors makes the prospects for DM searches at colliders and at direct detection experiments quite bleak (see, however, Ref.~\cite{Hooper:2019xss}). This is less true for indirect searches. In particular, during the EMD epoch that is predicted in many of these scenarios, density perturbations grow more quickly than otherwise expected, leading to a large abundance of sub-earth-mass DM microhalos (see Sec.~\ref{sec:microhalos} for details). Since the DM does not couple directly to the SM, the minimum halo mass is much smaller than predicted for weakly interacting DM, and the smallest halos could form during the RD era. In Ref.~\cite{Blanco:2019eij}, a calculation was performed of the evolution of density perturbations within the context of such hidden sector models, using a series of $N$-body simulations to determine the outcome of nonlinear collapse during a RD era. The resulting microhalos were found to be extremely dense, leading to very high rates of DM annihilation and to large indirect detection signals that resemble those ordinarily predicted for decaying DM. Existing measurements of the high-latitude gamma-ray background rule out a wide range of parameter space within this class of models.  The scenarios that are most difficult to constrain are those that feature a very long EMD epoch; if microhalos form prior to the decay of the unstable hidden sector matter, the destruction of these microhalos effectively heats the DM, suppressing the later formation of microhalos. This gravitational heating is discussed in more detail in Sec. \ref{sec:microhalos}.

Although we have only considered one representative scenario here, our treatment is without loss of essential generality. As long as the dark sector freeze out proceeds through 2 $\to$ 2 annihilation between the DM and the unstable long-lived species, the relic abundance will have the same parametric scaling as outlined above regardless of the spins of these species or  the Lorentz structures of their interactions;  variations relative to our benchmark case  only modify these results by order-one amounts \cite{Berlin:2016gtr,Berlin:2016vnh}.

\subsection{Phase transitions and baryon asymmetry (Authors: R. Allahverdi \& M. Lewicki)}
\label{sec:PTsandBaryogenesis}

\subsubsection{Phase transition dynamics}\label{sec:PTdynamics}

Let us begin with discussion of the modification of dynamics of a first order phase transition due to modified expansion history. While the modification itself can seem to be rather drastic, phase transitions prove to be largely insensitive. This is simply because the dynamics of a phase transition are largely dictated by the dynamics of fields undergoing the transition. For cosmological applications, in the most common case of a thermally driven transition of a scalar field, decay rate per unit volume is given by~\cite{Linde:1981zj,Linde:1980tt,Coleman:1977py,Callan:1977pt}
\begin{equation}\label{eq:PTdecaywidth}
\Gamma(T)\approx T^4 \exp\left(-\frac{S_3(T)}{T}\right),\quad S_3=4\pi \int {\rm d}r  r^2\left[\frac{1}{2}\left(\frac{{\rm d}\phi}{{\rm d}r}\right)^2+V(\phi,T)\right],
\end{equation}
where the action $S_3(T)$ depends on the potential of the field in question including thermal corrections.
Expansion history comes into play when we find the temperature at which bubbles begin nucleating defined by
\begin{equation}
\int_{0}^{t_n} {\rm d} t \frac{\Gamma(t)}{H(t)^3} = \int_{T_n}^{\infty} \frac{{\rm d} T}{T} \frac{\Gamma(T)}{ H(T)^4} =1 ,
\end{equation}
where we assumed adiabatic expansion, that is ${\rm d}t = -{\rm d}T/( T\,H(T))$.
Given the exponential dependence of the decay rate we can see that modified expansion history will impact the resulting temperature only logarithmically (see e.g. Ref.~\cite{Caprini:2019egz}). Thus, phase transitions are rather insensitive to expansion history and their course is instead linked to the (subdominant) radiation sector the fields in question are coupled to.   

With that said, it is of course true that possible observable consequences of such a transition can still be impacted a lot more severely due to the modified expansion. We will discuss these modification in terms of electroweak baryogenesis in the remainder of this section and the gravitational wave background in Sec.~\ref{sec:GWBGs}.

\subsubsection{Electroweak baryogenesis}

Let us now turn to electroweak baryogenesis \cite{Kuzmin:1985mm,Cohen:1993nk,Riotto:1999yt,Morrissey:2012db} in which the observed baryon asymmetry of the Universe is a result of a strong electroweak phase transition.
This requires a modification of the SM to make the electroweak phase transition first order and introduce extra sources of $CP$-violation. 
Our goal, however, is not to discuss any specific modification of SM aimed at satisfying these conditions but rather to discuss how they can be modified in nonstandard expansion histories.

\begin{figure*}
\centering
\includegraphics[width=0.37\textwidth]{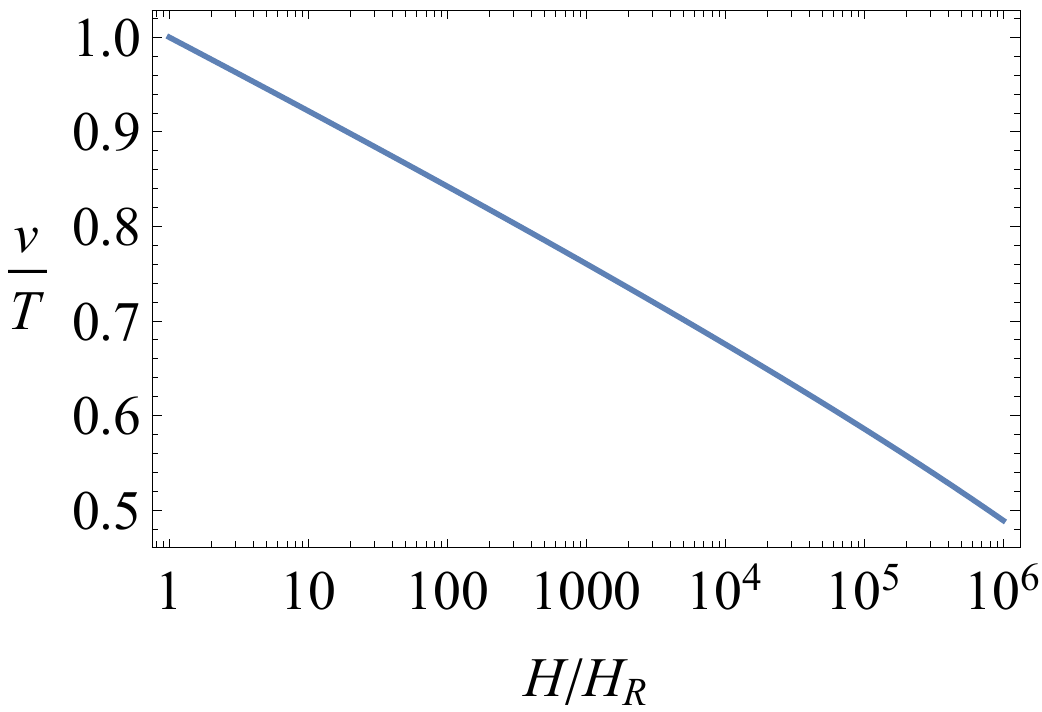} \hspace{4mm}
\includegraphics[width=0.382\textwidth]{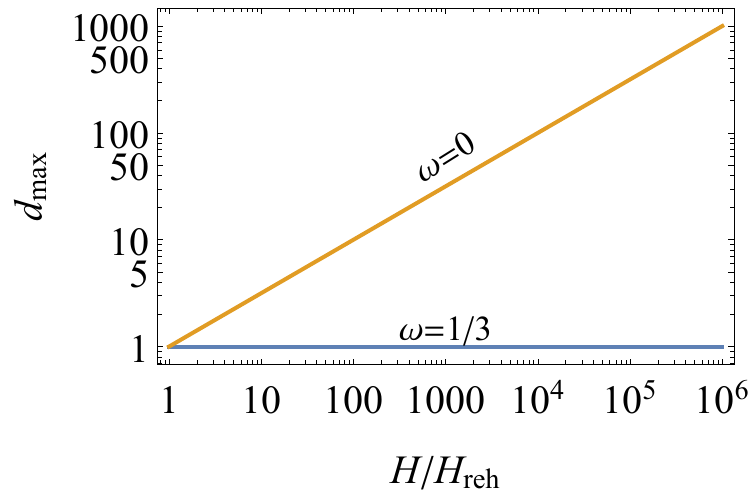}
\caption{Left panel: vev of the field over the temperature required to successfully decouple the electroweak sphaleron and avoid wash-out of the baryon asymmetry after the transition as a function of the Hubble rate during the transition normalized to the contribution to the expansion rate due to the subdominant radiation component. Right panel: Dilution of the baryon asymmetry upon reheating in models with expansion dominated by a component redshifting slower than radiation $(w<1/3)$ on the example of an EMD epoch with $w=0$. The result is shown as a function of the Hubble rate at the generation of the asymmetry normalized to the expansion rate at reheating temperature at which RD era resumes.}
\label{fig:vTsph}
\end{figure*}

We will focus on the main constraint typically put on the transition itself, that is the vacuum expectation value of the field over the temperature has to be large enough $v/T \gtrsim 1$. This comes from the requirement that the electroweak sphalerons providing required baryon number violation are suppressed after the transition to avoid washing out the asymmetry as the system goes back to thermal equilibrium.  
To be more specific, the sphaleron processes in question are SM $SU(2)$ gauge interactions and as a result are automatically suppressed once this symmetry is broken. 
The simplest criterion for their necessary decoupling translates to
the sphaleron rate being smaller than the Hubble rate~\cite{Joyce:1996cp,Joyce:1997fc}
\be
\Gamma_{\rm Sph} = T^4 \mathcal{B}_0 \frac{g}{4 \pi} \left(\frac{v}{T}\right)^7 \exp \left( -\frac{4 \pi}{g}\frac{v}{T} \right)\lesssim H(T),
\ee
where $v$ is the vev of the Higgs field after the transition. Calculation of the prefactor $\mathcal{B}_0$ is difficult and leads to a small spread of values used in the literature for the final bound on $v/T$ \cite{Katz:2014bha,Quiros:1999jp,Funakubo:2009eg}. Since our interest is mainly in the modification of this result in a nonstandard thermal history, we will simply use the value of $\mathcal{B}_0$ which leads to the standard bound $v/T \leq 1$ when the expansion is driven by SM radiation $H(T) = H_{\rm R}(T) \propto T^2$. The modification of the required $v/T$ is shown in Fig.~\ref{fig:vTsph} assuming the transition occurs at the typical temperature for electroweak symmetry breaking $T \approx 100$ GeV. While lowering the value of $v/T$ required to avoid wash-out by sphalerons might not seem like a huge modification, it has a significant effect on the parameter space of specific models~\cite{Lewicki:2016efe,Artymowski:2016tme}.

As discussed in Sec.~\ref{sec:causes}, a particularly well-motivated example of a nonstandard thermal history involves an EMD epoch. The EMD epoch ends upon completion of the decay and the Universe enters a RD phase with temperature $T_{\rm reh}$. Electroweak baryogenesis in such a scenario with $1\,{\rm GeV} \lesssim T_{\rm reh} \ll 100$\,GeV has been studied in Ref.~\cite{Davidson:2000dw}. There it was found that a faster expansion rate at the same temperature results in a bound $v/T \gtrsim 0.77 ~ (0.92)$ for $T_{\rm reh} = 1 ~ (10)$ GeV. Note that a modified expansion rate affects the out-of-equilibrium condition necessary for generation of baryon asymmetry (i.e. the third Sakharov condition~\cite{Sakharov}), as well as the efficiency of wash-out processes that could erase the asymmetry, in any baryogenesis scenario.

\subsubsection{Constraints on baryogenesis from dilution by reheating}

In general, a nonstandard thermal history can also affect baryogenesis by diluting the generated baryon asymmetry. This is due to entropy release during transition to the final RD phase. This can be readily seen by noting that the energy density of a component with equation of state $w < 1/3$ is redshifted more slowly than that of radiation. Thus its dominance keeps growing and there can be no transition to RD era unless it decays to SM radiation. This reheating process is accompanied by (a typically significant) increase in the entropy of the Universe. 

To demonstrate the effect of reheating on baryon asymmetry, we consider a nonstandard thermal history that involves a phase with equation of state $w < 1/3$ that is preceded and followed by RD era. Let us denote the Hubble rate at the onset of the nonstandard expansion phase or generation of the baryon asymmetry (whichever happened later) by $H$ and at the time of return to RD era by $H_{\rm reh}$. The number density of any quantity produced at $H > H_{\rm reh}$ is redshifted by a factor $(a_{\rm reh}/a)^3 = (H/H_{\rm reh})^{2/(1+w)}$ between $H$ and $H_{\rm reh}$. Applying this to the normalized baryon asymmetry $n_{\rm B}/s$, where $n_{\rm B}$ is the number density of baryon asymmetry, the maximum dilution factor during a nonstandard expansion phase with $w<1/3$ is found to be
\begin{equation} \label{dilution}
d_{\rm max} \sim \left ({H \over H_{\rm reh}}\right)^{(1-3w)/2(1+w)} \,.
\end{equation}
Fig.~\ref{fig:vTsph} shows the dilution factor in the case $w = 0$ as a function the Hubble rate normalised to the Hubble rate at reheating $H_{\rm reh}$.

To put things in perspective, let us consider an EMD epoch ($w = 0$) driven by some string modulus $\phi$ with mass $m_\phi$. We typically have $H \sim m_\phi$ at the onset of the EMD epoch and $H_{\rm reh} \sim m^3_\phi/M^2_{\rm P}$ in this case (see Sec. \ref{sec:moduli}), which results in $d_{\rm max} = {\mathcal O}(M_{\rm P}/m_\phi)$. Considering that $m_\phi \gtrsim {\cal O}(100 ~ {\rm GeV})$ is needed in order to to reheat the Universe before the onset of BBN (i.e. $T_{\rm reh} \gtrsim \mathcal{O}(1)$\,MeV), we find $d_{\rm max} \lesssim 10^{13}-10^{14}$. Such a large dilution can render any previously generated baryon asymmetry totally insignificant.             

Therefore, to obtain the observed value $(n_{\rm B}/s) \simeq 0.9 \times 10^{-10}$, we are left with two general options. First, to generate an initially large baryon asymmetry that matches the correct value after dilution. Second, to generate the correct baryon asymmetry after the transition to the final RD phase thereby avoiding dilution by reheating. Let us discuss these possibilities and specific baryogenesis mechanisms that can work in each case in some detail.     

Affleck-Dine (AD) baryogenesis~\cite{Davidson:2000dw,DRT} (for reviews, see Refs.~\cite{EM,Dine:2003ax,Allahverdi:2012ju}) is a mechanism that can naturally fit the first option. This scenario utilizes flat directions in the scalar potential of supersymmetric extensions of the SM. AD baryogenesis is known to be able to generate large values of baryon asymmetry, even ${\cal O}(1)$, in a standard thermal history. An EMD epoch (driven, for example, by string theory moduli) can be therefore helpful in this regard by regulating the baryon asymmetry and bringing it down to the observed value~\cite{CG,CGMO,K,Kane:2011ih}. A viable string embedding of AD baryogenesis has been discussed in Ref.~\cite{ACM} where the volume modulus in type IIB string models drives an EMD phase resulting in the observed baryon asymmetry for natural values of the underlying parameters. We note that dilution by late reheating can also help with other models of baryogenesis that generate a too large baryon asymmetry in scenarios with a standard thermal history.

The second route is to generate baryon asymmetry upon completion of transition to a RD Universe when the Hubble rate is $H_{\rm reh}$. Many scenarios with a nonstandard thermal history result in $T_{\rm reh} \ll 100$ GeV. In particular, in cases where an EMD epoch is responsible for nonthermal production of weak-scale DM, we may have $T_{\rm reh} \lesssim {\cal O}(1 ~ {\rm GeV})$. This severely constrains viable baryogenesis mechanisms that can work. For example, it rules out leptogenesis~\cite{ACM} (for reviews, see Refs.~\cite{Review1,Review2}) as sphalerons are out-of-equilibrium at these temperatures and hence unable to convert the generated lepton asymmetry to baryon asymmetry. Thermal baryogenesis is essentially ruled out as a possibility too because it will require $C$-, $CP$-, and $B$-violating interactions at such low scales.

Therefore, nonthermal post-sphaleron baryogenesis seems to be the available possibility for generating the observed baryon asymmetry at late times. An explicit model is discussed in Ref.~\cite{Allahverdi:2010im} in the context of EMD epoch driven by some string modulus $\sigma$. This model includes new particles $N$ with a mass $m_N \gg T_{\rm reh}$ that have $C$-, $CP$-, and $B$-violating interactions with the SM particles. $N$ quanta are produced in $\sigma$ decay, and their subsequent decay to the SM particles generates a baryon asymmetry given by
\begin{equation} \label{BAU}
{n_{\rm B} \over s} = Y_\sigma ~ {\rm Br}_N ~ \epsilon\,, \hspace{1cm} Y_\sigma \equiv {3 T_{\rm reh} \over 4 m_\sigma} .
\end{equation}
Here $Y_\sigma$ is the yield from $\sigma$ decay, ${\rm Br}_N$ is the branching fraction for production of $N$ from $\sigma$ decay and $\epsilon$ is the asymmetry generated per $N$ decay. One point to keep in mind is that $Y_\sigma$ is typically very small. For example, a modulus with mass $m_\sigma \sim 100-1000$ TeV gives $Y_\sigma \lesssim {\cal O}(10^{-6})$. Therefore, ${\rm Br}_N\epsilon$ cannot be too small in order to generate the correct value of $n_{\rm B}/s$. This implies an $\epsilon$ usually larger than that in thermal scenarios.

This nonthermal scenario may also address the baryon-DM coincidence problem (for a review, see Ref.~\cite{Davoudiasl:2012uw}) if $\sigma$ decay is the common origin of baryogenesis and DM~\cite{Allahverdi:2010rh,AB}. In addition, if $N$ particles are not much heavier than $\mathcal{O}(1)\,{\rm TeV}$, they could be produced at the LHC. The $B$-violating interactions of $N$ can also lead to observable low-energy phenomena like neutron-antineutron oscillations. Such a TeV scale baryogenesis scenario may therefore be tested by complementary probes at the cosmology, high energy, and low energy frontiers~\cite{Allahverdi:2017edd}.      

So far, our quantitative discussion of the effects of a nonstandard thermal history on baryogenesis has mainly focused on EMD epoch as an important and well-motivated case. However, the main considerations regarding the out-of-equilibrium condition to generate and preserve baryon asymmetry, and the entropy release during transition to the final RD phase are valid for any scenario with a modified expansion rate. Here, we mention in passing a number of studies on baryogenesis in more general nonstandard thermal histories. TeV scale leptogenesis in nonstandard thermal histories that involve a) an EMD epoch, and b) a phase with equation of state $w > 1/3$ has been discussed in Ref.~\cite{MB}. An important example of a scenario with $w > 1/3$ is kination where $w = 1$. Ref.~\cite{BR} considers AD baryogenesis in the context of a model that involves a post-inflationary kination phase. The effect of kination on the gravitational leptogenesis scenario has been investigated in Ref.~\cite{KKYY}. Gravitational leptogenesis for a general equation of state during inflationary reheating has been studied in Ref.~\cite{Adshead:2017znw}. It has been shown in Ref.~\cite{DFJN} that nonstandard cosmologies from scalar-tensor theories can open new paths to low-scale leptogenesis and thereby allow direct experimental tests of the scenario. Finally, it is possible to have more complicated situations that involve various sectors with different temperatures. A specific example of such a scenario is given in Ref.~\cite{Bernal:2017zvx} that considers thermal leptogenesis in the presence of a dark sector whose temperature is higher than that of SM radiation.                 

\vspace{1.5cm}
\subsection{Consequences for models of inflation (Authors: K. Freese \& T. Tenkanen)}
\label{sec:inflation}

Cosmic inflation can be regarded as a successful paradigm for two reasons. First, if inflation lasted long enough so that during it the Universe expanded at least by a factor\footnote{Assuming high scale inflation and standard expansion history. In nonstandard cosmologies, this number can range from $\sim e^{18}$ to $e^{68}$ \cite{Liddle:2003as,Dodelson:2003vq,Tanin:2020qjw}.} of $\sim e^{60}\sim 10^{26}$, inflation provides a plausible explanation for the classic horizon, monopole, and flatness problems (see e.g. Ref. \cite{Baumann:2009ds}). Second, inflation provides a natural way to generate the initial metric perturbations and explain their super-horizon correlations at the time of CMB decoupling. These correlations are usually described in terms of the scalar curvature power spectrum, which at least at the largest physical distance scales is measured to have a power-law form~\cite{Akrami:2018odb}
\begin{equation}
\label{Pzeta}
\mathcal{P}_{\zeta}(k) = \mathcal{A}\left(\frac{k}{k_*}\right)^{n_s-1}\,,
\end{equation}
to a high precision. Here $\zeta$ denotes the curvature perturbation and $k$ the wavenumber in a Fourier decomposition. The spectrum (\ref{Pzeta}) has the observed amplitude $\mathcal{A}\simeq 2.1\times 10^{-9}$ and spectral tilt $n_s\simeq 0.965$ at the pivot scale $k_*=0.05\, {\rm Mpc}^{-1}$~\cite{Akrami:2018odb}. The data are consistent with no running, ${\rm d}n_s/{\rm d\,ln}k$, or running of the running of the spectral tilt, ${\rm d}^2n_s/{\rm d(ln}k)^2$ \cite{Akrami:2018odb}, and therefore we will neglect them here for simplicity. While in principle solving the classic horizon, monopole, and flatness problems only requires exponential expansion at early times, any successful model of inflation also has to predict the measured values for observables, or be within the limits on them. In addition to depending on the underlying particle physics scenario, these predictions usually depend also on the post-inflationary expansion history, which is the topic of this section.

In the following, we will assume, for simplicity, that inflation was driven by a single scalar field $\phi$, which at early times dominated the total energy density of the Universe and drove inflation by rolling down in its potential $V$. In the slow-roll approximation, $|\ddot{\phi}|\ll 3H|\dot{\phi}|\,, \dot{\phi}^2 \ll V$, where $H\equiv \dot{a}/a$ is the Hubble parameter and the dot denotes derivative with respect to cosmic time, the inflationary dynamics can be characterized by the slow-roll parameters
\begin{equation}
\label{SRparameters1}
	\epsilon \equiv \frac{M_{\rm P}^2}{16\pi} \left(\frac{V'}{V}\right)^2 \,, \quad
	\eta \equiv \frac{M_{\rm P}^2}{8\pi} \frac{V''}{V} \,,
\end{equation}
where the prime denotes derivative with respect to $\phi$. In slow-roll $\epsilon\,, |\eta | \ll 1$, and inflation ends when $\epsilon =1$. In this case, then, the amplitude of the curvature power spectrum can be expressed in terms of the slow-roll parameters as~\cite{Lyth:1998xn}
\begin{equation}
\label{amplitude}
 \mathcal{A} = \frac{1}{24 \pi^2 M_{\rm P}^4} \frac{V(\phi_*)}{\epsilon(\phi_*)} ,
\end{equation}
where $\phi_*$ denotes the field value at the time when the scale $k_*$ exited the horizon. The leading order expression for the spectral tilt is
\begin{equation}
\label{ns}
n_s -1\equiv \left. \frac{{\rm d\,ln}\mathcal{P}_{\zeta}(k)}{{\rm d\,ln}k}\right|_{k=k_*} \simeq -6\epsilon(\phi_*) + 2\eta(\phi_*)\,.
\end{equation}
In addition to generating scalar curvature perturbations, fluctuations of the inflaton field also generate tensor perturbations, i.e. primordial gravitational waves. The observational constraints are usually expressed in terms of the tensor-to-scalar ratio
\begin{equation}
\label{r_16e}
r \equiv \frac{\mathcal{P}_{T}}{\mathcal{P}_{\zeta}} \simeq 16\epsilon\,,
\end{equation}
where $\mathcal{P}_{T}$ is the tensor power spectrum~(see e.g. Ref. \cite{Baumann:2009ds}) and the last expression applies at the leading order in slow-roll parameters. As primordial tensor perturbations have not been detected, observations of the CMB place an upper limit on tensor-to-scalar ratio, $r<0.06$ at the pivot scale $k_*=0.05\, {\rm Mpc}^{-1}$ \cite{Ade:2018gkx}. 

As can be seen by substituting the slow-roll parameters in Eq. (\ref{SRparameters1}) into Eqs. (\ref{amplitude}), (\ref{ns}), and (\ref{r_16e}), the inflationary observables depend solely on the shape of the inflaton potential. Constraints on the observables therefore limit different scalar field potentials, i.e. different models, at the energy scale where inflation occurred. However, judging whether a given model complies with observations is not entirely clear-cut. A given potential may predict wrong values for observables at some energy scale but get them exactly right if the distance scales where measurements are made exited the horizon when the inflaton was rolling in some other part of the potential where its shape was different and which therefore generated different values for observables. The time when a given scale must have exited the horizon is determined by the overall energy scale of inflation and the post-inflationary evolution of scales. Therefore, details of the entire post-inflationary expansion history and, in particular, any nonstandard period of expansion will affect inflationary model selection, as we will demonstrate below.

It is convenient to characterize the observables in terms of the number of e-folds between the horizon exit of the pivot scale and the end of inflation, $N\equiv \ln(a_{\rm end}/a_k)$ where $a_{\rm end}$ ($a_k$) is the value of the scale factor at the end of inflation (when the scale $k$ exited the horizon), which -- together with the parameters that characterize the post-inflationary evolution -- relate the pivot scale to the present Hubble scale $H_0$ as
\begin{equation}
\frac{k}{a_0 H_0} = \frac{a_k H_k}{a_0 H_0} = e^{-N}\frac{a_{\rm end}}{a_{\rm RD}}\frac{a_{\rm RD}}{a_0}\frac{H_k}{H_0}\,,
\end{equation}
where $H_k$ is the Hubble scale at $a_k$, $a_{\rm RD}$ is the scale factor at the time when the nonstandard period ended and the Universe became RD, and $a_0$ is the scale factor at present. Here we have assumed that there is exactly one nonstandard phase between the end of inflation and the subsequent RD period. Note that the nonstandard expansion period could have been caused by, for example, a kinetic energy-dominated kination phase \cite{Spokoiny:1993kt,Joyce:1996cp} or another period of inflation (see Sec. \ref{sec:extrainflation}) instead of conventional reheating discussed in Sec. \ref{sec:reheating}. Including another nonstandard expansion phase(s) that punctuated the subsequent RD era(s) is straightforward \cite{Liddle:2003as} but here we leave them out for simplicity. Assuming that the nonstandard phase can be characterized with a single effective equation of state parameter $w$, so that the total energy density scaled as $\rho \propto a^{-3(1+w)}$ during the nonstandard expansion epoch, we obtain
\begin{equation}
\label{Nresult}
N \simeq 57 + \frac12\ln\left(\frac{r}{0.1}\right) + \frac{1}{3(1+w)}\ln\left(\frac{\rho_{\rm RD}}{\rho_{\rm end}}\right) - \ln\left(\frac{\rho_{\rm RD}^{1/4}}{10^{16}\,{\rm GeV}}\right) - \ln\left(\frac{k}{0.05\,{\rm Mpc}^{-1}}\right)\,,
\end{equation}
where $\rho_{\rm RD}$ is the total energy density at the time the standard RD epoch began, $\rho_{\rm end}=\linebreak 3/(8\pi)H^2(a_{\rm end}) M_{\rm P}^2$ is the total energy density at the end of inflation, and we have fixed $a_0=1$, used $T_0=2.725\,{\rm K}$, and assumed, for simplicity, that the transitions between different epochs were instantaneous\footnote{Note that our treatment does not account for non-instantaneous transitions between epochs reminiscent of e.g. nonperturbative effects encountered in preheating (see Sec. \ref{sec:reheating}).}. Because quantities such as $w\,, \rho_{\rm RD}$ and $\rho_{\rm end}$ usually depend on the underlying particle physics model (see Sec. \ref{sec:causes}), the form of (\ref{Nresult}) is usually the most convenient choice to characterize the effect of post-inflationary evolution on models of inflation\footnote{See also Refs. \cite{Martin:2010kz,Mielczarek:2010ag,Martin:2014nya,Munoz:2014eqa,Creminelli:2014oaa,Creminelli:2014fca,Dai:2014jja,Cai:2015soa,Cook:2015vqa,Eshaghi:2016kne,Ueno:2016dim,Figueroa:2018twl,Ji:2019gfy} for scenarios where this reasoning was reversed so that the reheating temperature and the number of degrees of freedom were constrained by assuming a given model of inflation.}. Furthermore, by assuming instantaneous thermalization among particles at the time the RD epoch began, one can connect the energy density at this epoch to the reheating temperature, $\rho_{\rm RD} = \pi^2 g_* T_{\rm reh}^4/30$. This is usually a safe assumption, as e.g. the Standard Model plasma generically thermalizes in much less than one e-fold from its production, see e.g. Ref. \cite{McDonough:2020tqq}.

\begin{figure}
\begin{center}
\includegraphics[width=.5\textwidth]{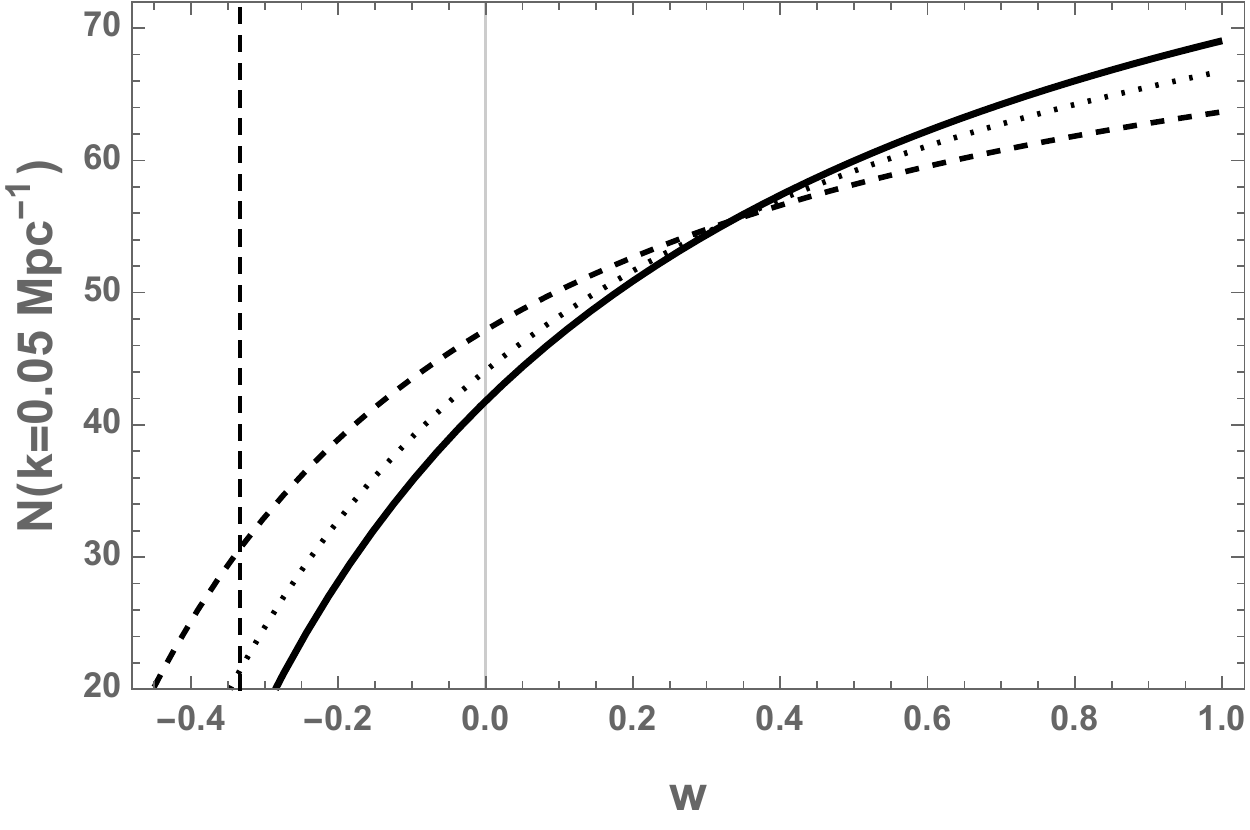}
\caption{The number of e-folds $N(k_*=0.05\,{\rm Mpc}^{-1})$ as a function of the EoS parameter $w$ as given in Eq. \eqref{NresultSimplified}. In this figure $r=0.01$ and $T_{\rm reh}= (0.01, 10, 10^5)$ GeV for the thick, dashed, and dotted curves, respectively. The vertical dashed line marks the threshold for accelerated expansion ($w=-1/3$). 
\vspace{10mm}}
\label{wplot}
\end{center}
\end{figure}

By assuming that the total energy density did not decrease much during the final e-folds of inflation\footnote{Because the first slow-roll parameter is $\epsilon=\dot{H}/H^2\ll 1$, it is usually a very good approximation that the total energy density did not change much during the range of e-folds that is of interest here. In particular, this is the case for plateau models that give the best fit to the CMB data \cite{Akrami:2018odb}.}, one can show that the result (\ref{Nresult}) becomes 
\begin{equation}
\label{NresultSimplified}
N \simeq 57 + \frac{1+3w}{6(1+w)}\ln\left(\frac{r}{0.1}\right) + \frac{1-3w}{3(1+w)}\ln\left(\frac{\rho_{\rm RD}^{1/4}}{10^{16}\,{\rm GeV}}\right) - \ln\left(\frac{k}{0.05\,{\rm Mpc}^{-1}}\right)\,,
\end{equation}
which is independent of $\rho_{\rm end}$ and only depends on $r$ (which is an observable) and $\rho_{\rm RD}$. For the range of $N$ as a function of $w$, see Fig. \ref{wplot}. Note that the distance scale $k_*^{-1}$ considered here is different from the current horizon $1/H_0$ and therefore the number of e-folds (\ref{NresultSimplified}) is not the one that solves the horizon and flatness problems. By choosing $k=H_0\simeq 0.0002\,{\rm Mpc}^{-1}$ and assuming high scale inflation with instant reheating, we obtain $N_{\rm horizon}\simeq 62$ in accord with Refs. \cite{Liddle:2003as,Dodelson:2003vq} (see also Ref. \cite{Tanin:2020qjw} for an upper limit on the number of e-folds, $N\lesssim 68$).

Let us then give few examples of how post-inflationary expansion history affects model selection (for a larger review of different models, see Ref. \cite{Martin:2013tda}). In slow-roll approximation the number of e-folds between end of inflation and the horizon exit of a given scale can be connected to the scalar potential through
\begin{equation}
\label{Nphi}
    N = \ln\left(\frac{a_{\rm end}}{a_k}\right) = \int_{t_k}^{t_{\rm end}} H(t){\rm d}t \simeq \frac{8\pi}{M_{\rm P}^2}\int_{\phi_{\rm end}}^{\phi_k}\frac{V}{V'}{\rm d}\phi
\,,
\end{equation}
where $\phi_{\rm end}$ is determined by $\epsilon(\phi_{\rm end})=1$ and $\phi_k\equiv \phi(a_k)$. Therefore, for e.g. a power-law potential $V\propto \phi^p$ one obtains at the leading order in the expansion in $1/N$ \cite{Dodelson:2003vq}
\begin{equation}
n_s^{\rm pl} - 1 = -\frac{2+p}{2N+p/2} \,, \quad r^{\rm pl}=\frac{16p}{p+4N}\,,
\end{equation}
where the superscripts refer to ``power-law". Other classic examples are the (cosine) natural inflation model with a potential of the form $V = \Lambda^4(1-\cos(\phi/f))$, where $\Lambda$ and $f$ are mass scales determined by the underlying high energy theory \cite{Freese:1990rb} (see also Ref. \cite{Pajer:2013fsa} for other axion inflation models), or the spontaneously broken supersymmetric (SB SUSY) inflation\footnote{While this model goes by the name ``SB SUSY'',  it is not necessarily related to e.g. supersymmetric extensions of the Standard Model, as the relatively general form of the potential shows.} with the potential $V = \Lambda_{2}^4(1+\gamma\ln(\phi/M_{\rm P}))$, where $\Lambda_2$ and $\gamma$ are free parameters \cite{Dvali:1994ms}. The predictions of these models can also be calculated in the usual way from Eqs. (\ref{ns}), (\ref{r_16e}), and (\ref{Nphi}). Further examples include the so-called $\xi$- and $\alpha$-attractor models \cite{Kaiser:2013sna,Kallosh:2013tua,Galante:2014ifa} which predict 
\begin{equation}
n_s^{\alpha-{\rm att.}} - 1 =  -\frac{2}{N}\,, \quad r^{\alpha-{\rm att.}}= \frac{12\alpha}{N^2}\,,
\end{equation} 
where $\alpha \ll N$ is a small parameter that is related to a pole in the kinetic term for the inflaton field. For example, for the Starobinsky model \cite{Starobinsky:1980te} $\alpha=1$ and Linde-Goncharov model \cite{Goncharov:1983mw} $\alpha=1/9$. Another notable example is Higgs inflation in metric gravity \cite{Bezrukov:2007ep} which also corresponds to $\alpha=1$, whereas for Higgs inflation in Palatini gravity $\alpha$ can be much smaller than this \cite{Bauer:2008zj} (although generally Palatini models are not $\alpha$-attractor models \cite{Jarv:2017azx}).

The above potentials are examples of models where the predictions are relatively sensitive to the number of e-folds, and thus also to the details of post-inflationary expansion history. While the effect enters only through logarithmic corrections to $N$ as in Eq. (\ref{Nresult}), our ignorance of the reheating energy scale $\rho_{\rm RD}$ allows their contribution to be substantial, and hence a nonstandard expansion phase can significantly affect model selection. However, there are also models where the value of $r$ is relatively insensitive to $N$. For example, in models where the gravity sector consists of a $\beta R^2$ term within Palatini gravity \cite{Enckell:2018hmo,Antoniadis:2018ywb}, the prediction is $r\simeq 1/\beta$ for large $\beta$, i.e. practically independent of $N$. As the energy scale of inflation is $V^{1/4}/M_{\rm P}\sim r^{1/4}$, this allows one to construct models where inflation occurs at a very low energy scale \cite{Tenkanen:2019wsd,Tenkanen:2020cvw}.

\begin{figure}
\begin{center}
\includegraphics[width=.5\textwidth]{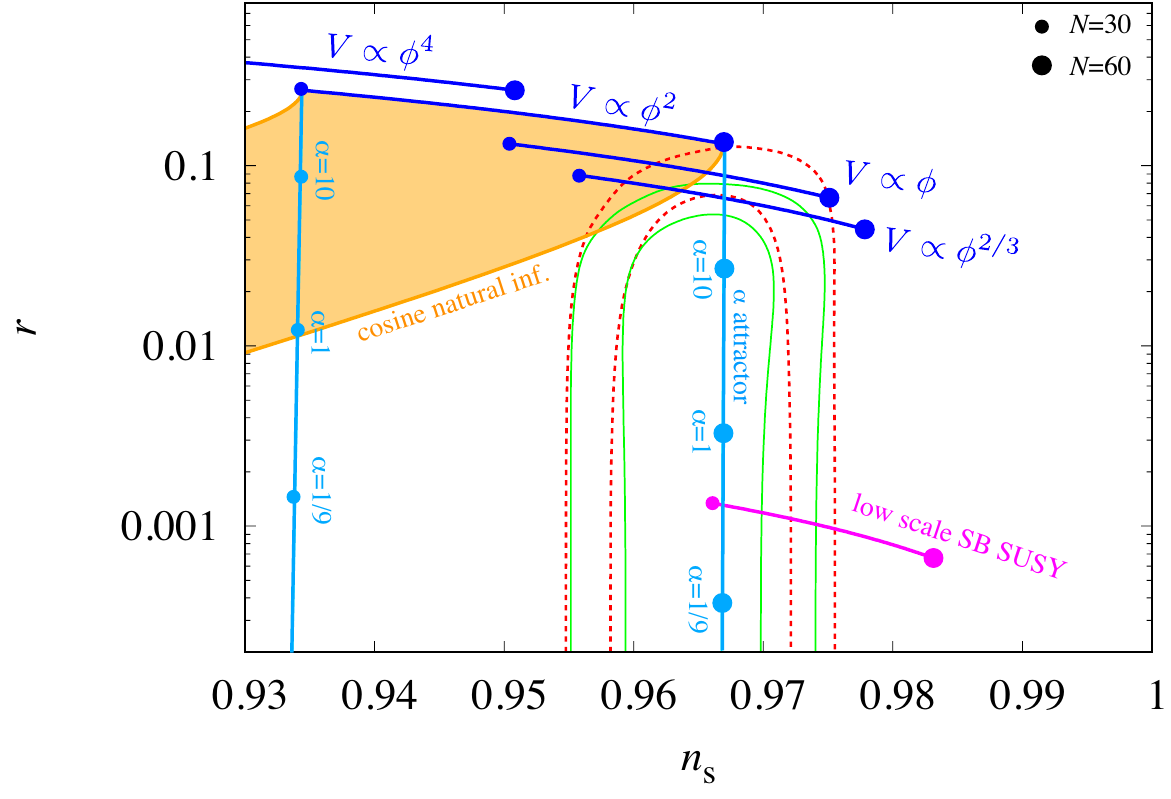}
\caption{The effect of a nonstandard expansion phase on the spectral tilt and tensor-to-scalar ratio for few example models. The red dotted and green solid contours correspond to
the constraints from Planck TT+TE+EE+lowE+lensing \cite{Ade:2018gkx} and Planck TT+TE+EE+lowE+lensing+BICEP2/Keck \cite{Array:2015xqh}, respectively. The inner and outer contours correspond to 68\,\% and 95\,\% CL, respectively. The range of e-folds shown here, $N=30-60$, is achievable within nonstandard expansion between inflation and BBN.}
\label{nsrplot}
\end{center}
\end{figure}

In Fig. \ref{nsrplot}, we show the effect of a nonstandard expansion phase on inflationary observables for few example models for a broad range of e-folds achievable within nonstandard cosmology, see Eq. (\ref{NresultSimplified}). We stress that the results only apply to scenarios where inflation was driven by a single scalar field that was in slow-roll at the time the measured perturbations were generated. The results can change substantially in the presence of e.g. a curvaton field (see Sec.~\ref{sec:curvaton}) or in models where multiple fields took part in inflationary dynamics, see e.g. Refs. \cite{Kaiser:2015usz,Bjorkmo:2017nzd,Carrilho:2018ffi} for some recent work on the topic. Also, we note that the best-fit regime for $n_s\,,r$ shown here was obtained from the Planck / Keck Array CMB data. Including other datasets can change the preferred values for these parameters, especially when the new dataset is in tension with the CMB measurements or when new physics is assumed to be involved. For example, the effects of nonstandard late-time expansion history (related to the so-called $H_0$ tension) have been discussed in Refs. \cite{Poulin:2018cxd,Liu:2019dxr,Guo:2019dui} and the effects of nonstandard neutrino interactions and presence of extra radiation on inflationary model selection have been discussed in Refs. \cite{Tram:2016rcw,DiValentino:2016ucb,Gerbino:2016sgw,Barenboim:2019tux}. These aspects are discussed in more detail in Sec. \ref{sec:postBBN} and Sec. \ref{sec:CMB_BBN_constr}.

\subsection{Curvaton scenario (Authors: C. Byrnes \& T. Takahashi)}
\label{sec:curvaton}

Remarkably, as discussed in the previous section, the statistical properties of the millions of temperature perturbation pixels observed in the CMB can be explained in terms of only two parameters -- an amplitude and scale dependence -- describing the primordial perturbations, Eq.~\eqref{Pzeta}. Extensive searches for additional primordial parameters, such as any form of non-Gaussianity, isocurvature perturbations, tensor perturbations or features in the primordial power spectrum have not detected any of these observables at high significance. Whilst these observations are consistent with some simple models of single-field inflation, it remains important to ask in which way the data could deviate from the simplest paradigm, and conversely, whether the absence of additional observables is evidence against any more complex model. 

If only a single light field was present during inflation, then the perturbations today are necessarily adiabatic, meaning that on large scales, the Universe locally looks the same everywhere, with the energy density of all different components having a fixed ratio. In this case, each part of the Universe shares the same history and the (adiabatic) density perturbation can be seen as a time shift between different, super-horizon sized regions. In contrast, multifield inflation necessarily involves the existence of isocurvature perturbations during inflation because the fields will fluctuate differently and hence local patches will contain different ratios of each scalar field. Whether these isocurvature perturbations persist until the present day is model dependent, with knowledge (or assumptions) of the entire reheating history required, from the end of inflation until the primordial components of baryons, DM, neutrinos and radiation have been produced.

\subsubsection{Curvaton perturbations and local non-Gaussianity}

The curvaton scenario \cite{Enqvist:2001zp,Lyth:2001nq,Moroi:2001ct} (see also Refs. \cite{Mollerach:1989hu,Linde:1996gt} for related older works) is arguably the most popular model that can generate large non-Gaussianity \cite{Lyth:2002my,Bartolo:2003jx,Lyth:2005du,Lyth:2005fi,Malik:2006pm,Sasaki:2006kq,Huang:2008ze,Multamaki:2008yv,Beltran:2008aa,Enqvist:2005pg,Enqvist:2008gk,Huang:2008zj,Enqvist:2009zf,Enqvist:2009eq,Enqvist:2009ww,Fonseca:2011aa,Dimopoulos:2003az,Kawasaki:2008mc,Chingangbam:2009xi,Kawasaki:2011pd} and/or isocurvature perturbations \cite{Moroi:2002rd,Lyth:2003ip,Moroi:2008nn,Takahashi:2009cx}, and it relies on the existence of a late decaying scalar field (e.g.~a modulus field \cite{Moroi:2001ct}) and is often connected to an EMD epoch. Both of these observational signatures are potential smoking gun signatures of multiple field inflation, and hence the curvaton scenario is a popular test case for studying inflation with more than one light scalar field present during inflation. The existence of multiple light fields is arguably natural in the context of high energy theories such as supersymmetric theories and string theory. We note that flat directions in such theories, which correspond to  light fields, can generically acquire a Hubble induced mass and become heavy. Such  fields cannot work as a curvaton, however we can still find candidates for the curvaton in such high energy theories (see e.g. Ref. \cite{Mazumdar:2010sa}). Even if one does not accept this motivation, unless the SM Higgs was the inflaton field (which requires a very large nonminimal coupling to gravity), at least two scalar fields were present during inflation.

The curvaton field $\sigma$ is posited to be a light and energetically subdominant field during inflation, which typically has essentially no impact on the background dynamics during inflation. However, provided that the curvaton mass is light compared to the Hubble parameter during inflation ($m_\sigma\ll H$), the curvaton field will be perturbed and its perturbations will form a quasi-scale invariant spectrum. 
In the simplest curvaton model, the curvature perturbation is given by 
\begin{eqnarray} 
\zeta = \frac{2}{3} r_{\rm decay}  \frac{ \delta \sigma_\ast}{\sigma_\ast}, 
\end{eqnarray}
where $r_{\rm decay} \equiv  3\rho_\sigma /( 4 \rho_r + 3 \rho_\sigma)|_{\rm decay}$ models the fraction of the background energy density in the curvaton field at the time it decays, 
with $\rho_\sigma$ and $\rho_r$ the energy densities of the curvaton and radiation evaluated at the time of the curvaton decay, respectively, and $\sigma_\ast$ and $\delta \sigma_\ast$ are the curvaton field and its fluctuations at the time of horizon exit of the scale of interest.
In order for the curvature perturbations generated from the curvaton to become a non-negligible part of the primordial curvature perturbation, for example observed via the CMB temperature perturbations, 
the background energy density of the curvaton must become a significant -- or even dominant -- fraction of the background before it decays, since $r_{\rm dec} \gg 10^{-5}$ given $\delta \sigma_\ast / \sigma_\ast \ll 1$.

Provided that the inflaton field decays first into a radiation bath, the background energy density of the Universe decays like $a^{-4}$ while the curvaton's energy density is initially frozen until $H\sim m_\sigma $, and afterwards decreases like $a^{-3}$ while it oscillates in a quadratic minimum \cite{Lyth:2001nq,Moroi:2001ct}, until the curvaton also decays into radiation. It is therefore a standard requirement of the curvaton scenario that it decays much more slowly than the inflaton field and that its potential has a quadratic minimum.

The curvaton model is a popular way of generating local non-Gaussianity, which measures the correlation between long and short-wavelength modes. It is defined via the bispectrum (related to the 3--point correlation function) $B_\zeta$ as
\begin{eqnarray} 
\langle \zeta({\bf k}_1)\zeta({\bf k}_2)\zeta({\bf k}_3)\rangle = (2\pi)^3\delta^3\left({{\bf k}_1+{\bf k}_2+{\bf k}_3}\right)B_\zeta(k_1,k_2,k_3), 
\end{eqnarray}
where the local bispectrum shape is related to local $f_{\rm NL}$ by
\begin{eqnarray}
B_\zeta^{\rm local}=\frac65f_{\rm NL} \left( P_\zeta(k_1)P_\zeta(k_2) +2\;{\rm permutations}\right).
\end{eqnarray}

In the simplest case, the curvaton is assumed to have a quadratic potential and negligible couplings to the inflaton field (any couplings could allow the inflaton to decay into the curvaton, which can have a big phenomenological impact given the initial subdominance of the curvaton \cite{Linde:2005yw,Sasaki:2006kq,Demozzi:2010aj,Byrnes:2016xlk}). In this case, analytic results exist for the spectral index of the curvaton perturbations and the local non-Gaussianity in terms of $r_{\rm decay}$
\begin{eqnarray}
&& n_s-1=-2\epsilon+2\displaystyle\frac{m_\sigma^2}{3H_\ast^2}, \\  
&& f_{\rm NL}= \displaystyle\frac{5}{4 r_{\rm decay}} - \frac53 - \frac{5 r_{\rm decay}}{6},   \label{eq:fnl}
\end{eqnarray}  
where $\epsilon = - \dot{H}/H^2$ is a slow-roll parameter which is determined by the inflaton potential. In the expression for $n_s$, $H_\ast$ is the Hubble parameter at the time of horizon exit.
Current constraints on the spectral index, which require $n_s<1$ at high significance, require the inflaton field to be of the large-field variety with sufficiently large $\epsilon$ since $ m_\sigma^2 / H_*^2$ always gives a positive contribution 
to the spectral index. This is the case for the simplest curvaton case. However, if one considers an axion-type potential for the curvaton or a self-interaction, the above prediction for $n_s$ and $f_{\rm NL}$ are modified \cite{Enqvist:2005pg,Enqvist:2008gk,Huang:2008zj,Enqvist:2009zf,Enqvist:2009eq,Enqvist:2009ww,Kawasaki:2008mc,Chingangbam:2009xi,Kawasaki:2011pd}.

Therefore, despite the inflaton field perturbations being subdominant (and perhaps completely unobservable, but see Refs. \cite{Langlois:2004nn,Lazarides:2004we,Moroi:2005kz,Moroi:2005np,Ichikawa:2008iq,Fonseca:2012cj,Enqvist:2013paa,Vennin:2015vfa,Fujita:2014iaa,Haba:2017fbi} for mixed models  in which both the inflaton and curvaton perturbations are important), the inflaton potential remains quite tightly constrained, and for example, the ``simplest'' curvaton scenario where both the inflaton and curvaton fields have quadratic potentials is now observationally excluded \cite{Bartolo:2002vf,Byrnes:2014xua}.  However, a quartic inflaton plus quadratic curvaton provides an excellent fit to the data unlike in the single-field case \cite{Enqvist:2013paa}, since, when the curvaton is a dominant source of density fluctuations, the tensor-to-scalar ratio is very suppressed \cite{Langlois:2004nn,Moroi:2005kz,Moroi:2005np,Ichikawa:2008iq} even if one assumes an inflaton potential that would predict a too large tensor-to-scalar ratio in the case that it also generated the observed perturbations\footnote{When both the inflaton and the curvaton contribute to primordial scalar perturbation, the tensor-to-scalar ratio can be written as 
\begin{equation}
    r = \frac{16\epsilon}{1+R}
\end{equation} 
where $R=P_\zeta^{\sigma}/P_\zeta^{\phi}$ with $P_\zeta^{\sigma}$ and $P_\zeta^{\phi}$ primordial power spectra generated from the curvaton $\sigma$ and the inflaton $\phi$. Furthermore, the consistency relation between the tensor spectral index $n_T$ and $r$ can be written as \cite{Langlois:2004nn} 
\begin{equation}
n_T = -(1+R) \frac{r}{8}    \,.
\end{equation}
As can be seen from this expression, when the curvaton perturbation is dominant, $r$ is greatly suppressed and the consistency relation is modified away from the (canonical) single-field inflation case. 
}.
Alternatively, the spectral index can be made red for small-field models of inflation if the curvaton potential is more complex and its effective mass is negative during inflation \cite{Kobayashi:2013bna}. 

Again considering the quadratic curvaton, one can see from Eq.~\eqref{eq:fnl} that large non-Gaussianity is possible if $r_{\rm decay}\ll1$. However, the current Planck non-Gaussianity constraint of $|f_{\rm NL}|\lesssim10$ already requires $r_{\rm decay}\gtrsim0.2$ \cite{Akrami:2019izv} assuming that the curvaton perturbations dominate, so the curvaton must have at least come close to dominating the background energy density of the Universe by the time it decays. Given the short additional time it takes for the curvaton to dominate, 
 $f_{\rm NL}\simeq -5/4$ is a rather natural prediction of the curvaton scenario, independently of the inflaton potential \cite{Hardwick:2015tma,Torrado:2017qtr}. A detection of this value of $f_{\rm NL}$, which may be marginally within reach of sky galaxy surveys such as SPHEREX, Euclid, SKA and so on within a decade \cite{Giannantonio:2011ya,Dore:2014cca,Yamauchi:2014ioa,Karagiannis:2018jdt,Ferraro:2019uce,Meerburg:2019qqi}, would be strong evidence that the Universe underwent an EMD epoch caused by curvaton domination.

By measuring additional non-Gaussianity parameters it is possible to test two of the key assumptions underlying the simplest curvaton scenario, namely whether the curvaton has a quadratic potential and whether all the perturbations were purely generated by a single curvaton field (see Refs. \cite{Byrnes:2014pja,Takahashi:2014bxa} for introductory references). The latter case can be tested by measuring the two trispectrum (4--point correlation function) parameters, $\tau_{\rm NL}$ and $g_{\rm NL}$, with $g_{\rm NL}$ defined by
\begin{eqnarray}
\zeta({\bf x})=\zeta_G({\bf x}) + \frac35 f_{\rm NL} \left( \zeta_G^2({\bf x})-\langle\zeta_G^2({\bf x})\rangle \right) + g_{\rm NL} \zeta_G^3 , 
\end{eqnarray}
while $\tau_{\rm NL}$, which has a different shape dependence than $g_{\rm NL}$, is defined in e.g. Refs.~\cite{Alabidi:2005qi,Seery:2006js,Byrnes:2006vq}. In the case that the perturbations of a single field generate the curvature perturbation, $\tau_{\rm NL}$ must satisfy the consistency relation $\tau_{\rm NL}=\left(6 f_{\rm NL}/5\right)^2$, whilst in general it will be larger \cite{Suyama:2007bg,Suyama:2010uj}. The quadratic curvaton predicts $g_{\rm NL}\sim f_{\rm NL}$ which is unobservably small, but nonquadratic models can (with tuning) produce $|g_{\rm NL}|\gg |f_{\rm NL}|$, as well as a potentially large scale-dependence of $f_{\rm NL}$ \cite{Byrnes:2010xd,Byrnes:2010ft,Huang:2010cy,Byrnes:2011gh,Kobayashi:2012ba,Fonseca:2012cj}. However, given the observational difficulty in determining the trispectrum parameters and/or the scale dependence of $f_{\rm NL}$ and given the current constraint on $f_{\rm NL}$, these higher-order non-Gaussian parameters are only expected to become detectable in quite a limited range of parameter space where the curvaton potential deviates significantly from being quadratic and/or in cases where it generates highly non-Gaussian perturbations but the inflaton perturbations dominate at the linear level \cite{Byrnes:2013qjy}. 

Before closing this section,  we make some comments on the relation of the curvaton to the SM Higgs field.  Although one may be tempted to consider the curvaton field as the Higgs,  the Higgs curvaton would not work within the standard thermal history case \cite{Kunimitsu:2012xx,Choi:2012cp,Tenkanen:2019cik} since the Higgs has a quartic potential, 
implying the oscillating curvaton energy density scales as $\rho_\sigma \propto a^{-4}$, making it impossible to have a large enough fraction of the energy density in the curvaton field to generate significant curvature perturbations.  Allowing an early period of kination alleviates this problem but it remains extremely challenging to make the Higgs work as a curvaton and/or provide the dominant energy source of reheating, even when allowing a nonminimal coupling to gravity \cite{Kunimitsu:2012xx,Figueroa:2016dsc,Figueroa:2018twl,Tenkanen:2019cik}.  The case where the curvaton is coupled to the Higgs has also been studied and in that case it may leave some observable signature \cite{Enqvist:2012tc,Enqvist:2013gwf}.

\subsubsection{Isocurvature perturbations and the evolution of non-Gaussianity after inflation}

The survival of isocurvature perturbations is not generic, because a complete thermalisation of the Universe during reheating, with no conserved quantum numbers, completely converts the initial isocurvature perturbations into adiabatic perturbations \cite{Weinberg:2004kf}.  Given the great uncertainty about how reheating proceeded, the time at which DM was created, the time of baryogenesis, and so on, no strong statements can be made about whether the persistence of isocurvature perturbations should be considered likely or fine-tuned. Hence, the observational absence of any variety of isocurvature perturbations cannot be -- in general -- interpreted as evidence against the existence of multiple light scalar fields during inflation \cite{Huston:2013kgl,Elliston:2013zja}.

If isocurvature perturbations were observed they could potentially provide many additional observables to probe both inflation and reheating, albeit with some degeneracy between the two. The Planck collaboration have searched for many types of isocurvature perturbations, including baryonic, DM, neutrino density and neutrino velocity isocurvature perturbations, in all cases measured relative to the primordial radiation curvature perturbation, and obtained severe constraints on all of these types \cite{Akrami:2018odb}. In all cases, these isocurvature perturbations can be partially or fully correlated (or anti-correlated) with the adiabatic perturbations, and they may also have a very different scale dependence. If all the fields are slowly-rolling during inflation it is reasonable to expect the isocurvature perturbations to also be quasi-scale invariant. However, if the fields that did not take part in driving inflation are in a regime where stochastic quantum fluctuations, rather than the classical slow-roll, dominated their evolution during inflation, the resulting isocurvature spectrum can be strongly scale-dependent \cite{Markkanen:2018gcw,Markkanen:2019kpv,Tenkanen:2019aij,Markkanen:2020bfc}.

In the curvaton scenario, isocurvature fluctuations can also be generated. For example, when DM freeze-out (or production of DM) and/or baryogenesis occur before the curvaton decay,  large (anti-correlated) baryon/DM isocurvature fluctuations can be 
produced \cite{Moroi:2002rd,Lyth:2002my,Lyth:2003ip,Beltran:2008aa,Moroi:2008nn,Takahashi:2009cx}, which are excluded by current data unless the curvature perturbation is  also  generated from the inflaton sector \cite{Moroi:2002rd,Moroi:2008nn,Takahashi:2009cx}.  We note that the issue of isocurvature fluctuations in the curvaton scenario should be carefully investigated especially when the curvaton decays after it dominates the Universe \cite{Kitajima:2017fiy}, which may well be the case given the observational constraint on $f_{\rm NL}$ \cite{Akrami:2019izv}. When DM and/or baryons are generated from the curvaton decay, correlated isocurvature fluctuations whose size depends on the $r_{\rm dec}$ parameter can be generated \cite{Lyth:2002my}. 

Non-Gaussianity of isocurvature fluctuations have also been studied \cite{Kawasaki:2008sn,Langlois:2008vk,Kawasaki:2008pa,Langlois:2011zz}, which can also be useful to constrain models with isocurvature fluctuations, including the curvaton scenario. In general, constraints from power spectrum observations are tighter than those from non-Gaussianities,  however, in some particular cases non-Gaussianities can give a competitive or more severe constraint (for instance, see Refs. \cite{Hikage:2012be,Hikage:2012tf} for the case of axion DM and the curvaton). Since isocurvature fluctuations can be correlated with adiabatic ones, several types of $f_{\rm NL}^{({\rm iso})}$ are defined 
and constraints on those have been investigated by using WMAP data \cite{Hikage:2008sk,Hikage:2012be,Hikage:2012tf} and Planck data \cite{Akrami:2019izv}. Their expected constraints from future observations have been studied too \cite{Langlois:2011hn,Langlois:2012tm}.  For discussion on the trispectrum from isocurvature fluctuations, see Refs. \cite{Kawakami:2009iu,Langlois:2010fe}.

One of the most important impacts of isocurvature perturbations, whether or not they persist until the present day, is that their existence allows the curvature perturbation to evolve on super-horizon scales \cite{Wands:2000dp,Wands:2002bn,Lyth:2004gb}. Therefore, even the predictions of the standard adiabatic perturbations may evolve and therefore, in contrast to single-field inflation, one cannot just match the predictions of the curvature perturbation calculated at horizon exit during inflation to the primordial curvature perturbation. This entails a significant additional complexity, rendering all predictions potentially dependent on the details of the full inflationary history and reheating. The  quadratic curvaton scenario represents a simple case study, a model where initially the curvature perturbation is Gaussian and generated by the inflaton, but as the late-decaying curvaton field starts to dominate the background energy density its perturbations may also dominate and the perturbations become non-Gaussian (and conserved after the curvaton field has decayed). In general, making analytic predictions is hard or even impossible, but extensive work has been done in the case of perturbative reheating following multiple (normally two-field) inflation at the level of the spectral index, bispectrum and trispectrum \cite{Dias:2011xy,Leung:2012ve,Leung:2013rza,Meyers:2013gua,Elliston:2014zea,Hotinli:2017vhx,Gonzalez:2018jax}. Some reasonably model-independent conclusions are that if there is non-negligible non-Gaussianity present at the end of inflation, then its amplitude often changes by an order unity factor until the end of reheating (although potentially switching sign), while the spectral index is limited to lie in between the values set by the smallest and largest values of the spectral indices of the individual scalar fields.

\subsection{Formation of microhalos (Author: A. Erickcek)}
\label{sec:microhalos}

As discussed in previous sections, the expansion history of the Universe affects the evolution of curvature perturbations and the eventual growth of structure.  For adiabatic perturbations, deviations from the standard post-inflationary epoch of radiation domination will only affect perturbations on scales that lie within the cosmological horizon during the period of nonstandard evolution.  The requirement that the Universe be radiation dominated at the time of neutrino decoupling \cite{Kawasaki:1999na, Kawasaki:2000en, Hannestad:2004px, Ichikawa:2005vw,IKT07,DeBernardis:2008zz} ensures that these scales are smaller than 200 pc \cite{Erickcek:2011us}. Since the signatures are confined to such small scales, only dark matter perturbations possibly retain a record of deviations from radiation domination in the early Universe.  Therefore, in this section, we will be primarily interested in the evolution of perturbations in the density of dark matter and how that evolution leaves an imprint on the matter power spectrum on sub-kiloparsec scales. Enhancements to the small-scale dark-matter power spectrum will lead to earlier-forming and more abundant gravitationally bound structures. The amount of dark matter present within the cosmological horizon at the time of neutrino decoupling implies that the masses of these structures are less than 1200 $M_\Earth$, and an earlier onset of radiation domination would confine the enhanced structures to even smaller masses \cite{Erickcek:2011us}.  Earth-mass halos are frequently called microhalos; we adopt that terminology here and extend it to include even smaller halos.

\subsubsection{Growth of density perturbations}
The evolution of dark matter density perturbations is determined by four factors: the expansion rate of the Universe, the evolution of curvature perturbations, interactions between the dark matter and other species, and the random motions of dark matter particles.  The first two factors are determined by the properties of the Universe's dominant component and can readily enhance the growth of dark matter perturbations, while the latter two factors depend on the properties of the dark matter particle and generally suppress the growth of dark matter density perturbations.  

The expansion rate of the Universe and the evolution of curvature perturbations are both determined by the pressure of the dominant component of the Universe.  If the Universe's dominant component is pressureless, then its density scales as $\rho \propto a^{-3}$, and the metric perturbation in Newtonian gauge $\Phi$ remains constant as perturbation modes enter the cosmological horizon \cite{Dodelson:2003ft}.  For perturbation modes with comoving wavenumber $k$, Poisson's equation states that $(k^2/a^2)\Phi = 4\pi G \rho \delta$, where $\delta$ is the fractional density fluctuation.  Since $\rho \propto a^{-3}$, the fact that $\Phi$ is constant implies that $\delta$ grows linearly with the scale factor.  The same perturbation evolution applies to oscillating scalar fields with quadratic potentials.  Since the pressure averages to zero over many oscillations, perturbations with $k/a < \sqrt{3Hm_\phi}$, where $m_\phi$ is the mass of the scalar field $\phi$, grow linearly with the scale factor \cite{Jedamzik:2010dq, Easther:2010mr} and can even form bound structures \cite{Jedamzik:2010hq,Musoke:2019ima,Niemeyer:2019gab}.  Furthermore, scalar fields with potentials that become shallower than quadratic away from their minima quickly form localized structures called oscillons that behave like massive particles \cite{Amin:2011hj, Amin:2019ums}.  Therefore an EMD epoch characterized by $\rho \propto a^{-3}$ and constant metric perturbations on subhorizon scales may be caused by massive particles or scalar fields.  In both cases, transitioning to radiation domination requires the dominant component of the Universe to decay into relativistic particles, at which point the subhorizon metric perturbations undergo decaying oscillations \cite{Erickcek:2011us}.

If dark matter does not interact with other particles, then the Boltzmann equation that governs the evolution of dark matter perturbations only depends on expansion rate and the metric perturbation $\Phi$, see e.g. Ref. \cite{Redmond:2018xty}:
\begin{equation}
a^2 \delta_\mathrm{DM}'' +\left[3+\frac{aH'}{H}\right]a\delta_\mathrm{DM}'= \frac{k^2}{a^2H^2}\Phi - 3\left[3+\frac{aH'}{H}\right]a\Phi'-3a^2\Phi'',
\label{eqn:deltaevol}
\end{equation}
where the prime denotes differentiation with respect to the scale factor $a$.
During an EMD epoch, $\Phi$ is constant and $H(a)\propto a^{-3/2}$, in which case Eq.~(\ref{eqn:deltaevol}) implies that $\delta_\mathrm{DM}$ is constant on superhorizon scales ($k\ll aH$) and grows linearly with the scale factor on subhorizon scales ($k\gg aH$).  This evolution is not altered if the dark matter is produced by the decay of the dominant component during the EMD epoch \cite{Erickcek:2011us}.  If dark matter is initially in chemical equilibrium with relativistic particles and then freezes out, dark matter annihilations alter the growth of perturbation modes that enter the horizon prior to freeze-out, but the dark matter particles' subsequent fall into the gravitational wells created by growing perturbations in the dominant component washes out this effect. If the freeze-out temperature is greater than twice the reheating temperature, there is no significant change in $\delta_\mathrm{DM}$ at the end of the EMD epoch compared to the noninteracting case \cite{Erickcek:2015jza}.  If dark matter is produced by the decay of the dominant component and then self-annihilates at the end of the EMD epoch, $\delta_\mathrm{DM}$ decreases by roughly a factor of ten at reheating relative to the noninteracting case due to the higher annihilation rate in overdense regions \cite{Fan:2014zua}.  A similar suppression occurs if dark matter is co-decaying \cite{Dror:2017gjq}.  For modes that enter the horizon well before the end of the EMD epoch ($k \gtrsim \sqrt{10} a_\mathrm{reh} H_\mathrm{reh}$, where $a_\mathrm{reh}$ and $H_\mathrm{reh}$ are the scale factor and Hubble rate, respectively, when the Universe becomes RD), the reduction in $\delta_\mathrm{DM}$ due to annihilations at reheating is less than the growth of the $\delta_\mathrm{DM}$ during the EMD epoch, implying that an EMD epoch still enhances the amplitude of density perturbations on these scales.
 
If the dominant component of the Universe has significant pressure, $\Phi$ rapidly decays after the perturbation mode enters the horizon \cite{Redmond:2018xty}.  The right-hand-side of Eq.~(\ref{eqn:deltaevol}) becomes negligible, implying that
\begin{equation}
\delta_\mathrm{DM} \propto \int^a \frac{da}{a^3H(a)}.
\label{eqn:drift}
\end{equation}
This solution has a simple physical interpretation.  Immediately after a perturbation mode enters the horizon, dark matter particles are gravitationally pulled toward overdense regions. When $\Phi$ decays, the particles stop accelerating and drift toward the initially overdense region with physical velocity $v\propto 1/a$.  The comoving displacement of these particles is $\int v \,dt/a = \int v \,da/(a^2H)$, and the convergence and divergence of these displacements results in a growing dark matter density perturbation.  Equation (\ref{eqn:drift}) recovers the well-known result that $\delta_\mathrm{DM}$ grows logarithmically with scale factor while the Universe is radiation dominated, but it also applies to any expansion phase that includes vanishing gravitational perturbations.  If the Universe is dominated by an energy source that is stiffer than radiation, then $\delta_\mathrm{DM}$ will grow faster than it does during radiation domination: $\delta_\mathrm{DM} \propto a^{(3w-1)/2}$, where $w$ is the equation of state parameter for the dominant component of the Universe.  Therefore, any expansion phase with $w > 1/3$ will enhance dark matter density perturbations on subhorizon scales.   An extreme example is a period of kination, during which the Universe is dominated by a fast-rolling scalar field ($w\simeq1$) and $\delta_\mathrm{DM}$ grows linearly with the scale factor \cite{Redmond:2018xty}.  Conversely, an expansion phase with $0<w<1/3$ will slightly suppress subhorizon perturbations because these modes will stagnate instead of growing logarithmically with the scale factor after horizon entry.

The impact of a nonstandard expansion phase on the matter power spectrum depends on the evolution of both $\delta_\mathrm{DM}(a)$ and $H(a)$.  The linear growth of $\delta_\mathrm{DM}(a)$ during both an EMD epoch and a period of kination implies that $\delta_\mathrm{DM}$ at the onset of radiation domination will be proportional to $a_\mathrm{reh}/a_\mathrm{hor}$, where $a_\mathrm{hor} = k/H(a_\mathrm{hor}) $ is the value of the scale factor when the mode enters the horizon.  During an EMD epoch, $a_\mathrm{hor} \propto k^{-2}$, whereas  $a_\mathrm{hor} \propto k^{-1/2}$ during kination.  In both cases, $\delta_\mathrm{DM}(k)$ is also proportional to the value of $\Phi(k)$ while the mode is outside the horizon, which has a dimensionless power spectrum proportional to $k^{n_s-1}$.  It follows that the dimensionless matter power spectrum $\Delta^2(k)$ will be proportional to $k^{3+n_s}$ for modes that enter the horizon during an EMD epoch and will be proportional to $k^{n_s}$ for modes that enter the horizon during kination.  For a general expansion phase with $w>1/3$, $\Delta^2(k) \propto k^{n_s+(3w-3)/(3w+1)}$ for modes that enter the horizon during this period.  All of these power spectra imply enhanced inhomogeneity on small scales compared to the $\Delta^2(k) \propto k^{n_s-1}\ln^2[k/(8k_\mathrm{eq})]$ scaling for modes that enter the horizon during radiation domination \cite{HS96}.  

\subsubsection{Cutting off microhalo formation}

Since $\Delta^2(k)$ increases as $k$ increases for modes that enter the horizon during an EMD epoch or during an expansion phase with $w\geq1/3$, smaller scales are increasingly inhomogeneous and could collapse to form bound structures at arbitrarily early times.  It is not possible for $\Delta^2(k)$ to continue monotonically increasing with $k$ indefinitely, however.  As an extreme limit, the power spectra derived above do not apply to modes that are so small that they never exit the horizon during inflation.  In most cases, the growth of the matter power spectrum cuts off at much larger scales than the horizon size at the end of inflation due to the dark matter particles' thermal motions or their interactions with other particles. The properties of the component that dominates the Universe during an EMD epoch can also suppress dark matter perturbations: if this component has significant pressure support prior to the EMD epoch, dark matter perturbations on scales that are affected by this pressure will not be enhanced during the EMD epoch \cite{Blanco:2019eij, Erickcek:2020wzd}.

If the dark matter interacts with relativistic particles, then perturbations that enter the horizon while the dark matter is kinetically coupled to these relativistic particles will experience damped oscillations between horizon entry and the decoupling of the dark matter, implying that $\Delta^2(k)$ decreases with increasing $k$ for such modes, see e.g. Ref. \cite{Bertschinger06}.  There is an exception to this generic behavior during an EMD epoch, however.  During an EMD epoch, the dominant matter component decays into relativistic particles, and once any pre-EMD epoch radiation has sufficiently diluted, the energy density in these particles scales as $\rho_r \propto a^{-3/2}$ due to the production of new relativistic particles \cite{Giudice:2000ex}.  At this point, density perturbations in these relativistic particles do not oscillate, but rather grow until they reach a steady state during which the production of new relativistic particles in increasingly overdense regions is balanced by the outflow of relativistic particles from these regions \cite{Erickcek:2011us, Fan:2014zua}.  If dark matter is in kinetic equilibrium with the relativistic particles, the dark matter particles are pulled out of the overdense regions by the outflow of relativistic particles, and the dark matter accumulates in underdense regions.  If the dark matter remains in kinetic equilibrium through reheating, the dark matter remains concentrated in these regions, resulting in large isocurvature perturbations on scales that enter the horizon during the EMD epoch \cite{2015PhRvL.115u1302C}.  If the dark matter kinetically decouples prior to reheating, the dark matter particles quickly fall into the gravitational wells created by density perturbations in the dominant component, and the dark matter power spectrum is not significantly different from the noninteracting case \cite{WEIinprep}.   Therefore, interactions between dark matter particles and the relativistic particles produced during an EMD epoch do not suppress perturbations on scales that enter the horizon prior to kinetic decoupling, and perturbations on these scales are still significantly enhanced by an EMD epoch.  

After dark matter particles kinetically decouple from all other particles, their random thermal motions allow them to free stream out of overdense regions.  Perturbations on scales smaller than the comoving free-streaming horizon are exponentially suppressed, with $\Delta^2(k) \propto e^{-k^2/k_\mathrm{fs}^2}$, where $k_\mathrm{fs}$ is the inverse of the comoving free-streaming horizon, see e.g. Ref. \cite{Bertschinger06}. If dark matter is a decay product of the species that dominates the Universe during an EMD epoch and never interacts with other particles, its free streaming will inhibit the formation of microhalos unless its initial velocity is much less than the speed of light \cite{Erickcek:2011us, Fan:2014zua, 2019PhRvD.100l3520M}. If dark matter interacts with other particles, the free-streaming horizon is determined by the velocity of the dark matter particles when they kinetically decouple. During a nonstandard expansion phase, the Hubble rate is larger than it is during a radiation-dominated phase with the same temperature, so DM particles will decouple from SM particles at a higher temperature than they would during radiation domination \cite{GG08, Visinelli_2015, Waldstein_2017}. Since the velocities of dark matter particles decrease as $v\propto 1/a$ after decoupling, an earlier decoupling reduces the free-streaming horizon.  If dark matter kinetically decouples during an EMD epoch, the free-streaming horizon is further reduced because the continual production of relativistic particles during an EMD epoch slows their cooling, leading to a greater reduction in the dark matter particles' velocities over a given temperature range.  \citet{GG08} showed that the minimum halo mass set by the free-streaming horizon can be reduced by several orders of magnitude if the dark matter kinetically decouples during an EMD epoch.   

Kinetic decoupling during an EMD epoch is more complicated than \citet{GG08} realized, however.  If dark matter is in kinetic equilibrium with the relativistic particles that are created during an EMD epoch, the slow cooling of these particles implies that the dark matter is still heated by its interactions with these particles long after the momentum transfer rate falls below the Hubble rate, which is the usual criterion for kinetic decoupling \cite{WEI17}.  The standard method of calculating the dark matter temperature \cite{Bertschinger06, 2007JCAP...04..016B} indicates that this ``quasi-decoupled" state persists until reheating, but this approach assumes that each collision causes a small fractional change in the dark matter particles' momenta. During an EMD epoch, this assumption ceases to hold at about the same time that the Hubble rate equals the collision rate \cite{WEIinprep}, which is roughly $m_\mathrm{DM}/T$ times the momentum transfer rate \cite{2007JCAP...04..016B}.  Since we do not know how the temperature of the dark matter evolves after this time during an EMD epoch, we cannot calculate the free-streaming horizon if dark matter kinetically decouples from SM particles during an EMD epoch.

If the nonstandard expansion phase lasts long enough for $\delta_\mathrm{DM}$ to grow larger than one, it is possible to form microhalos prior to reheating.  Structure formation during an EMD epoch would progress similarly to structure formation during matter domination: gravitational forces cause overdense regions to break away from the Hubble flow and collapse into bound structures when the value of $\delta$ predicted by linear theory exceeds 1.686.  If the collapsing region is sufficiently spherical and homogeneous, it will form a PBH, as will be discussed in Sec. \ref{sec:formationPBHs}. The more common outcome of gravitational collapse is the formation of a virialized microhalo that is primarily composed of whatever matter-like substance dominates the Universe during the EMD epoch, with just a small portion of the mass coming from dark-matter particles.\footnote{If dark matter is co-decaying \cite{Dror:2016rxc}, then the halos will be nearly half dark matter, but the vast majority of these particles will be annihilated during reheating.}  When the dominant component during the EMD epoch decays, these halos will become unbound, and the dark matter particles within these halos will be released.  These particles have higher velocities than unbound particles, and even more importantly, the directions of these velocities are randomized.  The subsequent free streaming of these gravitationally heated particles erases the perturbations that grow significantly during the EMD epoch, thus preventing the later formation of microhalos \cite{Blanco:2019eij}.  Structure formation while the Universe is dominated by a component with $w\geq 1/3$ is less understood because the dark matter particles are merely drifting towards each other, but \citet{Blanco:2019eij} demonstrated that it is possible to form dark matter halos during radiation domination if the convergence of the drifting particles causes a region to become locally matter dominated.  Unlike the halos that form during an EMD epoch, these halos would be composed solely of dark matter particles and would survive the transition to radiation domination.

\subsubsection{The microhalo population}
The first microhalos will form on scales with the largest values of $\Delta^2(k)$, therefore the peak scale in the power spectrum sets the mass of the first microhalos: $M\simeq (4\pi/3)\rho_\mathrm{DM,0}k_\mathrm{peak}^{-3}$, where $\rho_\mathrm{DM,0}$ is the present-day DM density.  Roughly speaking, these halos will form when $\Delta^2(k_\mathrm{peak})$ exceeds unity.  For halos that form during the matter-dominated era, when $\Delta^2(k) \propto a^{2}$, the typical value of the scale factor when halos form is proportional to $1/\sqrt{\Delta^2(k)}$.  Since $\Delta^2(k)$ increases very slowly with $k$ for modes that enter the horizon during radiation domination, a nonstandard expansion phase reduces the typical value of the scale factor at halo formation by roughly a factor of $\sqrt{\Delta^2(k_\mathrm{reh})/\Delta^2(k_\mathrm{peak})}$, where $k_{\rm reh} = a_{\rm reh} H_{\rm reh}$.  The peak scale is determined by the cut-off scale in the matter power spectrum, so the formation time of the first halos is most sensitive to the ratio $k_\mathrm{cut}/k_\mathrm{reh}$, with only a weak dependence on the reheat temperature \cite{Erickcek:2011us, Erickcek:2015jza}.  The formation time of a halo determines its density, and \citet{delos2019predicting} demonstrated that the density profiles of these first halos can be predicted directly from $\Delta^2(k)$.  

Looking beyond the first halos, the halo abundance predicted by Press-Schechter theory \cite{Press:1973iz} is enhanced for all scales with enhanced $\Delta^2(k)$.  The number density of such halos depends on their masses, but the total fraction of dark matter contained in halos at a given time is largely independent of $k_\mathrm{reh}$ and instead depends on the maximal enhancement of $\Delta^2(k)$, as determined by $k_\mathrm{cut}/k_\mathrm{reh}$, see e.g. Ref. \cite{Erickcek:2015jza}.  Following a nonstandard expansion phase that significantly enhances $\Delta^2(k)$ on small scales, over half of the dark matter can bound into microhalos at redshifts well above 100, long before structure is expected to form in standard cosmologies \cite{Erickcek:2011us, Erickcek:2015jza}.  It is even possible to form microhalos prior to matter-radiation equality \cite{Blanco:2019eij}.

\subsection{Formation of primordial black holes (Authors: T. Harada \& K. Kohri)}
\label{sec:formationPBHs}

Primordial black holes (PBHs) may have formed in the early Universe~\cite{Zeldovich:1967,Hawking:1971ei}. In the present Universe, PBHs can act as gravitational sources such as DM, or as GW sources. Furthermore, since the mass of the PBHs can be very small and  they can lose their mass by the Hawking radiation, PBHs provide us with a unique probe into high-energy physics and quantum effects of black holes. For this reason, PBHs have attracted intense attention not only in cosmology but also in other areas of fundamental physics.  

In previous sections, we have discussed mechanisms to produce curvature or density perturbations by models of inflation and curvaton. Because the quantum generation of fluctuations in such scenarios implies a small but nonvanishing probability of large amplitude of perturbations, the existence of PBHs is possible. Therefore, an interesting question is how many PBHs could have formed in the early Universe. In this section, we discuss PBHs formed through the collapse of overdense regions due to large density perturbations at small scales. For a recent review paper including other mechanisms for PBH formation, we refer to Ref.~\cite{Carr:2020gox}.

\subsubsection{Radiation-dominated universe}

In the early Universe, naively one would expect that the RD Universe was realized just after inflation, where the radiation temperature was larger than $\mathcal{O}(1)$\,MeV to be consistent with BBN (see Sec.~\ref{sec:CMB_BBN_constr}). To theoretically predict the abundance of PBHs, it is important to identify the conditions for PBH formation in terms of the amplitude of perturbations. A region with a large density perturbation $\delta$ which is enclosed by a length scale $R$ can collapse into a PBH. In 1975, B. J. Carr proposed a criterion by which the overdense region becomes a black hole if $R$ is larger than the Jeans radius at the time of its maximum expansion~\cite{Carr:1975qj}. His criterion gives $\delta > \delta_c \sim c_s^2$ with the sound speed $c_s^2 = w$ where $w$ is the EoS parameter. In a RD Universe, the criterion is $\delta_c \sim 1/3$. With more sophisticated methods in general relativity,  threshold values in the range $\delta_c \simeq  0.4 - 0.7$ have been reported. For example, spherically symmetric numerical relativity simulations report $\delta_{c}\simeq 0.42-0.66$~\cite{Musco:2008hv,Musco:2012au,Harada:2015yda,Musco:2018rwt}. Analytically, the authors of Ref.~\cite{Harada:2013epa} derived the formula
\be
\delta_c = \frac{3(1+w)}{5+3w}\sin^{2}\left(\frac{\pi\sqrt{w}}{1+3w}\right)\,,
\ee
which gives $\delta_c \approx 0.41$ in a RD Universe. However, it is worth remarking that the threshold actually depends on the spatial profile of the overdense region~\cite{Musco:2004ak,Musco:2008hv,Musco:2012au,Harada:2013epa,Musco:2018rwt,Germani:2018jgr,Shibata:1999zs,Harada:2015yda,Escriva:2019phb,Nakama:2013ica}. It should also be noted that near the threshold, there appear critical phenomena, such as the scaling law of the black hole masses and the self-similar solution at the  threshold~\cite{Niemeyer:1999ak,Shibata:1999zs,Musco:2012au}, which were originally discovered in the Einstein-scalar field system in an asymptotically flat  spacetime~\cite{Choptuik:1992jv}. The scaling law of PBH masses significantly deforms the observational bound on the power spectrum of the perturbation because the PBH mass can be much smaller than the mass enclosed within the Hubble horizon at the horizon entry of the perturbation~\cite{Niemeyer:1997mt,Yokoyama:1998xd}.

\begin{figure}
\begin{center}
\includegraphics[clip,width=8cm]{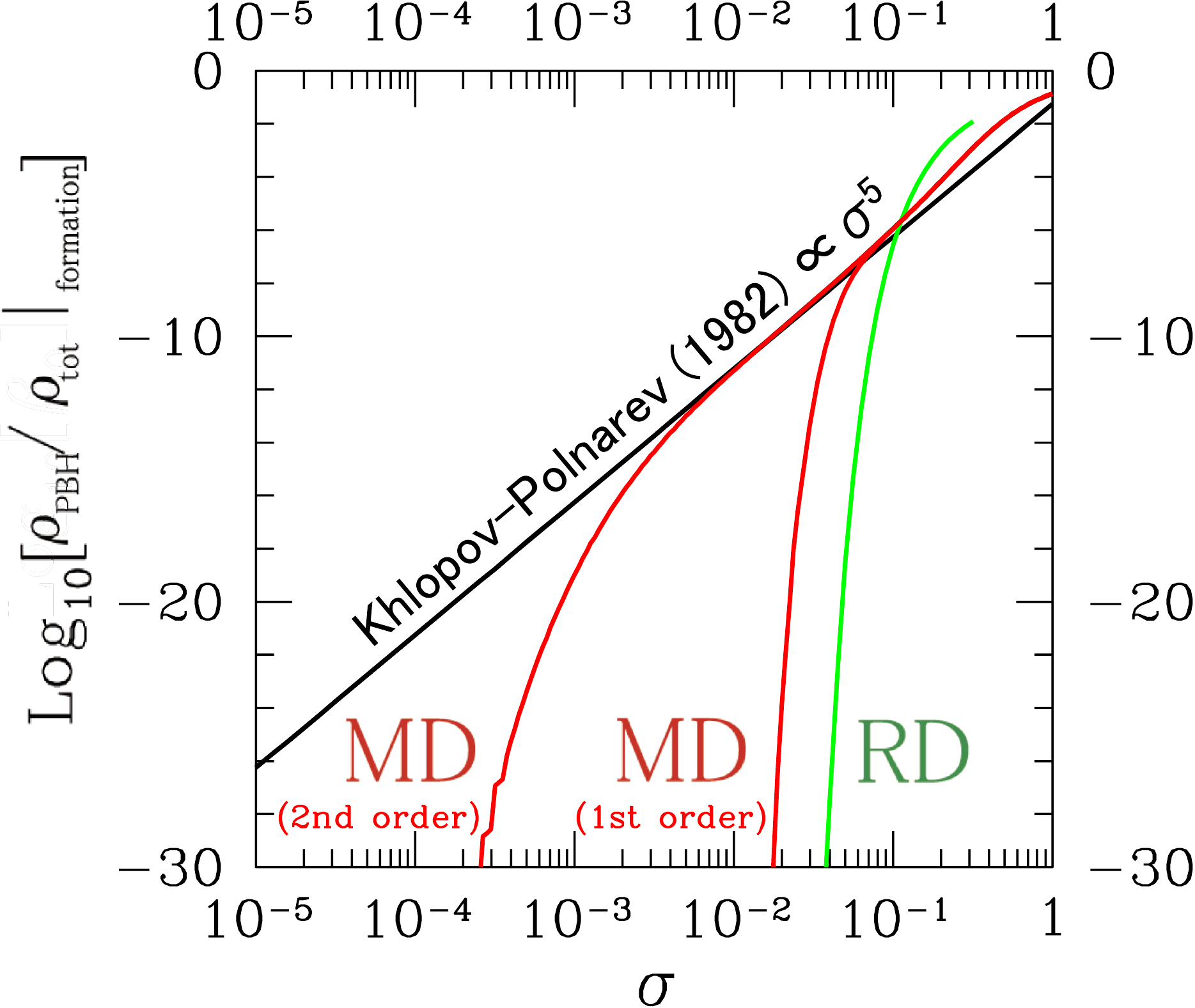}
\caption{The PBH abundance $\beta$ as a function of the mean variance $\sigma$ of the
  density perturbation $\delta$. The cases of the radiation domination
  (here: RD)~\cite{Harada:2013epa} and the early matter domination (here: MD) for
  both the first- and second-order effects~\cite{Harada:2017fjm} are
  plotted by the green and two red curves, respectively. The line of the
  power law $=0.05556 \sigma^5$ denotes the one without suppression
  by finite angular momentum in case of the early MD studied
  by~Refs.\cite{Khlopov:1980mg,Polnarev:1982,Harada:2016mhb}. The curves
  are from Ref.~\cite{Harada:2017fjm}.  }
\label{fig:beta_2cases} 
\end{center}
\end{figure}

To obtain the abundance of PBHs as a function of density perturbation or a variance $\sigma^2$ of it, for simplicity we can follow the Press-Schechter formalism~\cite{Press:1973iz}. A PBH can form just after a region with $\delta > \delta_c$ enters the Hubble horizon. By assuming that $\delta$ obey a Gaussian distribution with the variance $\sigma^2$, the probability of the PBH formation is given by
\be \label{eq:betaRDanalytic}
\beta = \int_{\delta_{c}}^{{\cal O}(1)}\frac{1}{\sqrt{2\pi\sigma^{2}}}\exp\left(-\frac{\delta^{2}}{2\sigma^{2}}\right)d\delta \simeq \mbox{erfc}\left(\frac{\delta_{c}}{\sqrt{2}\sigma}\right) \simeq \sqrt{\frac{2}{\pi}} \frac{\sigma}{\delta_{c}}\exp\left(-\frac{\delta_{c}^{2}}{2\sigma^{2}}\right).
\ee
It is notable that $\beta$ is nothing but $\rho_{\rm PBH}/\rho_{\rm tot}$ at the time of formation of the PBHs where $\rho_{\rm PBH}$ and $\rho_{\rm tot}$ are the energy densities of PBHs and the background, respectively.  In Fig.~\ref{fig:beta_2cases}, we plot this RD case with a green curve.

For simplicity, we show an approximate relation between the mass $m_{\mathrm{PBH}}$ and the cosmic time at the formation $t_{\rm form}$ for a PBH formed during a RD epoch:
\be  \label{eq:horizonMass}
  m_{\mathrm{PBH}} \sim m_{\rm P}^{2} t_{\mathrm{form}} \sim \frac{m_{\rm P}^{3}}{T_{\mathrm{form}}^{2}} \sim 10^{15} {\rm g}\left(\frac{T_{\mathrm{form}}}{3 \times 10^{8} \mathrm{GeV}}\right)^{-2} \sim 30 M_{\odot}\left(\frac{T_{\mathrm{form}}}{40 {\rm MeV}}\right)^{-2},
\ee 
where $m_{\rm P} = M_{\rm P}/\sqrt{8 \pi} \simeq 2.4 \times 10^{18}$GeV is the reduced Planck mass and $T_{\rm form}$ is the radiation temperature at the time of the PBH formation.

Compared to the above method, more sophisticated ones have been reported by Refs.~\cite{Yoo:2018esr,Kalaja:2019uju} for peak statistics, Ref.~\cite{Ando:2018nge,Young:2019osy,Tokeshi:2020tjq} for choices of the window function, Refs.~\cite{Yoo:2018esr,Germani:2018jgr,Kawasaki:2019mbl} for nonlinear relations between curvature and density perturbations and Refs.~\cite{Suyama:2019npc,Germani:2019zez,DeLuca:2020ioi,Young:2020xmk} for possible extensions of peak theory. As is discussed in Ref.~\cite{Yoo:2020dkz}, it is remarkable that the window-function dependencies vary more in the Press-Schechter formalism than those in peak theory.

\subsubsection{Matter-dominated universe}
\label{subsec:FromMD}

Next we discuss formation of PBHs during an EMD epoch. As was discussed in the previous sections, in modern cosmology it is known that an EMD epoch may have been realized just before the onset of the standard RD Universe. Considering the formation of PBHs during an EMD epoch, a severe problem is that there exist two conflicting effects: i) The EoS parameter $w$ is nearly zero since the cosmic pressure is negligible. Thus, as we can see in Eq.~\eqref{eq:betaRDanalytic} which gives $\delta_c = 0$, more PBHs could be produced. ii) On the other hand, density perturbation can evolve during a MD epoch, which means that also anisotropies of the density contrast should have evolved in the 3D space. In this case, a collapsed region may not be easily enclosed by the (event) horizon.

To calculate the formation rate of PBHs during a MD epoch, in Ref.~\cite{Harada:2016mhb} we have applied the Zel'dovich approximations to the collapsing objects anisotropically in the 3D space. To be a PBH, we judged if the region is enclosed by the apparent horizon by adopting the Hoop Conjecture for such non-spherical object. Due to a finite angular momentum induced by the anisotropic collapse, the abundance of PBHs produced in this case is exponentially suppressed~\cite{Harada:2017fjm}. This means that a produced PBH tends to be spinning nearly extremely. In other words, a more spinning object than an extremal one cannot become a PBH, but instead a kind of halo such as a CDM halo around a galaxy should form, as discussed in Sec.~\ref{sec:microhalos}. 

\begin{figure}
\begin{center}
\includegraphics[clip,width=6cm]{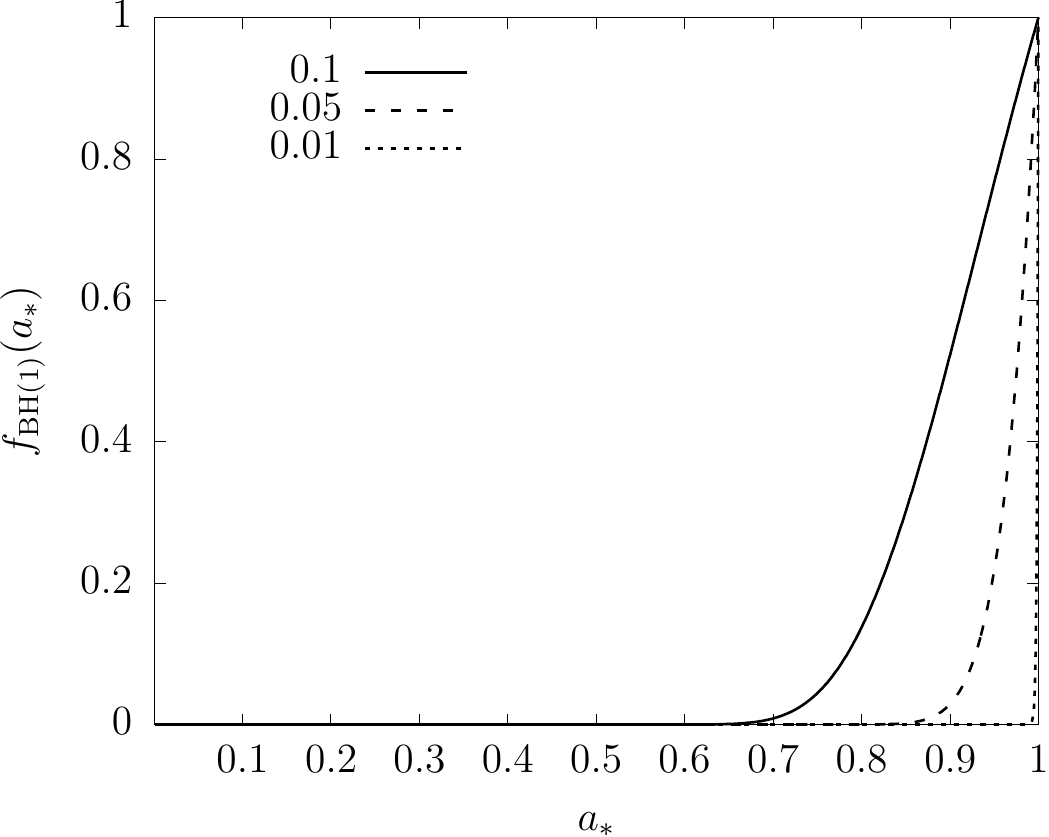} \hspace{1cm}
\includegraphics[clip,width=6cm]{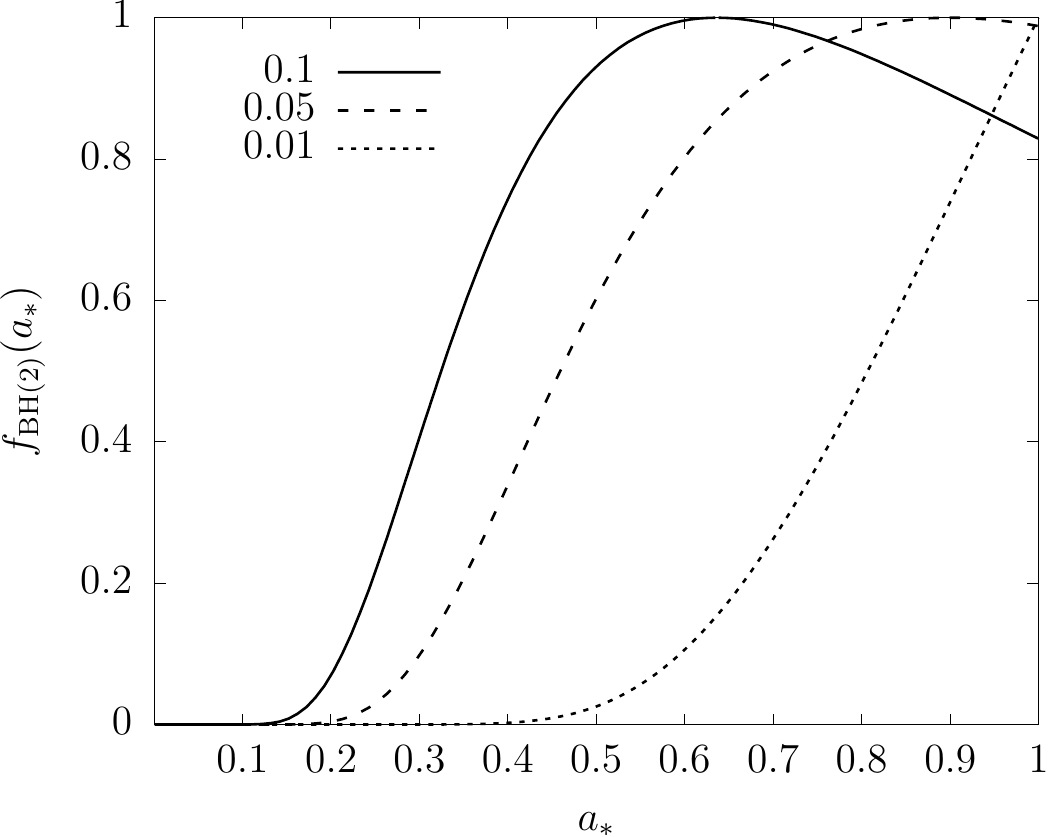}
\caption{Spin distribution of PBHs formed during an EMD epoch due to the first-order effect (left) and the second-order effect (right)~\cite{Harada:2017fjm}. The $x$-axis is the normalized Kerr parameter $a_*= L/(Gm_{\rm PBH}^2/c^2)$ with $L$ being the angular momentum. The curves denote the spin distribution functions normalized by their maximum values for the variance of density fluctuations, depending on $\sigma=0.1$, $0.05$, and $0.01$. This figure is from Ref.~\cite{Harada:2017fjm}.}
\label{fig:spin_1st} 
\end{center}
\end{figure}

In Fig.~\ref{fig:spin_1st}, we plot the spin distribution of PBHs formed in a MD epoch due to the first-order effect, discussed in Ref.~\cite{Harada:2017fjm}. The $x$-axis is the normalized Kerr parameter $a_*= L/(G m_{\rm PBH}^2/c^2)$ where $L$ is the angular momentum of the PBH. The curves denote the spin distribution functions normalized by their maximum values for the variance of density fluctuations, depending on $\sigma$. In Fig.~\ref{fig:beta_2cases}, we instead plot the abundance of PBHs formed during a MD epoch as a function of the mean variance $\sigma$ for two cases separately calculated by the first- and second-order effects reported by Ref.~\cite{Harada:2017fjm}. A finite angular momentum can be induced due to the first-order effect multiplying the position vector $\vec{x}$ by a perpendicular component of the peculiar velocity perturbation $\vec{u}$, and the second-order effect multiplying the density perturbation by $\vec{u}$. The first-order effect should dominate the second-order one unless there is a strong mechanism to make the collapsing region very spherical. The difference between these two lines can be understood to be the theoretical uncertainty in estimating the abundance of PBHs. In order to exclude parameters which appear in nonstandard theoretical models in which PBHs form during a MD epoch, one can use the first-order effect to obtain more conservative limits on the parameters. On the other hand, when one looks for the allowed regions of parameters in a theoretical model to fit observational data with PBHs produced during a MD epoch, the second-order effect is recommended to be used, as it fits data more conservatively with milder assumptions of the theoretical model. For details, see~\cite{Harada:2017fjm}. The curves obtained by the numerical computations can be fitted by the following semi-analytic expressions:
\begin{equation}
\beta_{(1)}\simeq 3.244\times 10^{-14} \frac{{q}^{18}}{\sigma^{4}}\exp\left[-0.004608 \frac{{q}^{4}}{\sigma^{2}}\right]
\label{eq:semianalytic_1st}
\end{equation}
for the first-order effect with $q=1/2$, and
\begin{equation}
\beta_{(2)}\simeq  1.921\times 10^{-7} \sigma^{2}\exp \left[-0.1474 \frac{1}{\sigma^{2/3}}\right]
\label{eq:semianalytic_2nd}
\end{equation}
for the second order effect~\cite{Harada:2017fjm}.

We also refer to Ref.~\cite{Kokubu:2018fxy} for another effect, which allows to simply multiply the expression for $\beta$ by an additional power $\propto \sigma^{3/2}$ induced by the initial inhomogeneity of the Universe. However, a complete analysis of generation of finite spin and suppression of $\beta$ with an initial inhomogeneity has not been fully done. Therefore, for the moment, we recommend the readers to discuss these two effects separately.

\section{Constraints on nonstandard expansion phases}
\label{sec:constraints}

In this section, we will discuss the known constraints on extra components that can affect the expansion of the Universe, as well as some future ways to put such scenarios into test. We will discuss BBN and CMB constraints and observational prospects for detecting microhalos and primordial black holes, as well as stochastic gravitational wave backgrounds.

\subsection{BBN and CMB constraints (Authors: M. Escudero \& V. Poulin)}
\label{sec:CMB_BBN_constr}

In this section we will discuss how observations of the CMB and the primordial element abundances can be used to obtain information about both the early and late (pre- and post-BBN) expansion history of the Universe, and how the topic has been discussed in the literature. To illustrate the constraining power of current cosmological observations, we outline in Table~\ref{tab:BBN_CMB_constraints} the up to date CMB and BBN constraints on key parameters constraining nonstandard expansion histories.

We start by discussing the BBN constraints. BBN represents the earliest epoch of the Universe from which we have data. The agreement between the observed primordial abundances of helium and deuterium with the predicted ones is a highly nontrivial success of the standard cosmological model. Equivalently, BBN serves as a powerful probe of new physics and nonstandard cosmological expansion histories. 

\subsubsection{A brief review of standard BBN} 

Big Bang Nucleosynthesis occurred when the Universe was about three minutes old. During BBN ($1\,\text{keV} \lesssim  T \lesssim 10\,\text{MeV}$), the Universe was radiation dominated and the only relevant species were electrons, positrons, photons, three neutrinos species, and a small number of protons and neutrons, $n_b/n_\gamma \sim 10^{-9}$. At temperatures $T \gtrsim 10\,\text{MeV}$ all of these species were tightly coupled via electromagnetic and weak interactions. However, the rate of weak interactions ($\Gamma \sim G_F^2 T^5$) fell below the expansion rate $H$ at a temperature $T \sim (G_F^2 M_{\rm P})^{-1/3} \sim 1\,\text{MeV}$. In particular, neutrino-electron interactions froze out at $T_{\nu}^{\rm dec} \sim 2\,\text{MeV}$~\cite{Dolgov:2002wy}. From then on, neutrinos evolved as a decoupled fluid while at $T_\gamma \sim m_e$ electrons and positrons started to annihilate into photons heating up the photon fluid with respect to the neutrino one, leading to temperature ratio $T_\gamma/T_\nu \simeq (11/4)^{1/3} \simeq 1.4$~\cite{Dodelson:2003ft}. Soon after neutrino decoupling, at $T_{\rm np} \sim 0.7\,\text{MeV}$, interactions interconverting protons to neutrons also froze out ($t_{\rm np} \sim 1\,\text{s}$). At temperatures $T \gtrsim 0.1\,\text{MeV}$, interactions between bounded nuclei and the plasma were highly efficient, and light nuclides were kept in kinetic equilibrium. During this epoch, given the high temperature of the plasma and the small baryon-to-photon ratio, every bounded nuclei was rapidly destroyed by energetic species in the plasma. This situation changed dramatically when the temperature of the Universe dropped below $T_\gamma^{\rm D} \simeq 0.073 \,\text{MeV}$ (corresponding to $t_{\rm D}\simeq 241\,\text{s}$). At this point, the Universe was cold enough for photons not to be capable of disassociating deuterium any more. Soon after this happened, a chain reaction was triggered and almost all free neutrons in the plasma formed Helium-4 (${}^{4}{\rm He}$). In particular, the standard BBN leads to a Universe in which roughly $\sim 75\%$ of the baryonic energy density is stored in the form of hydrogen and where the $\sim 25\%$ remaining baryonic mass is primarily in the form of (${}^{4}{\rm He}$). In addition, small (but relevant) abundances of heavier nuclei remained: a fraction $10^{-5}-10^{-3}$ of deuterium (${}^{2}{\rm H}$) and ${}^{3}{\rm He}$, and only very small ($\lesssim 10^{-9}$) abundances of heavier elements such as ${}^{7}{\rm Li}$ and ${}^{6}{\rm Li}$ were synthesised. 

Within the standard BBN, the only parameter needed to predict the light element abundances is the baryon energy density $\Omega_b h^2$, or interchangeably the baryon-to-photon number density ratio $\eta$: $\Omega_b h^2/0.02237 \simeq \eta/(6.112\times10^{-10}) $~\cite{pdg}. The ${}^{4}{\rm He}$ and deuterium primordial abundances are now measured with 1\% precision~\cite{pdg}. Their observed values are fully compatible with those predicted within the standard BBN and point to $0.021 < \Omega_bh^2 < 0.024$ at 95\,\% CL. This range is in excellent agreement with the values inferred from CMB observations within the framework of $\Lambda$CDM~\cite{Aghanim:2018eyx}. Hence, this agreement represents a clear triumph of the standard cosmological model.  

Only the measured abundance of ${}^{7} {\rm Li}$ does not match the the standard BBN prediction. The measured ${}^{7} {\rm Li}$ abundance is a factor of $\sim 3$ smaller than that expected from standard BBN assuming the baryon abundance measured from CMB observations. This is the so-called Lithium problem~\cite{pdg,Fields:2011zzb}. At present, there is no clear resolution to the problem, but the two most likely avenues to resolve it are: \textit{i)} a stellar depletion mechanism, or \textit{ii)} new physics capable of producing less ${}^{7} {\rm Li}$ than in the standard BBN. There have been several new physics attempts to solve the ${}^{7} {\rm Li}$ problem~\cite{Fields:2011zzb}. However, the scenarios where decays or annihilations of exotic particles were invoked to reduce the ${}^{7} {\rm Li}$ abundance also enhance the deuterium one (that matches the standard BBN prediction with 1\% precision) and are as of now not viable. A viable possibility is the absorption of nonstandard bosons in deuterium or ${}^{7} {\rm Be}$, see Ref.~\cite{Goudelis:2015wpa}. In the context of this review, however, we note that nonstandard expansion histories do not seem likely of being capable of ameliorating or solving the problem as they generically alter the abundances of all primordial light elements simultaneously~\cite{Fields:2011zzb,Mathews:2019hbi}.

Finally, we note that although the Lithium problem is an unresolved issue, the agreement between the observed and predicted abundances of ${}^4 \text{He}$ and $\text{D}$ are a success of the standard cosmological model and, as we discuss below, place stringent constraints on many scenarios beyond the standard cosmological model.

\begin{table}[t]
\center
{\def\arraystretch{0.9}
\begin{tabular}{llll}
\hline\hline
\textbf{Parameter}  $\,\,\,\,\,\,\,$          & \textbf{Description}  					& \textbf{Cosmological Data} $ \,\,\,$ 			& \textbf{Current Constraint} \\ \hline\hline
\multirow{3}{*}{$N_{\rm eff}$}  		& Number of effective  					 & Planck legacy 						 & $N_{\rm eff} = 2.99 \pm 0.34 $, $\Delta N_{\rm eff}< 0.30$~\cite{Aghanim:2018eyx}   \\
							 &  	relativistic neutrino species			&  BBN 								& $N_{\rm eff} = 2.92 \pm 0.54 $, $\Delta N_{\rm eff}< 0.33$~\cite{Fields:2019pfx}  \\
							 & 	($N_{\rm eff}^{\rm SM} = 3.045$~\cite{Mangano:2005cc,deSalas:2016ztq,Escudero:2020dfa,Akita:2020szl})	 & BBN+Planck	 legacy					&  $N_{\rm eff} =2.91 \pm 0.30 $, $\Delta N_{\rm eff}< 0.16 $~\cite{Fields:2019pfx}  \\ 
							 \hline
							 \multirow{2}{*}{$T_{\rm reh}$} 		 & \multirow{2}{*}{Reheating temperature}  		& Planck 2015 							 & $T_{\rm reh} >  4.7\,\text{MeV} $~\cite{deSalas:2015glj} ($\phi \to e^+e^-/\gamma \gamma$) \\
						 	&  									&  BBN 		 					& $T_{\rm reh} >  (4\!-\!5)\,\text{MeV} $~\cite{Hasegawa:2019jsa} ($\phi \to \text{hadrons}$)   \\ \hline
							 $w$   & Constant EoS of dark energy  & Planck+BAO+SN1a & $w =  -1.028 \pm0.062$~\cite{Aghanim:2018eyx}   \\ 
							 \hline
							 $w(a)=w_0 +(1-a)w_a$   & Time-evolving EoS of dark energy  & Planck+BAO+SN1a & $w_0 = -0.96^{+0.16}_{-0.15},~w_a = -0.29^{+0.58}_{-0.60}$~\cite{Aghanim:2018eyx}  \\ 
							 \hline
$z_{\rm MR}$   & Redshift at matter-radiation equality  & Planck  & $z_{\rm MR} =3400 \pm 50 $~\cite{Aghanim:2018eyx}   \\ \hline
$z_{\ast}$& Redshift at decoupling  & Planck  & $z_{\ast} =  1089.9 \pm 0.5$~\cite{Aghanim:2018eyx}   \\ \hline
  \hline
\end{tabular}
}
\caption{Summary of current constraints on various parameters constraining nonstandard cosmological expansion histories. All bounds or intervals are given at 95\,\%\,CL.}\label{tab:BBN_CMB_constraints}
\end{table}

\subsubsection{BBN constraints on nonstandard cosmologies} 

BBN has been widely used as a laboratory to constrain new physics (for reviews, see Refs.~\cite{Sarkar:1995dd,Iocco:2008va,Pospelov:2010hj}). Here, we will review the constraints that can be derived from BBN on some of the most popular nonstandard expansion histories. In particular, we will discuss cosmologies with dark radiation, EMD epoch (low reheating temperature), unstable massive relics, light interacting species, and modified theories of gravity. 

\textbf{Dark radiation:} Perhaps the single most studied nonstandard cosmology in the context of BBN is that featuring extra noninteracting massless species, i.e. dark radiation. Dark radiation is a generic prediction of many extensions of the SM, see e.g. Refs.~\cite{Chacko:2005pe,Cicoli:2012aq,Higaki:2012ar,Weinberg:2013kea}, and its energy density $\rho_{\rm rad}$ is typically parametrized by the number of effective relativistic neutrino species as relevant for CMB observations, 
\begin{equation}
\Delta N_{\rm eff} \equiv N_{\rm eff} - N_{\rm eff}^{\rm SM} \,, \qquad N_{\rm eff} \equiv \frac{8}{7} \left(\frac{11}{4}\right)^{4/3} \left(\frac{\rho_{\rm rad} - \rho_\gamma}{\rho_\gamma}\right)  \,,
\end{equation}
where\footnote{The small difference between $N_{\rm eff}^{\rm SM}$ and 3 mainly arises from the fact that neutrino decoupling (at $T \sim 2 \,\text{MeV}$) took place soon before electrons became nonrelativistic ($T \sim m_e$) and there were some residual $e^+e^-$ annihilations into neutrinos.} $N_{\rm eff}^{\rm SM} = 3.045$~\cite{Mangano:2005cc,deSalas:2016ztq,Escudero:2020dfa}. The main implication of $\Delta N_{\rm eff} > 0$ during BBN is to enhance the expansion rate of the Universe, since $H \propto \sqrt{\rho}$. $\Delta N_{\rm eff}$ is primarily constrained by the primordial ${}^{4}{\rm He}$ abundance because it is extremely sensitive to the time at which deuterium starts to form and it is only logarithmically sensitive to the baryon energy density. Pioneering BBN constraints set $\Delta N_{\rm eff} < 5$~\cite{Steigman:1977kc}, while for many years until the early 2010s, $\Delta N_{\rm eff} < 1$~\cite{Kolb:1990vq,Pospelov:2010hj} was quoted as a typical robust BBN bound. At present, and thanks to recent very precise measurements of the primordial ${}^{4}{\rm He}$ abundance~\cite{Aver:2013wba,Izotov:2014fga,Aver:2015iza}, state-of-the-art BBN analyses yield $\Delta N_{\rm eff} < 0.33$~\cite{Fields:2019pfx,Pitrou:2018cgg} at 95\% CL. This represents a very stringent constraint on many extensions of the SM. For example, it strongly disfavours light sterile neutrinos in thermal equilibrium during BBN~\cite{Dolgov:2003sg,Cirelli:2004cz,Hannestad:2012ky,Gariazzo:2019gyi,Hagstotz:2020ukm,Hasegawa:2020ctq}, and it precludes the existence of any number of massless species that were once in thermal equilibrium with the SM plasma but decoupled at $T \lesssim 100\,\text{MeV}$.

\textbf{EMD epoch/low reheating temperature:} As discussed in Sec.~\ref{sec:causes}, several extensions of the SM of particle physics predict the existence of heavy and long-lived particles, which may cause an EMD epoch. We denote these by $\phi$. Massive species dominating the early Universe should be unstable, as they would have otherwise overclosed the Universe. BBN is strongly sensitive to this EMD epoch and subsequent decay of massive particles, and the requirement of successful BBN typically leads to a constraint on the lifetime of species dominating the Universe, $\tau_\phi \lesssim 0.1\,\text{s}$. The exact description of the early Universe dominated by an unstable massive relic at $t \gtrsim 0.001\,\text{s}$ is far from trivial and there are only a handful of references dealing with this issue in detail~\cite{Kawasaki:1999na,Kawasaki:2000en,Hannestad:2004px,Ichikawa:2005vw,IKT07,deSalas:2015glj,Hasegawa:2019jsa}\footnote{These references account for the thermalization of the early Universe via $\phi$ decays but assume that somehow a baryon asymmetry is generated by the decays. Generating a baryon asymmetry at such low temperatures is not trivial. However, recently, several mechanisms capable of successfully generating the baryon asymmetry via $\phi$ decays have been put forward~\cite{Aitken:2017wie,Elor:2018twp,Nelson:2019fln,Kane:2019nes}.}. In these scenarios, the unstable particle dominating the Universe will decay into different states (typically within the SM) and will reheat\footnote{Note that, as discussed in Sec.~\ref{sec:dark_sectors}, particle decays cannot increase the temperature of the Universe~\cite{Scherrer:1984fd}.} the Universe to a temperature $T_{\rm reh} \simeq 0.7\,\text{MeV}\sqrt{\text{s}/\tau_\phi}$, defined via $\Gamma_\phi \equiv 3\,H(T_{\rm reh})$.  BBN represents a key probe to test these scenarios since observations are compatible with the standard RD picture of the early Universe. The precise implications for BBN of such type of scenarios depend upon the exact decay channels of the out-of-equilibrium decaying particle. Nonetheless, regardless of the dominant decay mode, these scenarios lead to a nonstandard expansion and to a nonthermal neutrino population which in turn modifies the proton-to-neutron rates via e.g. ${\nu}_e n \to e^- p $ interactions. In addition to these effects, if the $\phi$ species decays into hadrons a large number of mesons will be produced and they will also impact the proton to neutron conversion rate beyond the effect driven by a nonthermal neutrino population. The latest BBN analysis yields~\cite{Hasegawa:2019jsa} $T_{\rm reh} > (4\!-\!5)\,\text{MeV}$ ($\phi \to \text{hadrons}$) and $T_{\rm reh} > 1.8\,\text{MeV}$ ($\phi \to e^+e^-/\gamma \gamma$), both at 95\% CL. In addition, scenarios where $\phi$ possess a direct decay mode to neutrinos should fulfil $T_{\rm reh} > (1\!-\!4)\,\text{MeV}$~\cite{Hannestad:2004px} (for $\phi \to \bar{\nu}\nu/e^+e^-$) at 95\% CL. The ranges reflect the dependency of the bound on the decay mode and mass of $\phi$. Note that scenarios with a low reheating temperature are among the few situations exhibiting $N_{\rm eff} < N_{\rm eff}^{\rm SM}$. Since CMB observations are sensitive to $N_{\rm eff}$ they can also be used to constrain $T_{\rm reh}$. The latest CMB analysis, using Planck 2015 data, yields~\cite{deSalas:2015glj} $T_{\rm reh} > 4.7 \,\text{MeV}$ ($\phi \to e^+e^-$) at 95\% CL. 

To summarize, since $T_{\rm reh} \simeq 2.2\,\text{MeV}\sqrt{0.1\text{s}/\tau_\phi}$ and current bounds set $T_{\rm reh} \gtrsim 2\,\text{MeV}$, successful BBN bounds the lifetime of unstable relics dominating the early Universe to be $\tau_\phi \lesssim 0.1\,\text{s}$.

\textbf{Unstable massive species:} Species with small abundances at the time of BBN, $\rho/\rho_\gamma \ll 1 $, will not significantly impact the expansion of the Universe during BBN. Yet, they are subject to very stringent constraints from BBN~\cite{Scherrer:1987rr,Ellis:1990nb,Kawasaki:1994sc,Kawasaki:2004qu,Jedamzik:2004er,Jedamzik:2006xz,Poulin:2015opa}. BBN constraints on these species arise from the fact that the decay products  of relics with $m > 5\,$MeV can initiate electromagnetic or hadronic cascades (depending on their mass) which can alter the network of BBN reactions through e.g. photo-disassociation of already synthesized nuclei. Since the baryon-to-photon ratio $\eta \sim 10^{-9}$, the decay of even very small number densities of particles can still drastically alter the light element abundances. BBN bounds on unstable relics are used to constrain the abundance of a wide array of relics with lifetimes in the range $0.01\,\text{s} \lesssim \tau \lesssim 10^{12}\,\text{s}$. For state-of-the-art work constraining generic long-lived decaying particles with BBN, see Ref.~\cite{Kawasaki:2017bqm}. Examples of detailed BBN constraints on concrete and well-motivated particle physics scenarios include gravitinos~\cite{Kawasaki:2008qe}, dark scalars~\cite{Fradette:2017sdd,Fradette:2018hhl}, dark vector bosons~\cite{Fradette:2014sza,Coffey:2020oir}, MeV-scale dark sectors~\cite{Forestell:2018txr,Hufnagel:2018bjp,Kawasaki:2020qxm}, GeV-scale sterile neutrinos~\cite{Dolgov:2000jw,Ruchayskiy:2012si,Sabti:2020yrt}, and PBHs~\cite{Carr:2009jm}.

The take away message from these studies is that BBN represents a very stringent constraint on the abundance of unstable massive relics with lifetimes in the range $0.01\,\text{s} \lesssim \tau \lesssim 10^{12}\,\text{s}$.

\textbf{Light interacting species:} BBN is sensitive to the Universe in the temperature window $1\,\text{keV}\lesssim T \lesssim 10\,\text{MeV}$. This makes BBN a powerful tool to test extensions of the SM featuring particles with masses at the MeV scale and below. Examples of such particles include axions, majorons, dark photons, and generic dark sector states. The consequences on BBN of light ($m\lesssim 10\,\text{MeV}$) BSM states depend upon the exact properties of the given particle and its abundance at the time of BBN. Light particles in thermal equilibrium with the SM plasma at $T\lesssim 10\,\text{MeV}$ will have an energy density comparable to that of the SM plasma and will therefore alter the expansion history of the Universe. In addition, these particles will interact with SM species and eventually, at $T\sim m/3$, will release entropy into the SM plasma via their decays or annihilations. The cosmological consequences of such scenarios were first highlighted in Ref.~\cite{Kolb:1986nf}. The most recent BBN analysis shows that generic species in thermal equilibrium during BBN should have a mass $m > 0.4\,\text{MeV} $~\cite{Sabti:2019mhn} at 95\% CL, and that light species ($m < 0.1\,\text{MeV}$) in thermal equilibrium with the SM plasma are highly disfavoured by current cosmological observations. BBN represents a very stringent constraint on light species that were not in thermal equilibrium during BBN but that were still capable of modifying the expansion history of the early Universe. Relevant examples of these scenarios include axions~\cite{Cadamuro:2010cz,Blum:2014vsa}, axion-like particles~\cite{Millea:2015qra,Depta:2020wmr}, majorons~\cite{Dolgov:1996fp,Escudero:2019gfk}, and weakly coupled dark sectors~\cite{Vogel:2013raa,Berlin:2019pbq}.

\textbf{Modified gravity theories:} So far we have only discussed the impact on BBN of nonstandard cosmologies featuring BSM matter. However, modified gravity theories can also modify the expansion history of the early Universe. In this context, BBN has been used to constrain, for example, large extra dimensions~\cite{Cline:1999ts,Csaki:1999jh}, the value of Newton's constant during nucleosynthesis~\cite{Yang:1978ge,Copi:2003xd,Alvey:2019ctk}, and scalar-tensor theories of gravity~\cite{Campbell:1994bf,Coc:2006rt,DeFelice:2005bx}.

\textbf{BBN tools:} We finish by pointing the reader to publicly available state-of-the-art tools used to model the early Universe and BBN: \texttt{NUDEC\_BSM}~\cite{Escudero:2020dfa,Escudero:2018mvt}, \texttt{PArthENoPE}~\cite{Pisanti:2007hk,Consiglio:2017pot}, \texttt{AlterBBN}~\cite{Arbey:2011nf,Arbey:2018zfh}, and \texttt{PRIMAT}~\cite{Pitrou:2018cgg}. Given current and expected precision on BBN observables, these tools prove very useful in studying the impact of nonstandard cosmological expansion histories on BBN.

\begin{figure}
    \centering
    \includegraphics[scale=0.265]{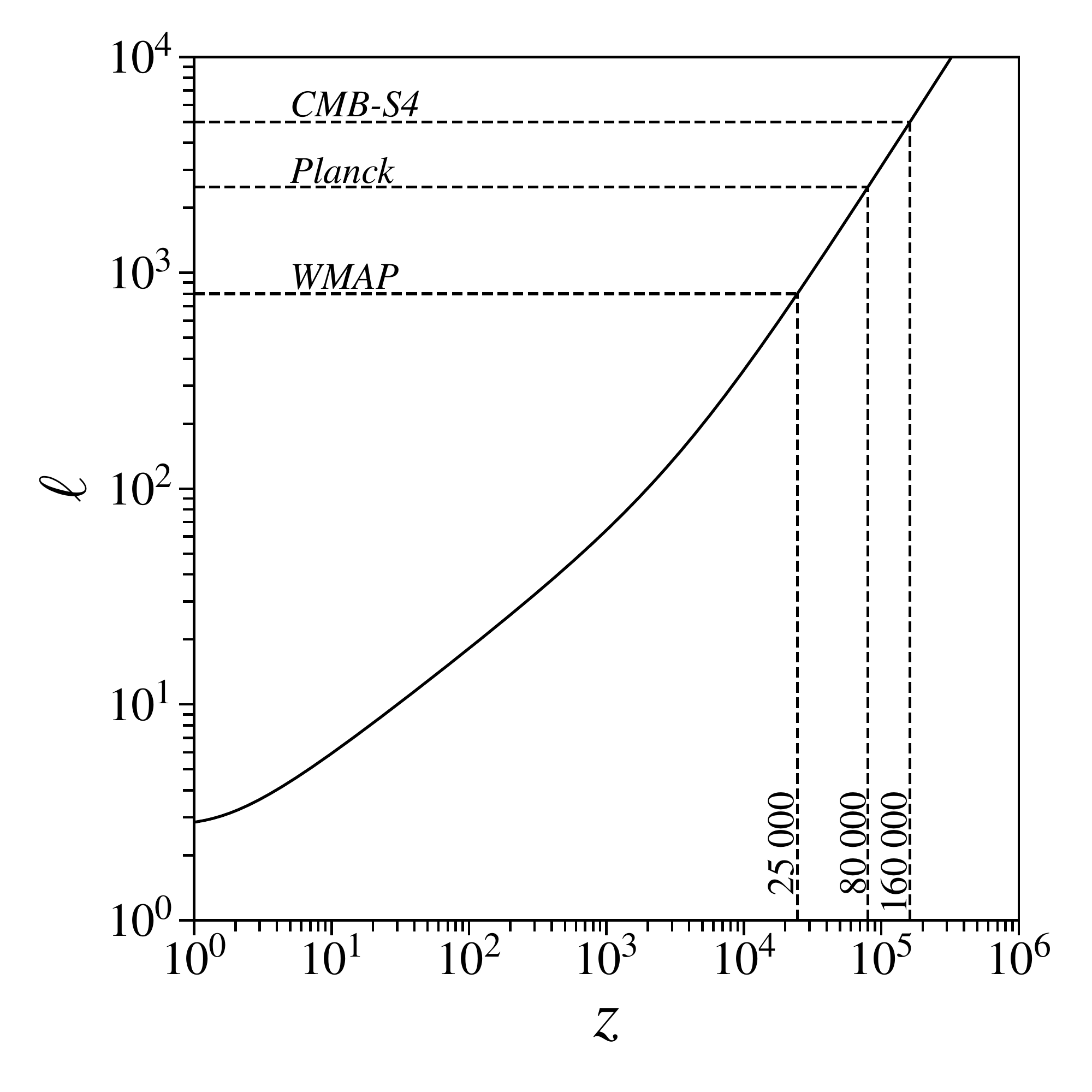}
    \includegraphics[scale=0.54]{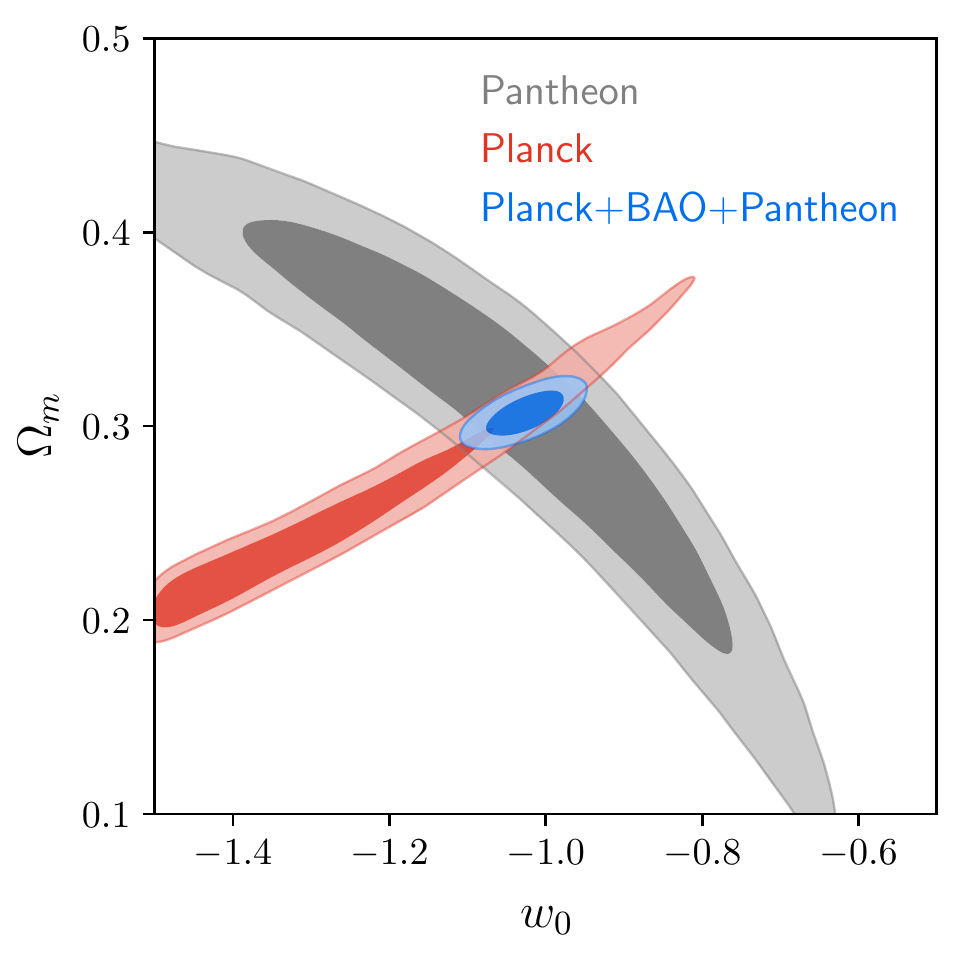}
    \includegraphics[scale=0.193]{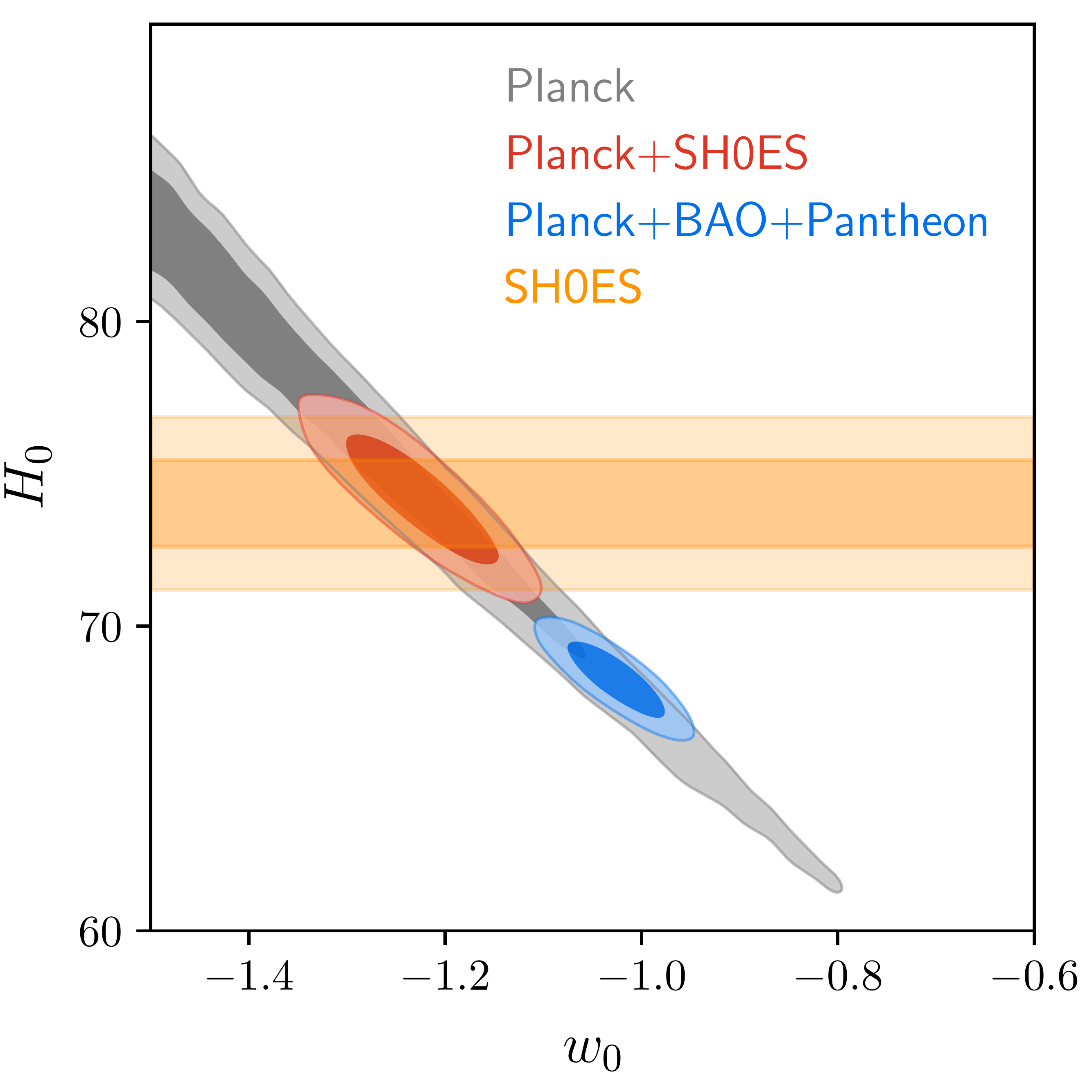}
    \caption{{\em Left panel}: Minimum multipole $\ell$ affected by a modified expansion history at $z$ in the standard $\Lambda$CDM cosmology. Reproduced from Ref.~\cite{Linder:2010wp}.  {\em Middle panel}: Constraints at 68\% and 95 \% CL on $w$ and  $\Omega_m=1-\Omega_{\rm \Lambda}$ obtained with CMB, BAO and SN1a data. Reproduced from Ref.~\cite{Scolnic:2017caz}. {\em Right panel}: Constraints at 68\% and 95 \% CL on $w$ and  $H_0$ obtained with CMB, BAO and SN1a data, compared with the SH0ES measurement of the Hubble constant $H_0$ \cite{Riess:2019cxk}.\vspace{1cm}}
    \label{fig:ell-vs-z}
\end{figure}

\subsubsection{Effects of nonstandard expansion rate on the CMB}

The tremendous increase in the precision of CMB observations over the last 20 years have led to a remarkably accurate determination of the $\Lambda$CDM parameters, some of which are nowadays known at precision better than 1\%. This is in particular true for the angular scale of sound horizon $\theta_s(z_{\rm rec})$, that has been determined at 0.04\% with Planck \cite{Aghanim:2018eyx}. The $\Lambda$CDM model teaches us that the Universe was dominated by radiation until redshift $z\sim3500$, then by matter until $z\sim0.7$, and finally by dark energy. Given such high accuracy measurements, it is natural to expect that strong constraints apply on deviations from the $\Lambda$CDM expansion history. 

In practice, however, the constraining power of the CMB greatly changes before and after decoupling. In fact, it is only by combining CMB with measurements of the Baryonic Acoustic Oscillations (BAO) (see e.g. Refs. \cite{Alam:2016hwk,Bautista:2017zgn,Bourboux:2017cbm,Agathe:2019vsu}) and Supernovae of type 1a (SN1a) (see e.g. Refs. \cite{Scolnic:2017caz}) at  $z\lesssim 2$ that one can constrain modifications to $\Lambda$CDM to high accuracy in the late Universe. As of yet, the era from (roughly) $2\lesssim z\lesssim 1000$ remains largely unconstrained. Future ground-based 21 cm surveys such as HERA \cite{Array:2019rzj} and SKA\footnote{\url{https://www.skatelescope.org/}}, and other intensity mapping technics, will allow to extend constraints to the cosmic dawn epoch at $z\lesssim 20$ \cite{Bull:2018lat,Kovetz:2017agg,Silva:2019hsh,Bernal:2019gfq,Bernal:2019jdo}, while future Moon-based experiments could probe the dark ages until $z\lesssim 200$ \cite{Silk:2018qet,Burns:2019zia,Furlanetto:2019jso}.  At $z>1000$, on the other hand, modifications of the expansion history affect the CMB spectra in ways that are much less subject to degeneracies. Still, the CMB cannot probe the expansion history up to arbitrarily high $z$. A multipole mode $\ell$ can be approximately mapped onto a physical wavenumber $k$ via\footnote{In reality, a given $\ell$ is related to $k$ via $\Theta_\ell(k,\eta_0)\equiv\Delta T/T|_\ell\sim\int d\eta S(k,\tau)j_\ell(k(\eta-\eta_0))$, where $S(k,\tau)$ is the so-called ``source function'' which encodes various contributions to the temperature anisotropies and $j_\ell$ is a Bessel function peaked around $k\eta_0$ (see e.g. Ref. \cite{Seljak:1996is}). While several $k$ modes can therefore contribute to a given $\ell$, we make the approximation that the Bessel function can be described by a Dirac delta function for the sake of the discussion. } $\ell \sim k\eta_0$ with $\eta_0\sim10^4$ Mpc the conformal time today. As a mode is mostly affected by an exotic expansion history if it is within the cosmological horizon, we expect that it will probe the expansion at times $\eta\gtrsim \eta_0/\ell$. We show the minimum multipole affected by a modified expansion history at (and below) $z$ in Fig.~\ref{fig:ell-vs-z}, left panel \cite{Linder:2010wp} and illustrate the maximum redshift probed by {\it WMAP}, {\it Planck} and a futuristic experiment like {\it CMB-S4} (assuming that no too big modifications to the expansion history are allowed when relating $\eta$ to $z$).  Concretely, Planck is sensitive to exotic histories up to $z_{\rm max}\sim  80,000$ ($T\sim 20$ eV), while an experiment like CMB-S4 would probe up to $z_{\rm max}\sim 160,000$ ($T\sim 40$ eV).

In the following,  we review how CMB measurements can be used to learn about the expansion history of the Universe from $z\sim 10^5$ to today. We discuss ``model-independent'' modifications\footnote{A detailed study of the sensitivity of the CMB to cosmological parameters can be found in e.g. Ref.~\cite{Hu:2000ti,Dodelson:2003ft}.} to $H(z)$ at $z\gg1000$ and $z\ll1000$. In practice, however, details of perturbations are important when deriving constraints on exotic expansion history. We will illustrate these subtleties with two specific scenarios: exotic neutrino properties and early dark energy.

\textbf{Pre-decoupling era:} Well before decoupling, photons and baryons form a tightly coupled fluid. In the Newtonian gauge (see e.g. Ref. \cite{Ma:1995ey} for definitions), super-horizon perturbations are frozen, while sub-horizon density perturbations  $\delta_\gamma\equiv \delta\rho_\gamma/\rho_\gamma = 3/4\delta_b$ experience driven acoustic oscillations whose dynamics is governed by\footnote{This solution is valid at length scales shorter than the sound horizon $r_s$, neglects the time dependence of the sound speed of the baryon-photon fluid, any sources of anisotropic stress, and the damping terms due to baryon friction.} $\delta_\gamma(x)=[\delta_\gamma(0)+\Phi(0)]\cos(x)+2\int_0^xdx'\Phi(x')\sin(x-x')$ where $\Phi$ is the Newtonian gravitational potential and $x = kc_s\eta$~\cite{Linder:2010wp}.

Therefore, well before decoupling exotic expansion histories can mainly influence (i) the sound horizon $r_s = c_s\eta$ and (ii) the time evolution of the gravitational potential. The most important scale in the CMB is the sound horizon at decoupling, which is the distance travelled by sound wave in the primordial plasma from reheating until decoupling, computed as (see e.g. Ref. \cite{Dodelson:2003ft})
\begin{equation}
    r_s(z_{\rm rec})=\frac{1}{1+z_{\rm rec}}\int_{z_{\rm rec}}^{\infty}\frac{dz}{H(z)} c_s(z)\,,~c_s(z) = \frac{ c}{\sqrt{3(1+R)}}\,,\label{eq:rs}
\end{equation}
where $c_s(z)$ is the sound speed in the tightly coupled photon-baryon fluid and $R=3\rho_b/4\rho_\gamma$. This scale dictates the typical size of inhomogeneous patches in CMB maps, and accordingly the position (or more accurately the spacing) of the acoustic peaks. Note that this scale is weighed towards smaller $z$, and is therefore more sensitive to change to $H(z)$ close to decoupling.

On the other hand, the decay of the gravitational potentials on sub-horizon scales during radiation domination lead to a boost in the amplitude of oscillations for modes $k>k_{\rm MR}\equiv H(z_{\rm MR})/(1+z_{\rm MR})$ \cite{Hu:2000ti}. This, in turn, increases the amplitude of the acoustic peaks at $\ell > \ell_{\rm MR}\sim k_{\rm MR}\eta_0$. It is mostly through this effect that CMB measurements allow to put strong constraints on the redshift of matter-radiation equality $z_{\rm MR} =3400 \pm 50 $ \cite{Aghanim:2018eyx}. Modifications to the standard time evolution of the gravitational potential at a given redshift $z$ therefore lead to change in the amplitude of acoustic peaks around the mode $\ell\sim k\eta_0$, where $k=H/(1+z)$ is the scale that entered the horizon around $z$. Interestingly, this scale-dependence can be used to trace the time at which the expansion rate is modified compared to $\Lambda$CDM.

Close to decoupling, the tight-coupling approximation breaks down and the photon mean free path $\lambda \equiv(n_e\sigma_T)^{-1}$ becomes significant. This leads to photon diffusion on a typical (comoving) distance $r_d^2(\eta_{\rm rec})\propto\int d\eta/\lambda$ (see e.g. Ref. \cite{Hu:1995em} for a more accurate definition). This defines the scale $k_d = r_d^{-1}$ above which modes $\ell > \ell_d \sim k_d\eta_0$ are exponentially suppressed. Change in the expansion history close to decoupling can therefore affect the damping tail that is well constrained by the data. This effect is particularly important to constrain extra relativistic degrees of freedom (although partially degenerate with change in the spectral index $n_s$ or the primordial helium fraction $Y_p$) \cite{Hou:2011ec}.

Finally, the decoupling process itself depends, in general, on the expansion history. However, the decoupling redshift $z_{\rm rec}$ is only mildly affected by change in $H(z)$. This can be understood by noting that $z_{\rm rec}$ can be approximately obtained by equating the Thomson interaction rate with the expansion rate  $H (z) \sim x_e(z) n_H(z)\sigma_T$. Hence, a higher $H$ leads to a higher residual free-electron fraction $x_e\equiv n_e/n_H$ but does not change $z_{\rm rec}$ to a first approximation.

\textbf{Post-decoupling era:} After decoupling, the main effect of changes in $H(z)$ is to change the distance that separates us from the last scattering surface, as known as the angular diameter distance to decoupling: 
\begin{equation}
    D_A(z_{\rm rec})=\frac{1}{1+z_{\rm rec}}\int_0^{z_{\rm rec}}\frac{dz}{H(z)}\,,
\end{equation}
or, equivalently, the conformal time today $\tau_0$. Once the pre-decoupling era is fixed, $D_A(z_{\rm rec})$ (or $\tau_0$) determines the overall location of the spectrum in multipole space. A change in $D_A(z_{\rm rec})$ will therefore shift the spectrum horizontally. It also dictates the multipole $\ell_d$ above which perturbations are damped due to diffusion and the multipole $\ell_{\rm eq}$ above which scales have been gravitationally enhanced by the gravitational potential decay during RD era. 

However, this pure geometric effect suffers strong degeneracy, since species with different $H(z)$ at late times can lead to the same $D_A(z)$.  There is, fortunately, a number of additional effects that help in breaking the degeneracy within CMB data. In particular, the late-time evolution of potential wells manifests itself as a sudden rise in the large scale anisotropies. Because different multipoles are affected at different times, as discussed above, this gives the CMB spectrum some sensitivity to $H(z<z_{\rm rec})$ (and not only to the integrated $H^{-1}(z\leq z_{\rm rec})$). Unfortunately, however, at small $\ell$ the power spectrum $C_\ell$ suffers a large theoretical uncertainty known as cosmic variance, which scales as $\Delta C_\ell/C_\ell=\sqrt{\ell/(2(\ell+1))}$ and severely limits the ability of the CMB to measure $H(z<z_{\rm rec})$.

\subsubsection{Combining CMB with BAO and SN1a} 

The CMB is not the only cosmological probe at our disposal in the post-decoupling era, and one can therefore break the geometric degeneracy that exists in CMB data. The canonical examples of such probes are SN1a, powerful standard(ized) candles that can be used to measure the luminosity distances $D_L(z)$ up to Gpc scales ($z\lesssim 1$ {currently), and have led to the discovery of modern accelerated expansion compatible with the Universe being dominated by a cosmological constant \cite{Perlmutter:1998np,Riess:1998cb}. Another important example is the BAO \cite{Eisenstein:1997ik}, which has been detected to high accuracy in the clustering of galaxies \cite{Cole:2005sx,Eisenstein:2005su},  and can be used to measure $D_A(z)$ (currently until $z\lesssim 0.5$  \cite{Alam:2016hwk}). More recently, the BAO has been observed in the (cross and auto-) correlation of quasar Ly-$\alpha$ forest, which allows to constrain the expansion history at $z\sim2.4$ \cite{Bautista:2017zgn,Bourboux:2017cbm,Agathe:2019vsu}. However, since we can only measure fluxes ($F\equiv L/4\pi D_L^2$) and angles ($\theta \equiv r_s(z_{\rm drag})/D_A$), observations of SN1a and detection of the BAO do not allow to measure an absolute expansion rate; rather, by themselves these measurements constrain the shape of the expansion history.  In practice, the CMB allows us to calibrate the BAO by providing a model-dependent measurement of the sound horizon at the redshift $z_{\rm drag}\sim 200$ at which baryons decouple from photons $r_s(z_{\rm drag})$)\footnote{Alternatively, it is possible to measure $r_s(z_{\rm drag})$ by combining BAO and the value of $w_b$ obtained from BBN \cite{Addison:2017fdm,Cuceu:2019for,Schoneberg:2019wmt}.}.  

The combination of CMB, BAO and (uncalibrated) SN1a data can be used to set constraints on a wide variety of dark energy and dark matter models which we do not attempt to review in detail here (a few canonical constraints are given in Table \ref{tab:BBN_CMB_constraints}).  The canonical example is the equation of state of DE, often parametrized as $w(a)=w_0+(1-a)w_a$ \cite{Chevallier:2000qy} that must nowadays satisfy $w_0 = -0.961\pm0.077 $ and $w_a = -0.28^{+0.31}_{-0.27} $ (68\% CL) \cite{Aghanim:2018eyx}. We show in the right panel of Fig.~\ref{fig:ell-vs-z} the constraints on the EoS $w_0$ (assuming $w_a=0$) and fractional matter density today $\Omega_m = 1 - \Omega_{\Lambda}$ obtained from a combination of {\em Planck} data \cite{Aghanim:2018eyx}, the Pantheon SN1a catalogue \cite{Scolnic:2017caz} and the BOSS DR12 BAO measurement \cite{Alam:2016hwk}. This clearly illustrates the geometric degeneracy that exists in CMB data: it is only by combining CMB with BAO and/or uncalibrated SN1a that one can unambiguously detect the presence of DE with an equation of state compatible with that of a cosmological constant. Another interesting example is that of DM decays on cosmological time-scales. A combination of CMB, BAO and LSS data sets a robust model-independent\footnote{These bounds were derive assuming massless products. Decay into warm DM lead to stronger constraints because of the impact of the warm component on structure formation \cite{Aoyama:2014tga}. Electromagnetic decay products change the thermodynamics history and are much more strongly constrained \cite{Slatyer:2016qyl,Poulin:2016anj}.}  bound on the lifetime of DM, $\tau > 158$ Gyr, and constrain the fraction of DM that has decayed between now and decoupling to be less than 4\% \cite{Audren:2014bca,Poulin:2016nat}.   Alternatively, it is possible to obtain model-independent constraints on the expansion history at late times from CMB, galactic BAO and uncalibrated SN1a. These restrict deviation from $\Lambda$CDM to be less than 10\% percent at $z\lesssim 1$ \cite{Bernal:2016gxb,Joudaki:2017zhq,Poulin:2018zxs}. However, there is a number of interesting tensions between Planck and probes of the late-time expansion rate that might start indicating deviation from a cosmological constant \cite{Zhao:2017cud}.

In particular, several methods have been proposed to calibrate the intrinsic luminosity of SN1a and obtain direct constraints on the Hubble constant $H(z=0)\equiv H_0$ (see e.g. Refs.~\cite{Riess:2019cxk,Freedman:2020dne}). In fact, the latest calibration of SN1a using cepheid stars \cite{Riess:2019cxk} has led to a measurement of $H_0$ with $1.4\%$ precision that is in $4.4\sigma$ (statistical) disagreement with what is predicted in the $\Lambda$CDM cosmology deduced from Planck CMB data \cite{Verde:2019ivm}. In fig.~\ref{fig:ell-vs-z}, right panel, we show that because of the geometrical degeneracy discussed previously, one can easily accommodate such a value of $H_0$ by allowing for a ``phantom'' equation of state of dark energy $w_0<-1$. However, the inclusion of BAO and Pantheon SN1a catalogue breaks that degeneracy, such that late-universe modifications as a resolution to the Hubble tension are tightly constrained \cite{Bernal:2016gxb,Poulin:2018zxs,Aylor:2018drw,Knox:2019rjx}. Barring the possible presence of systematics in the data (see e.g. Ref. \cite{Verde:2019ivm} for a discussion),  there is a growing concensus that it is necessary to decrease the value of the sound horizon compared to that deduced in the $\Lambda$CDM cosmology in order to resolve the tension rather than modifying the late-expansion history \cite{Bernal:2016gxb,Aylor:2018drw,Knox:2019rjx}. If confirmed, this discrepancy could therefore be a sign of an epoch of exotic expansion during the pre-decoupling era, as induced for instance by the presence of dark energy at early times \cite{Karwal:2016vyq,Poulin:2018cxd,Agrawal:2019lmo,Lin:2019qug,Niedermann:2019olb, Smith:2019ihp} or neutrinos with exotic properties \cite{Kreisch:2019yzn,Escudero:2019gvw}.

\subsubsection{Probing perturbation dynamics with CMB}

Until now, we have limited our discussion to modifications of the background expansion rate. However, the CMB is very sensitive to the detailed dynamics of perturbations in various fluids that contribute to the stress-energy tensor of the Universe. This is because, as argued previously, the time evolution of the gravitational potential ($\Phi$ in previous section)  leaves clear imprints on the CMB sky, and the evolution of the gravitational potential is related (via Einstein's equations) to the sum of the perturbed stress-energy tensor of each fluid. Therefore, species with the same background evolution but different perturbation dynamics can have very different signatures in the CMB. Moreover, in a perturbed Universe, it is not consistent to neglect perturbations in a fluid\footnote{The exception is a cosmological constant which has an equation of state $w=-1$ and does not develop perturbations.}, as it violates the conservation equations that follow from Bianchi identities. We illustrate this important subtlety for two specific scenarios, exotic neutrinos and early dark energy.

{\bf Neutrino properties:} As mentioned previously, the CMB can put strong constraints on neutrino properties, and more generally on any type of extra radiation density, often parametrized by $\Delta N_{\rm eff}$. However, bounds can strongly vary depending on the choice made for describing perturbations of the species, which are directly tied to the microphysical properties of neutrinos (or the extra species).  The standard constraint on $\Delta N_{\rm eff}$ is obtained by assuming that the extra radiation can free-stream, similarly to SM neutrinos. A free-streaming species leads to a specific phase-shift in the position and to small change in the amplitude of the acoustic peaks in the CMB~\cite{Bashinsky:2003tk}. In fact, the ``neutrino drag'' effect produced by SM neutrinos has been claimed to be robustly detected for the first time in Planck data \cite{Follin:2015hya}. On the other hand, a combination of CMB and BAO data leads to $\Delta N_{\rm eff}^{\rm fs} <  0.30 $~\cite{Aghanim:2018eyx} at 95\% CL. Interestingly, it is possible to completely change this constraint by relaxing the assumption of free-stream. In fact, CMB data are powerful in testing new interactions of neutrinos since these interactions affect the streaming properties of neutrinos. 

As a striking example, it has been found that the CMB temperature data can be equally well fit by postulating that neutrinos strongly self-interact through a contact interaction \cite{Cyr-Racine:2013jua,Oldengott:2017fhy,Lancaster:2017ksf,Kreisch:2019yzn}. In this cosmology, the phase shift of the acoustic peaks is completely erased, but the position of the peaks is kept constant by increasing $H_0$. In fact, Planck 2015 temperature data\footnote{Note that the inclusion of Planck 2015 polarization data restricts the success of the solution.} can fully accommodate 4 species of strongly interacting neutrinos, with $H_0 = 72.3\pm1.4$ km/s/Mpc \cite{Kreisch:2019yzn}. Such model would fully resolve the Hubble tension, while having interesting consequences for laboratory experiments. Unfortunately, the simplest realization of this model is strongly constrained \cite{Blinov:2019gcj}, but such degeneracy provide an interesting avenue to resolve the Hubble tension. In fact, scenarios with light mediators are substantially less constrained by laboratory data. These can also suppress neutrino free-streaming and therefore reduce the tension~\cite{Archidiacono:2014nda,Archidiacono:2015oma,Forastieri:2019cuf,Escudero:2019gvw,Escudero:2019gfk}. 

Looking forward, future CMB missions are expected to deliver high precision measurements of $N_{\rm eff}$. In particular, the upcoming Simons Observatory~\cite{Ade:2018sbj} is expected to constrain $\Delta N_{\rm eff} < 0.12$ at 95\% CL, and proposed experiments such as CMB-S4~\cite{Abazajian:2019eic} could constrain $\Delta N_{\rm eff} < 0.06$ at 95\% CL. These upcoming constraints could have a very important impact on many particle physics scenarios. In particular, if future measurements are in agreement with the SM expectation, CMB-S4 measurements could rule out massless dark radiation that decoupled from the SM plasma at temperatures above the QCD phase transition \cite{Aghanim:2018eyx,Baumann:2017gkg}.

{\bf Early dark energy:} The possibility of the presence of dark energy before last-scattering has been studied for more than a decade \cite{Doran:2000jt,Wetterich:2004pv}. However, these models have recently become increasingly interesting in the context of the Hubble tension mentioned above. Indeed, it has been shown in several studies \cite{Karwal:2016vyq,Poulin:2018cxd,Agrawal:2019lmo,Lin:2019qug,Niedermann:2019olb,Smith:2019ihp} that this tension could indicate the presence of a frozen scalar field acting like dark energy until $z\sim z_{\rm MR}$, where it contributes to $\sim10\%$ of the total energy density of the Universe, and subsequently dilutes like radiation or faster. Strikingly, Planck data (high-$\ell$ polarization in particular) not only constrain the background evolution of such field but also the dynamics of perturbations. Indeed, it has been found that the field must have an effective sound speed (in its rest frame) $c_s^2=0.8\pm0.1$, which can be achieved, for instance, through a noncanonical kinetic term \cite{Lin:2019qug} or an oscillating potential which flattens close to the initial field value \cite{Poulin:2018cxd,Smith:2019ihp}. Models with different perturbation dynamics can only lead to a milder alleviation of the tension \cite{Karwal:2016vyq,Agrawal:2019lmo,Niedermann:2019olb}.

\subsection{Observational probes of microhalos (Authors: M. S. Delos \& A. L. Erickcek)}
\label{sec:microhalo_constraints}

\subsubsection{Gravitational signatures}
The only guaranteed signatures of dark matter halos are gravitational.  Unfortunately, detecting the gravitational effects of the microhalos generated in nonstandard cosmologies is extremely challenging because microhalos are small and diffuse.  Since deviations from radiation domination only affect the growth of subhorizon perturbations, the masses of these microhalos must be less than the dark matter mass contained within the horizon when the radiation temperature was 3 MeV, which implies a maximum initial microhalo mass of 1200 $M_\Earth$ \cite{Erickcek:2011us}.  The radii of these microhalos depends on their formation time, but even the earliest-forming microhalos are still diffuse enough that their photometric microlensing signatures are far weaker than those generated by PBHs of the same mass \cite{ricotti2009ucmh, dror2019pta}.  

The small changes in the positions of background stars due to the motion of an intervening halo offer an alternative avenue for detection, but these astrometric microlensing signatures are only potentially observable for standard halos with masses greater than $10^5 M_\odot$ \cite{2011ApJ...729...49E, 2018JCAP...07..041V}.
Halos that form early in the matter-dominated era ($z\simeq1000$) from enhancements to the small-scale matter power spectrum are easier to detect because these ultracompact minihalos (UCMHs) \cite{ricotti2009ucmh} have denser central regions than later-forming halos.  Even so, an astrometric lensing search by Gaia is only sensitive to UCMHs with masses greater than a solar mass \cite{li2012ucmhlens}.\footnote{While halos with masses greater than a solar mass do not probe the expansion history between inflation and BBN, they provide valuable constraints on the primordial power spectrum generated during inflation \cite{Josan_2010,Bringmann_2012, li2012ucmhlens,Clark:2015tha, *10.1093/mnras/stw2305, Aslanyan:2015hmi, Yang_2017, Nakama_2018, Delos:2018ueo}.}
Moreover, this analysis assumed that UCMHs have density profiles with $\rho\propto r^{-9/4}$ at small radii, as predicted by \cite{ricotti2009ucmh}.  $N$-body simulations have revealed that this density profile is not realized in even the rarest and most isolated halos that form from Gaussian initial conditions \cite{gosenca20173d, delos2017ultracompact}.  Instead, $N$-body simulations have widely shown that the first halos develop radial density profiles that scale as $\rho\propto r^{-3/2}$ at small radii \cite{ishiyama2010gamma,anderhalden2013density,*anderhalden2013erratum,ishiyama2014hierarchical,polisensky2015fingerprints,ogiya2017sets,delos2017ultracompact,Delos:2018ueo}.  Moreover, successive mergers between halos with $\rho\propto r^{-3/2}$ inner density profiles gradually drive them toward shallower NFW density profiles \cite[]{ogiya2016dynamical,angulo2016earthmass,gosenca20173d}.   Shallower density profiles generate weaker astrometric microlensing signatures \cite{2011ApJ...729...49E}, so sub-solar-mass halos are well beyond the reach of astrometric microlensing \cite{2020arXiv200201938M}.   

Recently, two other gravitational probes have been proposed that have the potential to detect sub-earth-mass halos: pulsar timing and stellar microlensing.  A pulsar timing array can detect the presence of dark matter halos due to both the Shapiro time delay when light from a pulsar passes through a halo \cite{siegel2007pta} and the Doppler effect when a halo accelerates either the pulsar or the Earth \cite{seto2007pta}.  With twenty years of pulsar observations with SKA, the Doppler signal can be used to constrain the abundance of sub-earth-mass microhalos with NFW profiles and concentrations greater than 100 \cite{dror2019pta}.  Although such high concentrations are not expected in standard cosmologies \cite{Sanchez-Conde:2013yxa}, the early formation of microhalos following an EMD epoch may yield such profiles \cite{Erickcek:2015jza}.  Sub-earth-mass halos may also be detected by monitoring background stars as they are magnified by intracluster stars near lensing caustics within an intervening cluster: the presence of microhalos within the cluster would induce detectable magnification variations \cite{dai2020lens}.  While Ref. \cite{dai2020lens} only considered the signatures of halos formed from small-scale axion isocurvature perturbations, similarly dense and abundant halos can form from adiabatic initial conditions following an EMD epoch.

\pagebreak
\subsubsection{Dark matter annihilation within microhalos}

If dark matter is produced thermally, then it will self-annihilate.  These annihilations generally produce high-energy observable particles, and numerous attempts have been made to detect these indirect dark matter signatures \cite[e.g][]{Ackermann:2015zua,Fermi-LAT:2016uux,Cholis:2019ejx,Bergstrom:2013jra}.  Since the annihilation rate is proportional to the square of dark matter density, the presence of structure boosts the annihilation rate.  In standard cosmologies, the presence of structure enhances the cosmic annihilation rate by a factor of $\sim\!\!10^5$ \cite{Ackermann:2015tah} and the presence of substructure enhances the annihilation rate within galaxies by a factor of $\sim\!\!10$ \cite{Sanchez-Conde:2013yxa}.  Since the microhalos generated by nonstandard cosmologies are far denser and more abundant than microhalos in standard cosmologies, these boost factors are significantly higher in such scenarios: cosmic boosts following an EMD epoch can exceed $10^{16}$ \cite{Blanco:2019eij} and boosts within dwarf galaxies can exceed $10^6$ \cite{Erickcek:2015jza}.  These large boost factors can compensate for the reduced annihilation cross section required to generate the observed dark matter abundance when there is entropy production following dark matter freeze-out.  Therefore, searches for signatures of dark matter annihilation offer a promising way to constrain nonstandard thermal histories.

The first attempts to use gamma-ray observations to constrain nonstandard cosmologies used Fermi-LAT constraints on dark matter annihilations within dwarf spheroidal galaxies (dSphs) \cite{Erickcek:2015jza, Erickcek:2015bda}, but later analyses revealed that the isotropic gamma-ray background (IGRB) provides more powerful bounds on microhalo-dominated emission \cite{Blanco:2019eij,delos2019breaking}.  Following an EMD epoch, most of the dark matter is bound into early-forming microhalos that track the dark matter density, and the dark matter annihilation rate within these dense microhalos far exceeds the annihilation rate outside these microhalos.  If the central regions of these microhalos are not significantly affected by interactions with other structures, the annihilation rate per dark matter mass is homogeneous and constant after the microhalos form.  Consequently, microhalo-dominated annihilation can be cast as an effective dark matter lifetime: $\tau_\mathrm{eff} = \left[2m_\mathrm{DM}(\Gamma/M)\right]^{-1}$, where $m_\mathrm{DM}$ is the mass of the dark matter particle and $\Gamma/M$ is the annihilation rate per dark matter mass.  Observational lower bounds on $\tau_\mathrm{eff}$ correspond to lower bounds on the dark matter lifetime for particles with mass $2 m_\mathrm{DM}$ \cite{Blanco:2019eij}.  

\begin{figure}
\centering
\includegraphics[width=0.56\textwidth]{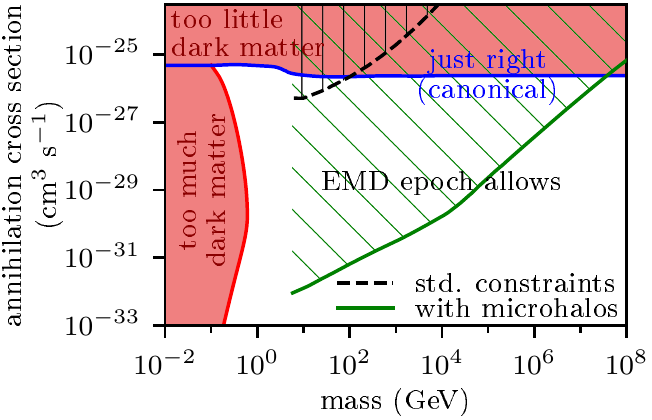}
\caption{Constraints on dark matter annihilation in EMD scenarios with $T_\mathrm{reh} = 10$ MeV. Thermal relics with annihilation cross sections well below the canonical value of \mbox{$\langle \sigma v \rangle \simeq 2\times 10^{-26}$ cm$^3$/s} \cite{Steigman:2012nb} can achieve the observed dark matter abundance in these scenarios: all points in the white region are viable if the EMD epoch is preceded by a radiation-dominated era, while its low-mass boundary (red curve) is viable without prior RD. The black dotted line shows standard upper bounds on dark matter annihilation \cite{Ackermann:2015zua}.  Including EMD epoch-generated microhalos dramatically extends these bounds: the green line shows how bounds on the dark matter lifetime from the IGRB \cite{Blanco:2018esa} restrict annihilation within microhalos for $k_\mathrm{cut}/k_\mathrm{reh} = 40$ \cite{delos2019breaking}.}
\label{Fig:AnnCon}
\end{figure}

The annihilation rate per mass depends on the microhalos' density profiles and their abundance. Ref. \cite{delos2019predicting} showed that microhalo profiles and their abundance can be predicted in a simple way from properties of the precursor linear density field, enabling rapid computation of $\Gamma/M$ in a given cosmological scenario \cite{delos2019breaking}. Other works \cite{Erickcek:2015jza, Erickcek:2015bda, Blanco:2019eij} have employed Press-Schechter theory \cite{Press:1973iz} to characterize microhalo populations and assumed that these microhalos have NFW profiles with low concentrations at formation.  The two approaches yield $\Gamma/M$ values that differ by a factor of $\sim\!\!3$ \cite{delos2019breaking}, but in both approaches, $\Gamma/M$ is strongly dependent on the formation time of the first microhalos (as determined by the ratio of the cut-off scale to the reheating horizon scale) and largely insensitive to the reheat temperature.  

Figure \ref{Fig:AnnCon} illustrates how gamma-ray observations can constrain thermal dark matter production in nonstandard cosmologies.  This figure shows constraints on the dark matter velocity-averaged annihilation cross section and the dark matter particle mass for an EMD epoch with a reheat temperature of $10$ MeV.  The shaded regions cannot generate the observed dark matter abundance, and the blue line shows the annihilation cross section required to generate the observed dark matter abundance in a standard cosmology.  All points in the white region can yield the observed dark matter abundance if the EMD epoch is preceded by a radiation-dominated era, while this region's low-mass boundary produces the observed abundance without prior radiation domination \cite{delos2019breaking}.\footnote{If dark matter freezes out during or before an EMD epoch, a smaller-than-canonical annihilation cross section is required to achieve the observed dark matter abundance because the entropy injected by the decay of the dominant $\phi$ component dilutes the dark matter relic abundance (see Sec.~\ref{sec:dark_sectors}). The extent of this dilution is reduced if the EMD epoch is preceded by an RD epoch and dark matter freezes out before this entropy injection alters the evolution of the radiation density; see Ref.~\cite{delos2019breaking} for details.}  The black dashed line shows limits on dark matter annihilation given standard structure formation \cite{Ackermann:2015zua}, while the green solid line shows how microhalos' contribution to the IGRB can enhance these constraints \cite{delos2019breaking}. Ref.~\cite{Blanco:2019eij} demonstrated how observations of the IGRB severely restrict hidden-sector dark matter models that include EMD periods, provided that these EMD periods are too short to permit structure formation prior to reheating.

Finally, we note that early-forming microhalos would enhance the dark matter annihilation rate prior to reionization, when the energy injection from dark matter annihilations can affect CMB temperature and polarization anisotropies and the temperature of the intergalactic medium (IGM).  The CMB already establishes powerful limits on dark matter annihilation, see e.g. Ref.~\cite{Slatyer:2015jla}.  For standard models of structure formation, dark matter halos form too late to significantly impact these constraints \cite{2013JCAP...07..046L}, but if most of the dark matter is bound into microhalos at $z\gtrsim 800$, the CMB could provide a new avenue for their detection.  Similarly, dark matter annihilations within microhalos can dramatically increase the IGM temperature, leading to 21cm emission (as opposed to absorption) at $z\gtrsim50$, long before conventional astrophysical sources of high-energy photons are expected to exist \cite{PSL15}.

\subsubsection{The microhalo population within galaxies}

Since the microhalos' gravitational signatures and annihilation signatures both depend on the microhalos' density profiles and abundance, we must understand how subsequent structure formation affects the microhalo population.  Most microhalos eventually become subhalos within much larger, later-forming halos, such as those that surround galaxies. Within the deep potential wells of galactic halos, microhalos are subjected both to tidal forces and to energetic encounters with other objects.\footnote{Subhalos are also subject to dynamical friction \cite[]{chandrasekhar1943dynamical}, which decelerates them. However, as this deceleration is proportional to the mass of the subhalo, it is expected to be negligible for microhalos.} The vast difference in scales between these microhalos and their hosts precludes direct numerical simulation of this picture, but a variety of analytic and semianalytic treatments have been developed.

To predict the impact of tidal forces, the simplest and most widely used method involves the tidal radius, the radius within the subhalo beyond which the host's tidal forces exceed the subhalo's self-gravity.  Any subhalo material lying beyond the tidal radius is expected to be stripped \cite[for a review of tidal radius definitions, see][]{bosch2017disruption}.  A refinement to this approach involves conditioning tidal stripping on the energy of subhalo material instead of its radius \cite{Drakos_2020}. Beyond this tidal truncation picture, material below the tidal radius is also heated by tidal forces, and the subhalo undergoes internal dynamics in response. No simple treatments exist for these effects, and models that predict tidal evolution are tuned to match numerical simulations. These models include \cite{taylor2000dynamics}, \cite{penarrubia2004effects}, \cite{kampakoglou2006tidal}, \cite{gan2010improved}, \cite{pullen2014nonlinear}, \cite{jiang2014statistics}, and~\cite{delos2019tidal}, which describe the time evolution of subhalos; and \cite{hayashi2002structural}, \cite{penarrubia2010impact}, and~\cite{green2019tidal}, which describe how their density profiles respond to this evolution. The impact of tidal effects on microhalos depends strongly on the microhalos' characteristic internal density, see e.g. Ref. \cite{delos2019breaking}, but as long as they posses divergent central density, tidal evolution is expected to be a continuous process that never fully destroys subhalos \cite[]{errani2019tides}.

In addition to tidal evolution, microhalos inside larger structures are also subject to encounters with other orbiting objects, such as stars or other microhalos. Owing to the high virial velocities within galactic halos, these encounters are expected to be impulsive; that is, the encounter timescale is much shorter than the timescale for the microhalo's internal dynamics. In this limit, the energy injected by the encounter may be computed readily, see e.g. Refs. \cite{spitzer1958disruption,gnedin1999tidal,donghia2010quasiresonant}. Additionally, a general argument suggests that the microhalo's density profile should relax to $\rho\propto r^{-4}$ at large radii \cite[]{jaffe1987envelopes}. However, the microhalo's full response to the impulse remains complicated. The response of a collisionless system to an impulsive shock has been widely studied in $N$-body simulations, originally in the context of galaxies, see e.g. Refs. \cite{aguilar1985tidal,aguilar1986density} and star clusters \cite{moore1993upper,gieles2006star,webb2018systematic} and later in the context of microhalos themselves \cite{goerdt2006survival,green2006minihalo,angus2006analysis,delos2019evolution}. Ref. \cite{delos2019breaking} found that for microhalos produced by an EMD epoch, the impact of encounters with other microhalos is subdominant to that of encounters with stars. Due to the relative rarity of stars, the impact of stellar encounters is highly stochastic, with the few closest encounters dominating a microhalo's evolution \cite[]{delos2019breaking}. Additionally, unlike tidal evolution, a penetrative encounter can disrupt a microhalo's divergent central density cusp \cite[]{delos2019evolution}, potentially leading to its destruction.

Finally, we remark that the impact of these evolutionary effects can discriminate between microhalo-dominated annihilation and dark matter decay. Microhalos near the centers of galactic systems suffer greater disruption compared to other microhalos, giving $\Gamma/M$ spatial dependence. Focusing on the Draco dwarf galaxy, \cite{delos2019breaking} found that the gamma-ray signal morphology from annihilation within microhalos can differ significantly from that of dark matter decay because the former is suppressed by tidal evolution and encounters with stars. This discrepancy would be even more pronounced when comparing different systems. For instance, the Galactic Center's high stellar density would greatly suppress any signal from annihilation within microhalos therein, whereas the isotropic gamma-ray background suffers minimal suppression. Consequently, a comparison between the gamma-ray signals from both systems could distinguish between microhalo-dominated dark matter annihilation and dark matter decay. Since the efficiency of both tidal evolution and impulsive encounters is highly sensitive to the internal density of the microhalos, such comparisons could even serve to measure detailed microhalo properties.

\subsection{Observational probes on primordial black holes (Authors: T. Harada \& K. Kohri)}

In this section we briefly review the current status of observational constraints on the abundance of PBHs. We also discuss how the curvature power spectrum can be constrained by using the PBH constraints. We begin by discussing PBH evaporation and its consequences.

\subsubsection{PBH evaporation and its consequences}
\label{subsub:PBHdom}

By emitting Hawking radiation~\cite{Hawking:1974rv} a PBH can evaporate. Their lifetime is given by
\be \label{eq:lfeteimePBH}
\tau_{\mathrm{PBH}} \sim \frac{m_{\mathrm{PBH}}^{3}}{M_{\rm P}^{4}}  \sim t_0 \left(\frac{m_{\mathrm{PBH}}}{5.1 \times 10^{14} \,\mathrm{g}}\right)^{3}  \sim 10^{64} \,{\rm yrs} \left(\frac{m_{\mathrm{PBH}}}{1 M_{\odot}}\right)^{3}.
\ee
where $t_0 = 13.8$ Gyrs is the current age of the Universe~\cite{Aghanim:2018eyx} and $M_{\odot} \simeq 2 \times 10^{33} $~g denotes Solar mass. According to the Hawking process, a black hole has a finite temperature,
\be \label{eq:BHtemperature}
T_{\mathrm{PBH}} \sim \frac{M_{\rm P}^{2}}{m_{\mathrm{PBH}}} \sim 0.1 \,\mathrm{MeV}\left(\frac{m_{\mathrm{PBH}}}{10^{15} \,\mathrm{g}}\right)^{-1} \sim 10^{-9} \,{\rm K}\left(\frac{m_{\mathrm{PBH}}}{1 \mathrm{M}_{\odot}}\right)^{-1} \,,
\ee
which means that we may observe $\gamma$-rays or cosmic rays (electron, positron, neutrino, etc.) produced by a presently evaporating PBH. Thus, one can severely constrain the abundance of PBHs with masses around $10^{15}$\,g by the non-detection of $\gamma$-rays or cosmic-ray positrons.

Due to Hawking evaporation, PBHs with masses smaller than $5.1 \times 10^{14}$\,g should have evaporated by the current age of the Universe. Therefore, such PBHs cannot constitute DM. An unstable PBH whose mass is smaller than $\sim 10^9$\,g (with lifetime being ${\cal O}(1)$\,s) cannot be constrained by observations as strongly even by adopting the constraint from BBN which has a good sensitivity at around $m_{\rm PBH} \gtrsim 10^9$~g, see Refs.~\cite{Carr:2009jm,Carr:2020gox} and the discussion below.\footnote{In the SM of particle physics, to maintain stability of the Higgs field at its vacuum expectation value, emission of high-energy particles from evaporating PBHs is dangerous. This can be used to constrain the evaporating PBHs even for $m_{\rm PBH}<10^9$~g. However, a concrete bound on the number density of PBHs has not been obtained~\cite{Kohri:2017ybt}.} In principle, however, evaporating PBHs can be constrained by production of hypothetical massive DM relics, such as the lightest supersymmetric particle (LSP) or Planck mass relics at which the evaporation terminates. From a nonthermal production of the LSP due to the evaporation, one can obtain an upper bound on $\beta$ in order not to exceed the observed DM density~\cite{Green:1999yh,Khlopov:2004tn,Lemoine:2000sq}:
\be \label{eq:betaLSP}
\beta \lesssim 10^{-18} \left(\frac{m_{\rm PBH}}{10^{11} \,\mathrm{g}} \right)^{-1/2} \left(\frac{m_{\mathrm{LSP}}}{100 \,\mathrm{GeV}}\right)^{-1} \left(m_{\rm PBH}<10^{11}\,\mathrm{g} \left(\frac{m_{\mathrm{LSP}}}{ 100\, \mathrm{GeV}}\right)^{-1} \right) \,,
\ee
with $m_{\mathrm{LSP}}$ being the mass of the LSP of the order of the weak scale. Quite recently, it was reported that this bound can be strengthened for lighter DM by considering free-streaming and an erase of small-scale fluctuations due to its high initial momentum~\cite{Lennon:2017tqq,Baldes:2020nuv}. Additionally, from a possible production of Planck mass relics at the last stage of the evaporation process, one can obtain an upper bound on $\beta$ in order not to exceed the DM density:
\be \label{eq:PlanckMassrelic}
\beta \lesssim 10^{20} \left(\frac{m_{\rm PBH}}{{\rm g}}\right)^{3/2} \,,
\ee
which applies for $1\,{\rm g}\lesssim m_{\rm PBH} \lesssim 10^7$\,g~\cite{Carr:1994ar}.

The above bounds depend on the detailed nature of DM, which is currently unknown. Therefore, if one omits these bounds due to such uncertainty, PBHs can be allowed to even dominate the early Universe for a short amount of time before they completely vanish away. Other possible probes of light PBHs are provided by GWs, that can be produced by nonlinear effects related to the large curvature perturbation at small scales which produced the PBHs~\cite{Inomata:2020lmk}, through Hawking radiation~\cite{Anantua:2008am,Dolgov:2011cq,Dong:2015yjs,Inomata:2020lmk} or by a coalescence of binary PBHs before the evaporation~\cite{Anantua:2008am,Dolgov:2011cq,Zagorac:2019ekv,Hooper:2020evu}. We also note that in some models, baryogenesis through nonthermal leptogenesis may be realized successfully by right-handed neutrinos produced by the huge amount of evaporating PBHs~\cite{Fujita:2014hha}.

\subsubsection{Observational constraints on PBHs}

PBHs can naturally form in the early Universe by large density/curvature fluctuations generated during inflation or in other scenarios. To obtain a theoretical prediction for the abundance of PBHs, the physics of PBH formation must be understood.  This gives a theoretical basis to use the observational data about PBHs to constrain the specific cosmological scenarios.  In Fig.~\ref{fig:fPBH} we plot a summary of observational constraints on the abundance of PBHs as a function of the PBH mass. The $y$-axis is the fraction of the mass density of PBHs to that of CDM, i.e., $f = \Omega_{\rm PBH} / \Omega_{\rm CDM}$.  The curves in this figure are from Ref.~\cite{Carr:2020gox}. The reason we did not adopt the constraints coming from 1) femtolensing, 2) neutron stars, and 3) white dwarfs are also discussed in Ref.~\cite{Carr:2020gox}. The quantities $\beta$, $f \equiv \Omega_{\rm PBH} / \Omega_{\rm CDM}$, and the curvature perturbation power spectrum ${\cal P}_{\zeta}$ are related to each other in the following way:
\be \label{eq:beta-fraction}
f \equiv \frac{\Omega_{\text {PBH }}}{\Omega_{\text {CDM}}} \sim\left(\frac{\beta}{10^{-18}}\right)\left(\frac{m_{\text {PBH }}}{10^{15} \,{\rm g}}\right)^{-1 / 2} \sim\left(\frac{\beta}{10^{-8}}\right)\left(\frac{m_{\text {PBH }}}{30 M_{\odot}}\right)^{-1 / 2} \sim 10^{8}\left(\frac{m_{\text {PBH }}}{30 M_{\odot}}\right)^{-1 / 2} \sqrt{{\cal P}_{\zeta}} \exp \left[-\frac{1}{18 {\cal P}_{\zeta}}\right].
\ee
Here we have assumed radiation had fully dominated after the end of the primordial inflation. By this relation (see Ref.~\cite{Carr:2020gox} for more precise relations), one can transfer the bounds on $\beta$ to the bounds on $f$ or ${\cal P}_{\zeta}$.  Next, we introduce each constraint shown in Fig.~\ref{fig:fPBH} on the abundance of PBHs.

\begin{figure}
\begin{center}
\includegraphics[clip,width=11cm]{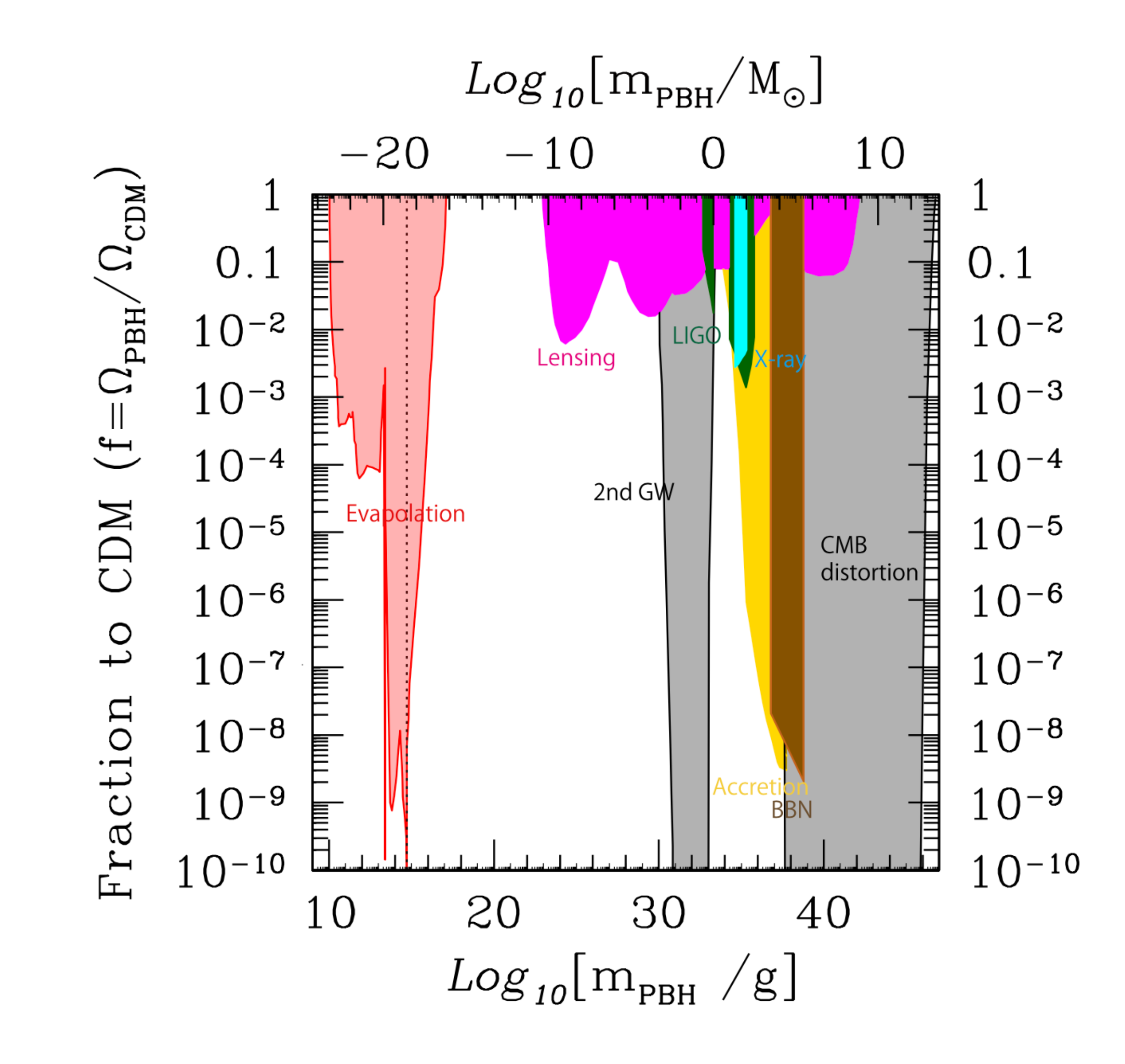}
\caption{Summary of observational constraints on the abundance of PBHs as a function of the PBH mass. The $y$-axis is the fraction of PBHs normalized to the energy density of CDM.  The red region excludes the evaporating PBHs through the observations of gamma-ray background~\cite{Carr:2009jm}, annihilating electron-positron pairs~\cite{DeRocco:2019fjq}, positrons by Voyager 1~\cite{Boudaud:2018hqb}, light element abundances with BBN~\cite{Carr:2009jm,Keith:2020jww}, cosmological 21~cm signals~\cite{Carr:2009jm,Mack:2008nv}, and CMB anisotropy~\cite{Carr:2009jm}. The magenta region excludes PBHs through gravitational lensing observations by  HSC~\cite{Niikura:2017zjd,Smyth:2019whb}, OGLE~\cite{Niikura:2019kqi}, EROS~\cite{Tisserand:2006zx}, MACHO~\cite{Allsman:2000kg} and Icarus~\cite{Oguri:2017ock}. From the observations of pulsar timing, one can constrain PBHs by the non-detection of GWs nonlinearly-produced by curvature perturbation~\cite{Saito:2008jc,Chen:2019xse,Carr:2020gox}.  Through the $\mu$-distortion of the global spectrum of CMB observations, one can obtain an upper bound on the curvature perturbation and the abundance of PBHs~\cite{Chluba:2012gq,Chluba:2012we,Kohri:2014lza,Inomata:2016rbd,Carr:2020gox}. The green regions exclude PBHs by the non-detection of GWs by LIGO/Virgo~\cite{Authors:2019qbw,Ali-Haimoud:2017rtz} (see also Ref. \cite{Vaskonen:2019jpv}). Because accretion of gas in to a PBH can emit radiation at radio and X-ray frequencies, observational data exclude PBHs masses $\mathcal{O}(10) M_{\odot}$~\cite{Gaggero:2016dpq} (see also Ref. \cite{Hektor:2018rul}).  The cosmological accretion in to PBHs is severely constrained by the observations of CMB polarization~\cite{Poulin:2017bwe,Serpico:2020ehh}. The dissipation of curvature perturbation induces dilution of baryon-to-photon ratio which affects BBN~\cite{Jeong:2014gna,Nakama:2014vla,Inomata:2016uip}. The lines in this figure are from Ref.~\cite{Carr:2020gox}.}
\label{fig:fPBH} 
\end{center}
\end{figure}

The red region excludes the evaporating PBHs through the following observations: One can constrain the abundance of PBHs through observations of the gamma-ray background of isotropic components~\cite{Carr:2009jm} (see also Refs. \cite{Ballesteros:2019exr,Arbey:2019vqx}), galactic components~\cite{Carr:2009jm,Carr:2016hva}, line emissions from annihilating electron-positron pairs~\cite{DeRocco:2019fjq,Laha:2019ssq,Dasgupta:2019cae}, and cosmic-ray positrons by Voyager 1~\cite{Boudaud:2018hqb}. These observations have excluded the possibility for PBHs to be the majority of DM for the masses of $10^{14.5}\,{\rm g} \lesssim m_{\rm PBH} \lesssim 10^{17}\,{\rm g}$.

The evaporating PBHs emit high-energy particles whose electromagnetic energy can modify the recombination history and background temperature. This potentially affects the observations of CMB anisotropy~\cite{Carr:2009jm,Poulin:2016anj,Stocker:2018avm,Poulter:2019ooo,Lucca:2019rxf,Acharya:2020jbv} and cosmological 21~cm line signals~\cite{Carr:2009jm,Mack:2008nv,Clark:2018ghm}. By the non-detection of those features one can constrain the abundance of PBHs for the masses of $10^{13.5}\,{\rm g} \lesssim m_{\rm PBH} \lesssim 10^{14.5}\,{\rm g}$.

Emissions of high-energy hadrons, gamma-rays, and charged leptons are dangerous for light elements which are produced in BBN. That is because they induce hadronic and electromagnetic showers through subsequent scatterings off the background particles, and then the produced daughter particles can destroy light elements through inelastic hadron scattering and photodissociation~\cite{Kawasaki:2004qu}. Comparing theoretical predictions of BBN with observational abundance of light elements, one can exclude the masses of $10^{9} {\rm g} \lesssim m_{\rm PBH} \lesssim 10^{13.5} {\rm g}$ as a considerable fraction of DM~\cite{Carr:2009jm} (see also Refs.~\cite{Poulin:2016anj,Acharya:2020jbv} for possible modifications only on photodissociation).

Nowadays there is strong motivation to study PBHs because the rate of GW detected by aLIGO~\cite{Abbott:2016blz} can be fitted by the event rate of coalescence of binary PBHs with masses $\sim 30\,M_{\odot}$~\cite{Sasaki:2016jop} for the fraction $f \sim {\cal O} (10^{-3})$ (see also Refs.~\cite{Bird:2016dcv,Raidal:2017mfl,Sasaki:2018dmp}). In other words, one can simultaneously constrain the number density of PBHs by observing GWs produced by binary PBHs with arbitrary masses. The two green regions are excluded due to the non-detection of such GWs by the current sensitivities of aLIGO/aVirgo for $0.2 M_{\odot} \lesssim m_{\rm PBH} \lesssim 1 M_{\odot}$ and $10 M_{\odot} \lesssim m_{\rm PBH} \lesssim 3 \times 10^{2} M_{\odot}$~\cite{Authors:2019qbw,Ali-Haimoud:2017rtz} (see also Refs.~\cite{Raidal:2018bbj,Vaskonen:2019jpv}). Future detectability of GWs from lighter binary PBHs and its consequences on a bound on PBH abundance are discussed in Ref.~\cite{Wang:2019kaf}.

Observations of gravitational lensing can potentially give a constraint on an object passing through the line of sight towards a light source. The magenta region is excluded through gravitational lensing by the observations of HSC~\cite{Niikura:2017zjd,Smyth:2019whb}, OGLE~\cite{Niikura:2019kqi}, EROS~\cite{Tisserand:2006zx}, MACHO~\cite{Allsman:2000kg} and Icarus~\cite{Oguri:2017ock}. From Fig.~\ref{fig:fPBH}, we see that the masses of $10^{-10} M_{\odot} \lesssim m_{\rm PBH} \lesssim 2 \times 10^{4} M_{\odot}$ and $6 \times 10^{4} M_{\odot} \lesssim m_{\rm PBH} \lesssim 10^{9.5} M_{\odot}$ are excluded by the non-detection of lensing events caused by PBHs. On the other hand, we note that one event in HST~\cite{Niikura:2017zjd} and six events in OGLE~\cite{Niikura:2019kqi}  can be fitted by PBHs.

As discussed below, upper bounds on curvature perturbation ${\cal P}_{\zeta}$ can be transformed into the ones on the abundance of PBHs ($f$ or $\beta$). That is because PBHs are expected to form in the early Universe by the collapse of inhomogeneous regions, which are characterized by the curvature perturbation. The dissipation of density perturbation at small scales in the early Universe induces dilution of baryons which affects the freeze-out value of neutron-to-proton ratio $n/p$ during BBN. Then, the brown region is excluded by comparing the theoretical predictions of light element abundances with the observations for $10^{3.5} M_{\odot} \lesssim m_{\rm PBH} \lesssim 10^{5.5} M_{\odot}$~\cite{Jeong:2014gna,Nakama:2014vla,Inomata:2016uip}. Through observations of the CMB $\mu$-distortion caused by the dissipation of density perturbation, one can obtain an upper bound on curvature perturbation and, simultaneously, the abundance of PBHs for $10^{4.5} M_{\odot} \lesssim m_{\rm PBH} \lesssim 10^{13.5} M_{\odot}$~\cite{Kohri:2014lza,Inomata:2016rbd,Carr:2020gox}. From the observations of pulsar timing, one can constrain PBHs by the non-detection of the GW background secondarily-induced by curvature perturbation~\cite{Saito:2008jc,Kohri:2018awv,Chen:2019xse,Cai:2019elf,Carr:2020gox}.

Radio waves and X-rays are emitted through accretion of gas into a PBH. From observations, one can obtain an upper bound on the abundance of PBHs with masses ${\cal O}(10)$ -- ${\cal O}(10^2) M_{\odot}$, which is denoted by the cyan region~\cite{Gaggero:2016dpq}. Cosmological accretion is a relatively new subject. Considering a finite relative velocity between baryon and CDM/PBHs,  disk accretion of baryons into PBHs must have occurred inevitably at high redshifts from $z \sim {\cal O}(10^2)$ to $z \sim {\cal O}(10^3)$. The photons emitted through the accretion can change the recombination history of baryons, which is severely constrained by observations of the CMB anisotropy and polarization. The yellow ocher region is excluded by the current observations of accretion into a PBHs and their CDM halo systems~\cite{Poulin:2017bwe,Serpico:2020ehh} (see also Refs.~\cite{Ricotti:2007au,Ali-Haimoud:2016mbv,Hektor:2018qqw}).

\subsubsection{Constraints on the power spectrum of curvature perturbation}
\label{subsub:constRD}

In earlier sections, we have discussed inflation and curvaton models which produce curvature perturbations. Actually, so far a lot of inflationary models have been proposed in which PBHs are produced at small scales while simultaneously fitting the CMB normalization at large scales. For example, there are models of double inflation~\cite{Kawasaki:1997ju,Inomata:2016rbd}, hilltop inflation~\cite{Kohri:2007gq,Kohri:2007qn,Alabidi:2009bk}, preheating~\cite{Taruya:1998cz,Martin:2019nuw}, curvaton~\cite{Kawasaki:2012wr,Kohri:2012yw}, thermal inflation~\cite{Dimopoulos:2019wew}, false vacuum bubbles~\cite{Garriga:1992nm,Kusenko:2020pcg}, Q-ball formation~\cite{Cotner:2019ykd,Kawasaki:2019iis}, and so on. These models predict that curvature perturbation can be ${\cal P}_{\zeta} \sim {\cal O}(0.1)$ at small scales due to special features in the spectrum, e.g. a blue-tilted spectrum, a sharp peak structure, and so on.

In particular, motivated by the blue-tilted spectrum predicted by some of the inflation models above, as an example, here we introduce a simple parametrization of large curvature perturbation at small scales as follows:
\be \label{eq:parametrization}
{\cal P}_{\zeta}(k)=\mathcal{A}\left(\frac{k}{k_{*}}\right)^{n_{s}-1+\frac{\alpha_{s}}{2} \ln \left(\frac{k}{k_{*}}\right)+\frac{\beta_{s}}{6}\left(\ln \left(\frac{k}{k_*}\right)\right)^{2} + \cdots} \,,
\ee
where $\mathcal{A}$ is the normalization of the amplitude, called the ``CMB normalization''~\cite{Ade:2015xua} at the pivot scale of the wavenumber $k=k_* =0.05$ Mpc$^{-1}$, $n_s$ is the spectral index, $\alpha_\text{s}$ is its running, and $\beta_\text{s}$ is the running of the running defined as
\be \label{eq:Defs}
n_{s} -1 \equiv \frac{\rm{d} \ln P_{\zeta}}{\rm{d} \ln k}, \quad \alpha_{{s}}\equiv\frac{\rm{d} n_{{s}}}{\rm{d}\ln k}, \quad \beta_{{s}}\equiv\frac{\rm{d}^{2} n_{{s}}}{\rm{d}(\ln k)^{2}} \,.
\ee
This kind of parametrization considering the higher-order corrections up to $\beta_\text{s}$ was used by the Planck collaboration~\cite{Ade:2015lrj}. It also extends the definition used in Sec. \ref{sec:inflation}. The Planck 2018 observations (with the TT, TE, EE$+$lowE$+$lensing dataset)  give
\be
\begin{aligned}
n_{\text{s}}= & 0.9625 \pm 0.0048 , \\
\alpha_s =& 0.002 \pm 0.010, \\
\beta_s =& 0.010 \pm 0.013,
\end{aligned}
\ee
with $\ln(10^{10}\mathcal{A}) = 3.045 \pm 0.0016$ at 68$\%$ CL~\cite{Aghanim:2018eyx,Akrami:2018odb}. Actually some inflation models predict such a blue-tilted spectrum with these quantities, see e.g. Refs.~\cite{Kohri:2007gq,Kohri:2007qn,Alabidi:2009bk}.

In Fig.~\ref{fig:pzetak}, we plot the constraints on the curvature power spectrum ${\cal P}_\zeta$ as a function of wavenumber $k$. Here we have a simple relation between the power spectrum of the curvature perturbation ${\cal P}_{\zeta}(k)$ and the density perturbation, ${\cal P}_{\zeta}(k) \sim \sigma^2 {(5 + 3 w)^2}/{2(1+ w)^2}$ with $\sigma^{2}$ being the mean variance of density perturbation~\cite{Kohri:2018qtx} (see also ~Ref.~\cite{Yoo:2018esr} for a more accurate relation). Allowed within 95$\%$ CL, we can see that the predicted curve of curvature perturbation with $n_s = 0.96$, $\alpha_s = 0.010$, and $\beta_s = 0.020$ can reach the edge of the bound from PBHs with the e-folding number $N\sim 17$.  This means that a PBH with the mass 30\,M$_{\odot}$, which can fit the LIGO events, can be produced in this parametrization~\cite{Kohri:2018qtx} at $k=k_{\rm LIGO} \sim 10^6 {\rm Mpc}^{-1}$. See also Ref.~\cite{Byrnes:2018txb} for similar analyses by using only $n_s$ and Refs.~\cite{Carr:2017edp,Carr:2018nkm} by using only $n_s$ and $\alpha_s$. Possible uncertainties about this kind of parametrizatios are also discussed in Ref.~\cite{Green:2018akb}. Here we used the relation between the wavenumber $k = k_{\rm RD}$, at which the mode entered into the horizon when the PBH was produced, and the mass of the PBH 
\be  \label{eq:kRD}
k_{\rm RD} \sim  6 \times 10^{15} \,{\rm Mpc}^{-1}  \left(\frac{m_{\rm PBH}}{10^{15}\, {\rm g}}\right)^{-1/2} \,.
\ee

\begin{figure}
\begin{center}
\includegraphics[width=8cm]{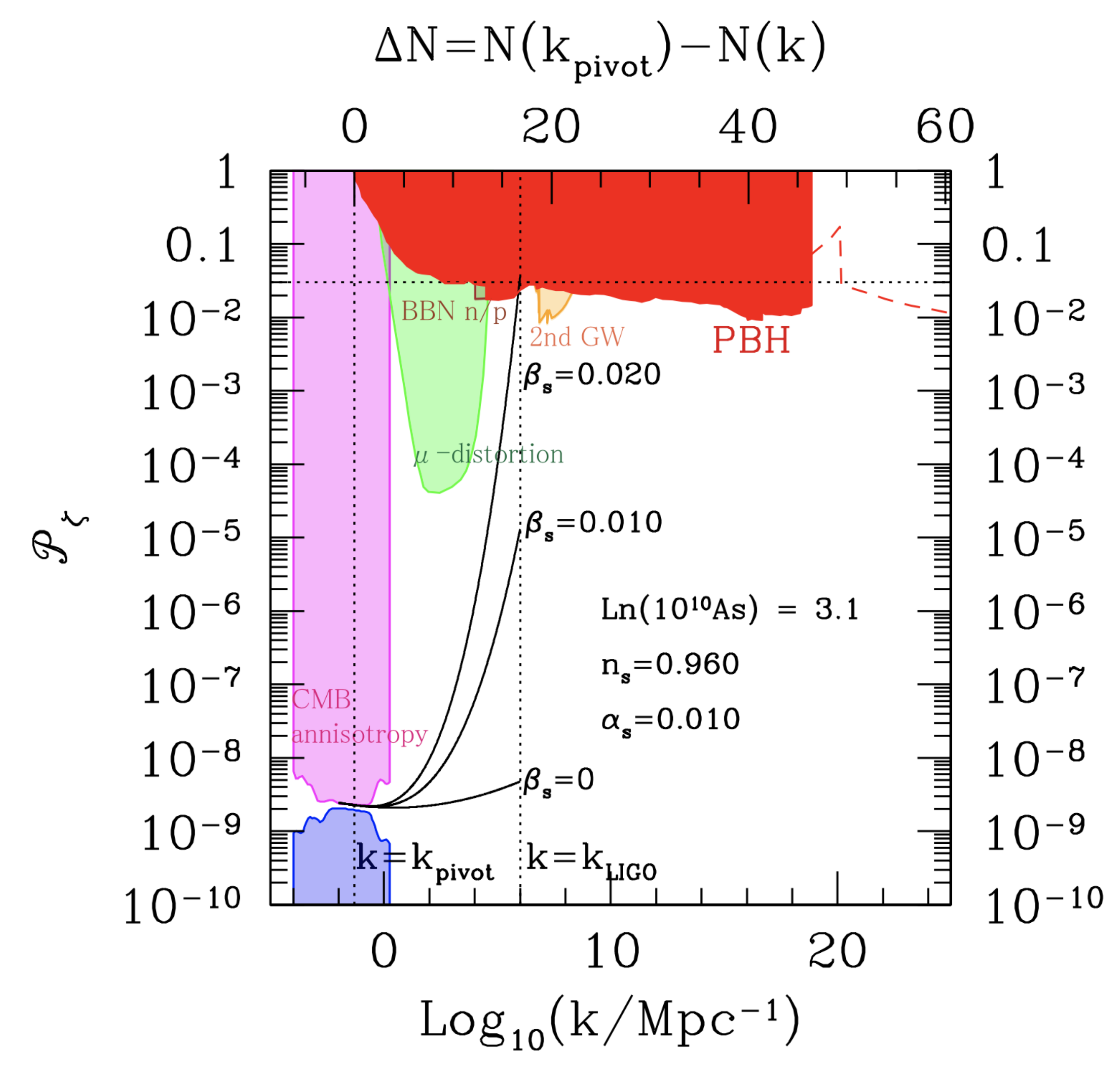}
\caption{Constraints on curvature perturbation as a function of wavenumber (comoving momentum) $k$ in units of Mpc$^{-1}$. The red region is excluded by the constraints on PBHs~\cite{Carr:2020gox}. The dashed curve denotes the upper bound in case of nonstandard scenarios, cf. Eqs.~\eqref{eq:betaLSP} and \eqref{eq:PlanckMassrelic}. From the CMB observations one obtains the allowed region at the pivot scale~$k=k_*=$0.05\,Mpc$^{-1}$~\cite{Aghanim:2016yuo,Aghanim:2018eyx}. The green, orange, and brown regions are excluded by the $\mu$-distortion of the CMB spectrum~\cite{Chluba:2012gq,Chluba:2012we,Kohri:2014lza,Inomata:2016rbd,Carr:2020gox}, secondary GWs~\cite{Saito:2008jc,Kohri:2018awv,Chen:2019xse,Carr:2020gox}, and the dilution of baryons (or the $n/p$ ratio) after BBN~\cite{Jeong:2014gna,Nakama:2014vla,Inomata:2016uip}, respectively. The lines in the figure are from Ref.~\cite{Carr:2020gox}.}
\label{fig:pzetak} 
\end{center}
\end{figure}

\subsubsection{Constraints on curvature perturbations from PBH formation during matter dominance}
\label{subsub:constMD}

Finally, we discuss how one can constrain the abundance of PBHs produced during an EMD epoch. An EMD epoch may end by the decay of a massive particle $X$ -- such a the inflaton or curvaton field -- which had dominated the Universe and its subsequent thermalization. This transition is characterized by the reheating temperature $T_{\rm reh}$ defined by $\Gamma_X \equiv 3H(T=T_{\rm reh})$] with $\Gamma_X$ being the decay rate of the massive particle $X$ (see e.g. Sec.~\ref{sec:dark_sectors} for details). The wavenumber which crosses the horizon at $T=T_{\rm reh}$ is defined by
\be \label{eq:kTH}
k_{\rm R} \equiv a H(T=T_{\rm reh}) \,.
\ee 

If a PBH was produced during an EMD epoch, the relation between the energy fraction $\beta = \beta_{\rm MD}$ at the time of formation time and the energy fraction in terms of CDM at present, $f$, is modified to~\cite{Kohri:2018qtx,Dalianis:2018ymb}
\be  \label{eq:betaMD}
\beta_{\rm MD} \sim 6 \times 10^{-18} f \left(\frac{T_{\rm reh}}{10^{10}\,{\rm GeV}}\right)^{-1} \,.
\ee
After reheating, the standard RD epoch was realized. According to Eq.~\eqref{eq:beta-fraction}, we see that the abundance of PBHs produced during the RD epoch is 
\be \label{eq:betaRD}
\beta_{\rm RD} \sim 10^{-18} f \left(\frac{m_{\rm PBH}}{10^{15}\, {\rm g }}\right)^{1/2},
\ee
In addition, the relation between $k$ and $m_{\rm PBH}$ for PBHs produced during a MD epoch is different from Eq.~\eqref{eq:kRD}, as now~\cite{Kohri:2018qtx,Dalianis:2018ymb}
\be  \label{eq:kMD}
k_{\rm MD} \sim  2 \times 10^{16}\,{\rm Mpc}^{-1} \left( \frac{m_{\rm PBH}}{10^{15}\, {\rm g}}\right)^{-1/3}\left( \frac{T_{\rm reh}}{10^{10} \,{\rm GeV} }\right)^{1/3} \,.
\ee
Then, we can use the wavenumber
\be \label{eq:kaf}
k = \left\{\begin{array}{ll}k_{\mathrm{MD}}, & \text { for } \quad k \gtrsim k_{\mathrm{R}} \\ 
k_{\mathrm{RD}} & \text { for } \quad k \lesssim k_{\mathrm{R}}\end{array}\right. \,,
\ee
and the abundance
\be  \label{eq:betaaf}
\beta = \left\{\begin{array}{ll}\beta_{\mathrm{MD}}, & \text { for } \quad k \gtrsim k_{\mathrm{R}}  \\ 
\beta_{\mathrm{RD}} & \text { for } \quad k \lesssim k_{\mathrm{R}}\end{array}\right. \,.
\ee

In Fig.~\ref{fig:PRk_comb}, the constraints on the power spectrum of the curvature perturbation $P_{\zeta}$ are plotted, as an example, without the spin suppression discussed in Sec.~\ref{sec:formationPBHs}. The figure is from Ref.~\cite{Carr:2017edp}, where the authors adopted the relation $\beta = 0.056 \sigma^5$.  The constraint is highly dependent on the reheating temperature, obeying a simple scaling $P_{\zeta} \propto \sigma^2 \propto \beta_{\rm MD}^{2/5} \propto T_{\rm reh}^{-2/5}$. We stress the importance of the effects of the exponential suppression of $\beta$ as a function of $\sigma$ due to a finite spin (see Sec. \ref{sec:formationPBHs}). In Fig.~\ref{fig:PMD2ndOrder}, we show the constraints on $P_{\zeta}$ depending on the reheating temperature, $T_{\rm reh}=10^4$\,GeV (left panel) and $T_{\rm reh}=10^9$\,GeV (right panel) by adopting the relation between $\beta$ and $\sigma$ for PBHs produced during an EMD epoch. Here we have used the result for the second-order effect shown in Fig.~\ref{fig:beta_2cases}. As noted in Sec.~\ref{subsec:FromMD}, we recommend using this curve when looking for allowed regions of parameters in theoretical models which produce PBHs during an EMD epoch because it fits data more conservatively with milder assumptions of the theoretical model.

\begin{figure}
\center
\includegraphics[width=6cm]{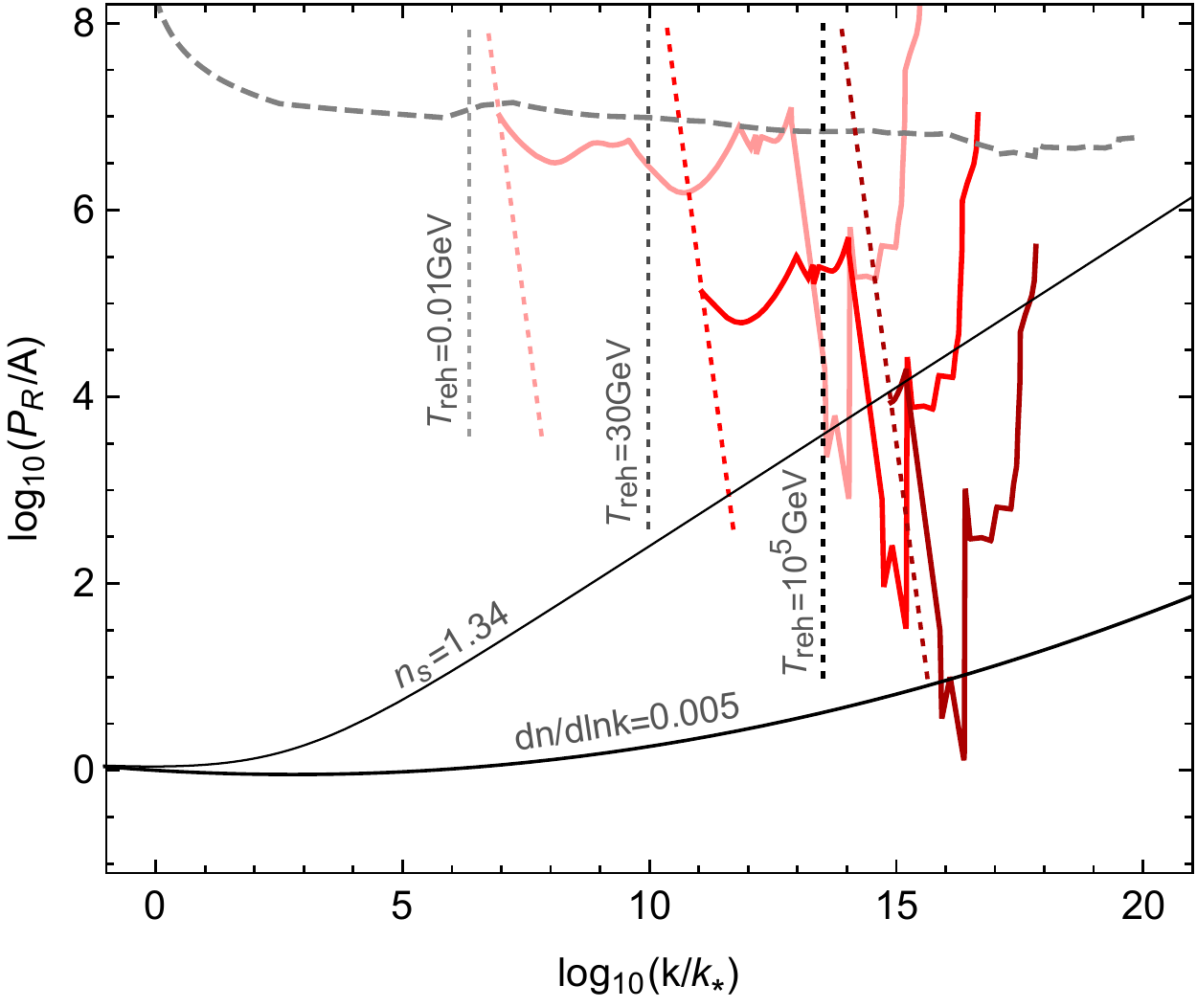}
\caption{Constraints on $P_R = P_{\zeta}$ in various cases for reheating temperature $T_{\rm reh}$ by using the relation $\beta = 0.056 \sigma^5$ and omitting the suppression caused by a finite spin~\cite{Khlopov:1980mg,Polnarev:1982,Harada:2016mhb}. Here $A = \mathcal{A}$ denotes the CMB normalization. The dashed line denotes the bound in the RD case. The vertical dotted line denotes $k_{\rm R}$, and the oblique dotted line denotes the wavenumber above which a PBH can be produced by the gravitational nonlinearity within an early MD epoch. The figure is from Ref.~\cite{Carr:2017edp}.}
\label{fig:PRk_comb} 
\end{figure}

\begin{figure}
\centering
\includegraphics[width=0.4\columnwidth]{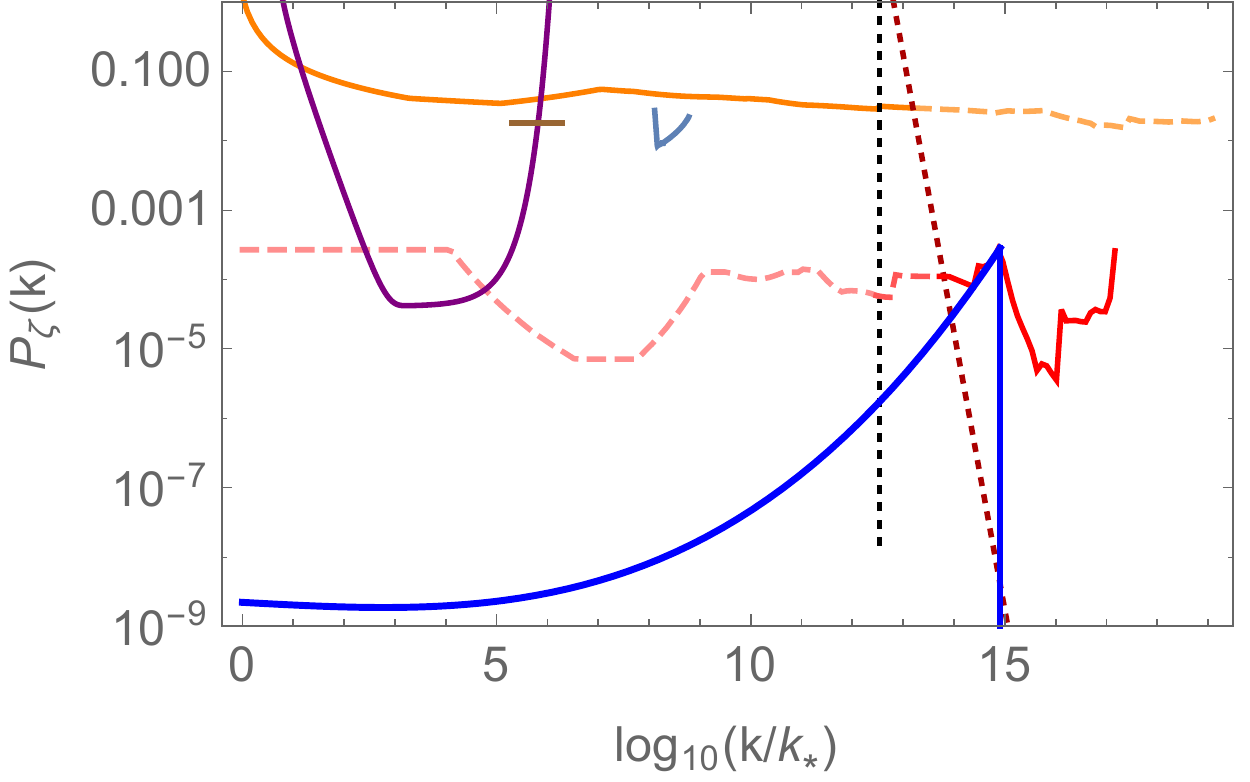} \hspace{6mm}
\includegraphics[width=0.4\columnwidth]{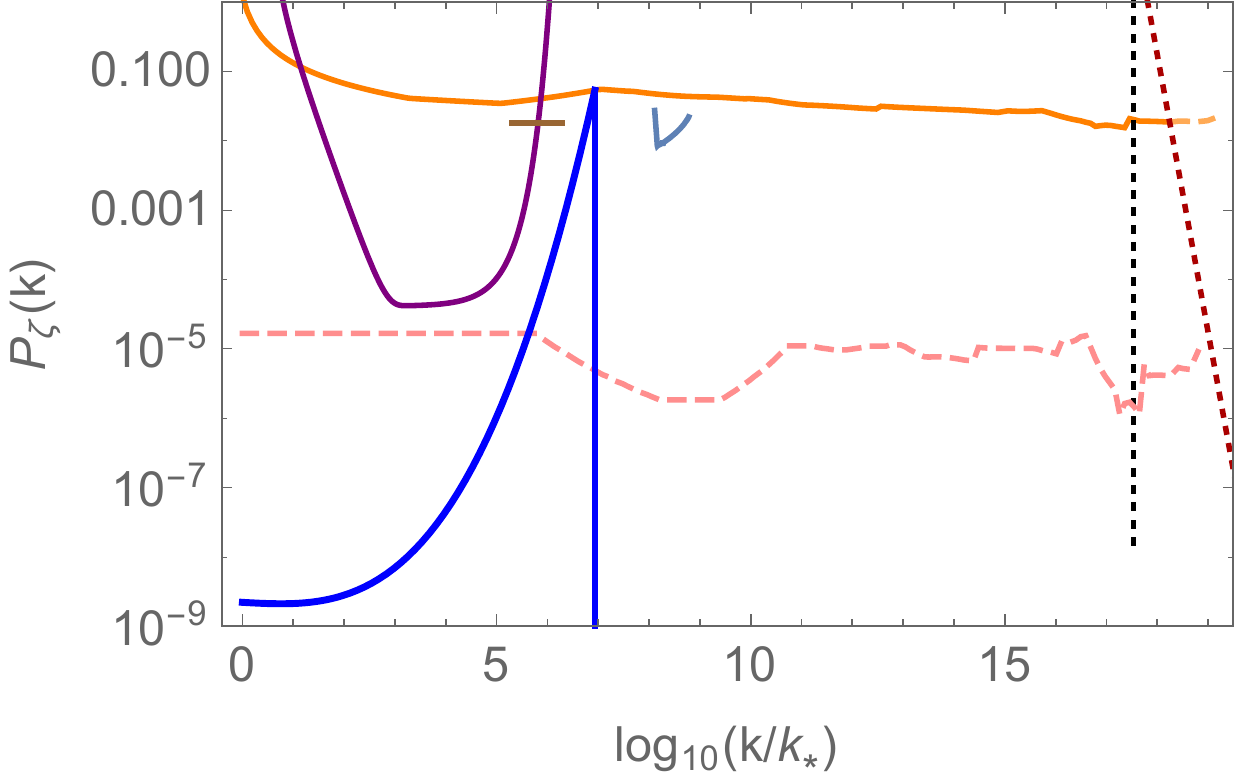}
\caption{\emph{Left panel:} An example of the power spectrum $P_\zeta (k)$ (blue solid line) which realizes 100$\%$ of DM in PBHs ($f=1$) with $m_{\rm PBH} \sim 10^{17}$\,g. The parameters characterizing the running spectral indices are taken to be $n_s=0.96, \alpha_s = 0, \beta_s = 0.0019485$.  The orange and red lines are the constraints on $f$ in RD and MD cases, respectively. Other solid lines show the constraints from CMB $\mu$-distortion, BBN $n/p$, and second order GWs by PTA, from left to right. The black dotted line denotes reheating at $T_{\rm reh}=10^4$\,GeV, and the dark red dotted line shows the minimum $k$ which becomes nonlinear during the MD era. \emph{Right panel:} The same as the left panel except for the following points: (i) $f=1.2 \times 10^{-3}$ to fit the GW signal observed by LIGO with $m_{\rm PBH} \sim$ 30 $M_{\odot}$~\cite{Sasaki:2016jop,Sasaki:2018dmp}, (ii) $n_s=0.96, \alpha_s = 0, \beta_s = 0.026$, and (iii) $T_{\rm reh} =10^9$\,GeV. The figures are from Ref.~\cite{Kohri:2018qtx}.}
\label{fig:PMD2ndOrder}
\end{figure}

\vspace{3.5cm}
\subsection{Gravitational wave backgrounds (Authors: D. G. Figueroa \& M. Lewicki)}
\label{sec:GWBGs}

In this section we discuss three potential cosmological contribution to the stochastic GW background (SGWB), inflation, cosmic strings and phase transitions, and how a nonstandard expansion phase can effect the GW spectrum from these sources.

\subsubsection{Inflation}

During the inflationary period, tensor perturbations, originated out of quantum fluctuations, are spatially stretched to scales exponentially larger than the inflationary Hubble radius. This results in a (quasi-)scale invariant tensor power spectrum at super-horizon scales~\cite{Grishchuk:1974ny,Starobinsky:1979ty, Rubakov:1982df,Fabbri:1983us},
\be
\left<h_{ij}(t,\mathbf{x})h^{ij}(t,\mathbf{x})\right>\equiv\int \frac{dk}{k} \Delta_{h,{\rm inf}}^2(t,k) \quad \Longleftrightarrow \quad \Delta_{h,\text{inf}}^2(k)\simeq \frac{2}{\pi^2}\left(\frac{H_{\text{inf}}}{ m_{\text{p}}}\right)^2\left({k\over k_*}\right)^{n_t} \label{inf}\,;\hspace{0.5cm}n_t \simeq -2\epsilon \simeq -{r_*\over8} \,,
\ee
where $k_{*}$ is a pivot scale and $H_{\rm inf}$ is the inflationary Hubble scale when the mode $k_*$ crossed the Hubble radius during inflation. Slow-roll inflation leads typically to a slightly red ``slow-roll suppressed" tilt, which can be expressed as proportional to the tensor-to-scalar ratio $r_{*}$ (evaluated at $k_*$). As the latter is constrained (at $k_*/a_0 = 0.002~{\rm Mpc}^{-1}$) by the B-mode in the CMB as $r_* \leq 0.06$~\cite{Akrami:2018odb,Ade:2018gkx}, this leads to an upper bound on the scale of inflation $H_{\rm inf} \lesssim H_{\rm max} \simeq 6.6\times 10^{13}\,{\rm GeV}$, implying also that the red tilt can only be very small $-n_t \leq 0.008 \ll 1$. 

Following inflation the tensor modes re-enter successively back inside the horizon. After crossing the horizon, each tensor mode starts oscillating with decreasing amplitude $h_{ij} \propto 1/a$, independently of the expansion rate~\cite{Caprini:2018mtu}. Since different modes re-enter the horizon at different moments of the cosmic history, the may propagate through different periods of the evolution of the Universe after becoming sub-horizon. In order to characterize the spectrum of the SGWB today, we introduce the energy density spectrum, normalized to the critical density $\rho_{\text{crit}}=3m_p^2H^2$, as~\cite{Caprini:2018mtu}
\be
\begin{aligned} \label{transferfunc}
&\Omega_{\text{GW}}(t, k) \equiv\dfrac{1}{\rho_{\text{crit}}}\dfrac{d\rho_{\text{GW}}(t,k)}{d\ln k} = \frac{k^2}{12a^2(t)H^2(t)}\Delta_h^2(t,k)\,,\\  
&\Delta_h^2(t,k)\equiv T_h(t,k) \Delta_{h,\text{inf}}^2(k)\,,~~~~ T_h(t,k) \equiv {1\over 2}\left(a_k\over a(t)\right)^2\,,
\end{aligned}
\ee
where $T_h(t,k)$ is a transfer function~\cite{Boyle:2005se} that characterizes the expansion history between the moment of horizon re-entry $t = t_k$ of a given mode $k$ (defined as $a_kH_k \equiv k$, with $a_k \equiv a(t_k)$, $H_k\equiv H(t_k)$), and a later moment $t > t_k$. For tensor modes crossing during RD epoch, the resulting GW energy density spectrum retains the original shape 
\be \label{eq:InfGWtodayRD}
 \Omega_{\rm GW}^{(0)}{\Big |}_{\rm inf}  \simeq \mathcal{G}_k { \Omega_{\rm rad}^{(0)}\over12\pi^2}\left(H_{\rm inf}\over m_p\right)^2\left(H_{\rm inf}\over H_{\rm max}\right)^2\left(f\over f_{\rm p}\right)^{n_t}  \simeq 10^{-16}\left(H_{\rm inf}\over H_{\rm max}\right)^2\left(f\over f_{\rm p}\right)^{n_t}\,,
\ee
with $n_t$ given by the original inflationary tilt. 

If there is, however, a period of expansion after inflation for which the Universe energy budget is not dominated by relativistic species, the (quasi-)scale invariance of the original tensor spectrum is broken, as the inflationary GW energy density spectrum develops a significant tilt within the frequency range corresponding to the modes crossing the horizon during such period. We may detect or constrain in this way the post-inflation expansion history (and hence the properties of the matter fields driving the expansion), by attempting to detect the relic SGWB of inflationary origin, see e.g. Refs.~\cite{Giovannini:1998bp,Giovannini:1999bh,Boyle:2005se, Watanabe:2006qe, Boyle:2007zx,Kuroyanagi:2008ye,Kuroyanagi:2010mm,Kuroyanagi:2014qza}. In order to see this, let us consider a cosmic epoch between the end of inflation and the onset of RD era, with effective EoS parameter $w \neq 1/3$. 

In scenarios where the inflaton potential is $V(\phi) \propto \phi^p$, the inflaton oscillates around the minimum of $V(\phi)$, so that an effective EoS (averaged over inflaton oscillations) emerges following the end of inflation~\cite{Turner:1983he,Lozanov:2016hid}, e.g.~ within the range $0 < w < 1/3$ for $p < 4$. For $p = 2$, it is well known that $w \simeq 0$ emerges, as oscillations of a massive free field lead to the energy density redshifting on average (over one oscillation cycle) as  $\left\langle\rho_\phi \right\rangle_{\rm osc} \propto 1/a^3$, analogous to that of nonrelativistic particle species. It is also possible that there is a stiff stage with {\small$1/3 < w < 1$}. The latter case can be actually achieved quite naturally if following the end of inflation the kinetic energy of inflaton dominates, either through inflaton oscillations~\cite{Turner:1983he} under a steep potential (e.g. $V(\phi) \propto \phi^p$ with $p > 4$), or simply by an abrupt drop of the inflationary potential. 

Propagating the inflationary tensor modes through a nonstandard epoch starting immediately after the end of inflation leads to a GW energy density spectrum today, expressed as a function of present-day frequencies $f=k/(2\pi a_0)$, as~\cite{Figueroa:2019paj}
\begin{eqnarray}\label{eq:GWfullSpectrumInstant}
\Omega_{\text{GW}}^{(0)}(f) = \Omega_{\rm GW}^{(0)}{\Big |}_{\rm inf} \times \mathcal{W}(f/f_{\rm RD}) \times \mathcal{C}_{w}\,\left({f\over f_{\rm RD}}\right)^{2\left({3w-1\over 3w+1}\right)} \simeq \Omega_{\rm GW}^{(0)}{\Big |}_{\rm inf}
\times\left\lbrace
\begin{array}{crl}
1 & \,, & ~~f \ll f_{\rm RD} \vspace*{0.3cm}\\
\mathcal{C}_{w}\,\left({f\over f_{\rm RD}}\right)^{2\left({3w-1\over 3w+1}\right)} & \,, &~~f \gg f_{\rm RD} \\
\end{array}
\right.\,,
\end{eqnarray}
with $\mathcal{C}_{w}$ just a numerical prefactor ranging $1 < \mathcal{C}_{w} < 2^{5/2}/\pi \simeq 1.80$ for $1/3 < w < 1$, $f_{\rm RD} \equiv k_{\rm RD}/(2\pi a_0)$ the frequency corresponding to the horizon scale at the onset of RD era, $k_{\rm RD} = a_{\rm RD}H_{\rm RD}$, and $\mathcal{W}(x)$ a window function connecting the two branches of the resulting spectrum, corresponding to modes crossing during the nonstandard cosmology period ($f \gg f_{\rm RD}$) and modes crossing during RD era ($f \ll f_{\rm RD}$). One can easily understand the result for the $f\gg f_{\rm RD}$ branch intuitively as follows: when tensor modes become sub-horizon they behave as proper GWs with an energy density scaling as $(d\rho_{\rm GW}/d\log k) \propto 1/a^4$. The ratio -- for a fixed mode $k$ -- of the GW energy density spectrum to the EoS-$w$-background, scales therefore as $\rho_{\rm tot}^{-1}(d\rho_{\rm GW}/d\log k) \propto a^{3w -1}$, which is a decreasing (growing) function for $w < 1/3$ ($w > 1/3$). As successive modes cross the horizon, the spectrum becomes $\rho_{\rm tot}^{-1}(d\rho_{\rm GW}/d\log k) \propto a(\tau_k)^{3w-1} \propto \tau_k^{\alpha(3w-1)} \propto k^{-2(3w-1)/(3w+1)}$, where we have used $a(\tau) \propto \tau^{\alpha}$ with $\alpha = 2/(1+3w)$ and $\tau$ the conformal time, and the horizon crossing condition $k = a_kH_k = {\alpha \tau_k^{-1}}$.

Given the tilt in Eq.~(\ref{eq:GWfullSpectrumInstant}), scenarios with a stiff fluid dominated epoch between inflation and the onset of RD era are actually very appealing from an observational point of view, since they develop a blue tilt for tensor perturbations, $0 < n_t < 1$, at large frequencies~\cite{Giovannini:1998bp,Giovannini:1999bh,Riazuelo:2000fc,Sahni:2001qp,Tashiro:2003qp,Boyle:2007zx,Giovannini:2008zg,Giovannini:2008tm,Caprini:2018mtu,Bernal:2019lpc,Figueroa:2019paj}, opening up the possibility of detection by upcoming direct detection GW experiments. The possibility of a stiff epoch is actually well-motivated theoretically in various early Universe scenarios, like Quintessential inflation~\cite{Peebles:1998qn,Peloso:1999dm,Huey:2001ae,Majumdar:2001mm,Dimopoulos:2001ix,Wetterich:2013jsa,Wetterich:2014gaa,Hossain:2014xha,Rubio:2017gty}, gravitational reheating~\cite{Ford:1986sy,Spokoiny:1993kt} (though with the caveats explained in Ref.~\cite{Figueroa:2018twl}), Higgs-reheating scenario~\cite{Figueroa:2016dsc} and  variants~\cite{Dimopoulos:2018wfg,Opferkuch:2019zbd}. The shape of the GW spectrum in these scenarios is controlled by $w$, $f_{\rm RD}$, and $H_{\rm inf}$, and the parameter space compatible with a detection by various experiments has been recently analyzed in Refs.~\cite{Bernal:2019lpc,Figueroa:2019paj} (see also Ref. \cite{Tanin:2020qjw}). Consistency with upper bounds on stochastic GW backgrounds rules out, however, a significant fraction of the observable parameter space. This renders the signal unobservable by Advanced LIGO, independently of the parameter space $\lbrace w, f_{\rm RD}, H_{\rm inf}\rbrace$~\cite{Figueroa:2019paj}. The GW background remains detectable in experiments like LISA, but only in a small island of parameter space, corresponding to contrived scenarios with a a low EoS $0.46 \lesssim w \lesssim 0.56$, and a high inflationary scale $H_{\rm inf} \gtrsim 10^{13}$ GeV but low frequency $10^{-11}~{\rm Hz} \lesssim f_{\rm RD} \lesssim 3.6\times10^{-9}~{\rm Hz}$ (corresponding to a low reheating temperature $1~{\rm MeV} \lesssim T_{\rm RD} \lesssim 150$ MeV).

If there is an intermediate phase of MD (or, in general, a period with EoS ranging $0 \leq w < 1/3$), the transfer function of the inflationary GW spectrum in Eq.~(\ref{eq:GWfullSpectrumInstant}), develops instead a red-tilted high-frequency branch, as modes propagate through that phase. If the inflationary background could be observed by GW direct detection experiments, this could led to determine the reheating temperature $T_{\rm reh}$, since the end of the matter-like era would be determined by the start of RD era, and hence the pivot frequency $f_{\rm RD}$ separating the two branches of the GW spectrum would inform us directly about the energy scale at the onset of RD epoch, see e.g. Refs.~\cite{Nakayama:2008wy,Kuroyanagi:2008ye, Kuroyanagi:2010mm,DEramo:2019tit}. Unfortunately, given the current upper bound on the scale of inflation, only futuristic experiments like BBO or DECIGO~\cite{Crowder:2005nr,Seto:2001qf,Sato:2017dkf} may have a chance to measure the inflationary GW background today, as its amplitude is very suppressed, c.f.~Eq.~(\ref{eq:InfGWtodayRD}). Furthermore, unless upcoming CMB B-mode experiments make a detection of the inflationary background, the current upper bound on the scale of inflation is expected to be lowered by roughly an order of magnitude. Taking into account that, together with considering a natural reddening of the tilt at small scales (simply due to the progressive breaking of slow-roll as inflation evolves), it is very likely that the inflationary GW background will never be detected by direct detection GW experiments. Thus, the possibility to infer $T_{\rm reh}$ with this technique seems rather difficult, unless B-mode experiments manage to detect the inflationary tensor modes within the next decade.

As discussed in Sec. \ref{sec:reheating}, an EMD epoch following immediately after the end of inflation, can actually be considered a rather generic phenomenon, as long as the inflaton oscillates after inflation around its potential minimum, with the potential curvature dominated (at sufficiently small values) by a quadratic term. If this is the case, GWs produced by second order effects of the primordial curvature perturbation~\cite{Ananda:2006af,Baumann:2007zm,Assadullahi:2009jc,Saito:2009jt}, can be enhanced at small scales, simply because of gravitational instability of scalar perturbations (as long as the EMD era lasts long enough, see e.g. Ref.~\cite{Musoke:2019ima} for recent numerical simulations on such scenarios). As a matter of fact, as discussed in Sec. \ref{sec:formationPBHs}, very large density perturbations at small scales may lead to the formation of PBHs, which potentially might explain the origin of (a fraction of) the DM. Early computations assuming a sudden transition between early MD and the onset of RD epoch~\cite{Alabidi:2013wtp}, indicated that the induced GW background could be within the range of BBO/DECIGO, as long as the MD phase lasted until the reheating of the Universe at some low temperature $T_{\rm reh} \sim 10^9$ GeV. Ref.~\cite{Inomata:2019zqy} has, however, revisited the idea by incorporating the evolution of the gravitational potentials during a gradual transition from an early MD period to the RD era, as expected in more realistic modelings of a smooth transition with a timescale comparable to the Hubble time at those moments (as it is the case in perturbative reheating scenarios with constant decay rate). They conclude that the presence of an early MD era does not enhance the induced GW background as much as originally thought. For a sudden transition between MD and RD eras, however, an enhanced production of GWs occurs just after the reheating transition concludes, due to fast oscillations of scalar modes well inside the Hubble horizon as shown in Ref.~\cite{Inomata:2019ivs}. They claim that this enhancement mechanism just after an early MD era is in fact much more efficient than any previously known enhancement mechanism during such MD period, leading to a detectable GW signal by BBO/DECIGO if the reheating temperature is in the range $T_{\rm reh} < 7\times 10^{-2}$\,GeV or 20\,GeV $< T_{\rm reh} < 2\times 10^7$\,GeV. In summary, the GW-induced effects due to an early MD period may strongly depend on how the reheating of the Universe took place. Also, more generally, the thermal history of the early Universe can be probed from the second order induced GW background~\cite{Hajkarim:2019nbx,Domenech:2019quo,Domenech:2020kqm}. Going back to the case of a stiff epoch preceding the standard RD era, it is worth stressing that Refs.~\cite{Domenech:2019quo,Bhattacharya:2019bvk} have pointed out that in such a case the amplitude of the second order induced GW spectrum can be greatly enhanced, leading to a potentially observable signals at LISA and other planned observatories.

\subsubsection{Cosmic strings}

Cosmic strings are topological defects that can be produced in gauge theories undergoing phase transitions~\cite{Nielsen:1973cs,Kibble:1976sj} or fundamental strings stretched to cosmological sizes~\cite{Dvali:2003zj,Copeland:2003bj}. While their evolution is in general very complicated, it does exhibit a scaling solution~\cite{Vilenkin:2000jqa} in which the the total string density is a fixed small fraction of the total density. This behaviour is a consequence of the string network being chopped up into loops, which then oscillate and decay into GWs, though intercommutation. This behaviour was confirmed in numerical simulations~\cite{BlancoPillado:2011dq,Blanco-Pillado:2013qja,Blanco-Pillado:2017oxo,Blanco-Pillado:2019tbi} which were also crucial in fixing the loop number density which, in turn, is key in calculation of the resulting GW spectrum dominantly produced by loops.

For a recent review of the general formalism used to calculate the GW spectrum from cosmic strings we refer the reader to Ref.~\cite{Auclair:2019wcv}. The key feature of these spectra is that a flat plateau develops, produced by the strings continuously emitting throughout the RD period. In the case of strings evolving in a nonstandard period of expansion~\cite{Cui:2017ufi,Cui:2018rwi,Gouttenoire:2019kij,Gouttenoire:2019rtn} the spectra will change similarly to the case of an inflationary plateau discussed above. The modification will be visible above a characteristic frequency corresponding to the reheating temperature $T_{\rm reh}$~\cite{Cui:2018rwi}
\begin{equation} \label{eqn:fdeltaforlargealpha}
f_{\rm reh}=
  (8.67\times 10^{-3} \, {\rm Hz})\,
\left(\frac{T_{\rm reh}}{\rm GeV} \right)
\left(\frac{ 10^{-11}}{G\mu}\right)^{1/2}
  \left(\frac{g_*(T_{\rm reh})}{g_*(T_0)}\right)^\frac{8}{6} \left(\frac{g_{*,s}(T_0)}{g_{*,s}(T_{\rm reh})}\right)^{\frac{7}{6}}\, ,
\end{equation}
where $G\mu$ is the tension of the strings forming the network. If the modification is simply an expansion driven before reheating by an energy constituent with a barotropic parameter $w$ the spectrum at higher frequencies will approximately follow Eq.~\eqref{eq:GWfullSpectrumInstant}. This simple result is modified by summation of emission modes for $w<1/3$~\cite{Blasi:2020wpy,Cui:2019kkd,Gouttenoire:2019kij}. The resulting spectrum will be less steep depending on the main emission mode. For example with $w=0$, if the emission is dominated by cusps, the high frequency spectrum would follow $f^{-\frac13}$.
More subtle modification such as a change in the number of degrees of freedom will also leave a distinguishable feature in the spectrum at frequency~\eqref{eqn:fdeltaforlargealpha}~\cite{Cui:2018rwi,Gouttenoire:2019rtn}.
A short period of inflation would leave an imprint on the spectrum similar to the case of EMD~\cite{Guedes:2018afo,Gouttenoire:2019kij}, however, strings produced close to the beginning of inflation can have a much more striking phenomenology, as the stochastic background can be severely diluted, leaving recently produced GW bursts as the main detection prospect~\cite{Cui:2019kkd}. Let us also note that GWs could be generated by short-lived  cosmic defects in scenarios where an EoS $w \simeq 1$ follows after inflation and nonminimally coupled spectator fields are present~\cite{Bettoni:2018pbl}.

We show a few examples of possible standard and modified GW spectra in Fig.~\ref{fig:modGWs} together with expected sensitivities of planned GW observatories including
LISA~\cite{Audley:2017drz}, ET~\cite{Punturo:2010zz,Hild:2010id}, 
AION/MAGIS~\cite{Graham:2016plp,Graham:2017pmn,Badurina:2019hst},
AEDGE~\cite{Bertoldi:2019tck} and  SKA~\cite{Janssen:2014dka}, as well as the existing 
PPTA~\cite{Shannon:2015ect} and LIGO~\cite{TheLIGOScientific:2016wyq} observatories. At first glance cosmic string spectra bear some resemblance to an inflationary spectrum due to the presence of flat plateau at high frequencies which is produced during radiation domination. Upon closer inspection, however, it becomes clear that features in the spectrum such as the peak produced during recent matter domination and slight steps at higher frequencies produced by variation in the number of degrees of freedom appear at very different frequencies in the two cases. It should therefore be possible to distinguish between the inflationary and cosmic strings spectra.

\begin{figure*}
\centering
\includegraphics[width=0.44\textwidth]{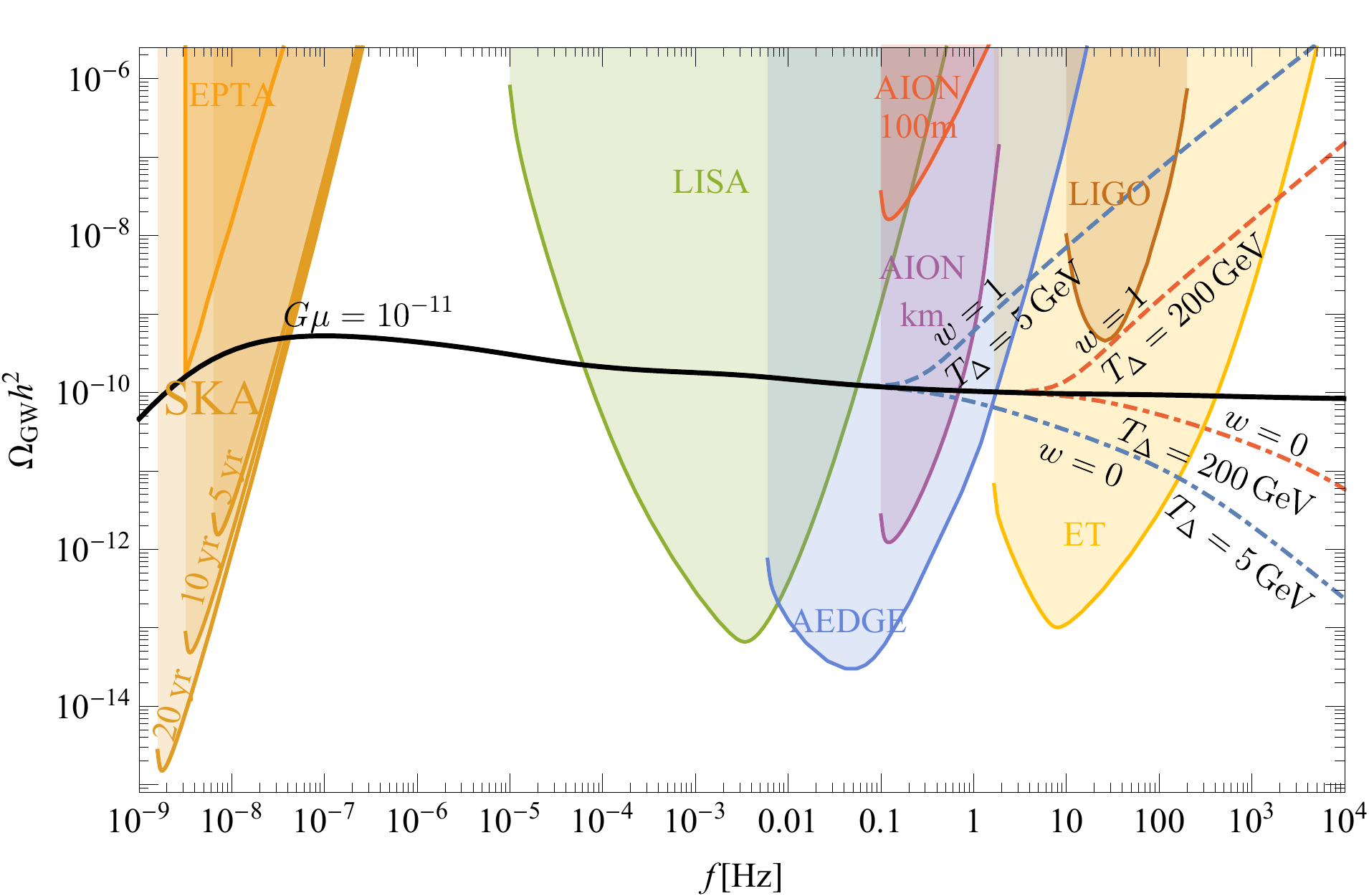} \hspace{4mm}
\includegraphics[width=0.44\textwidth]{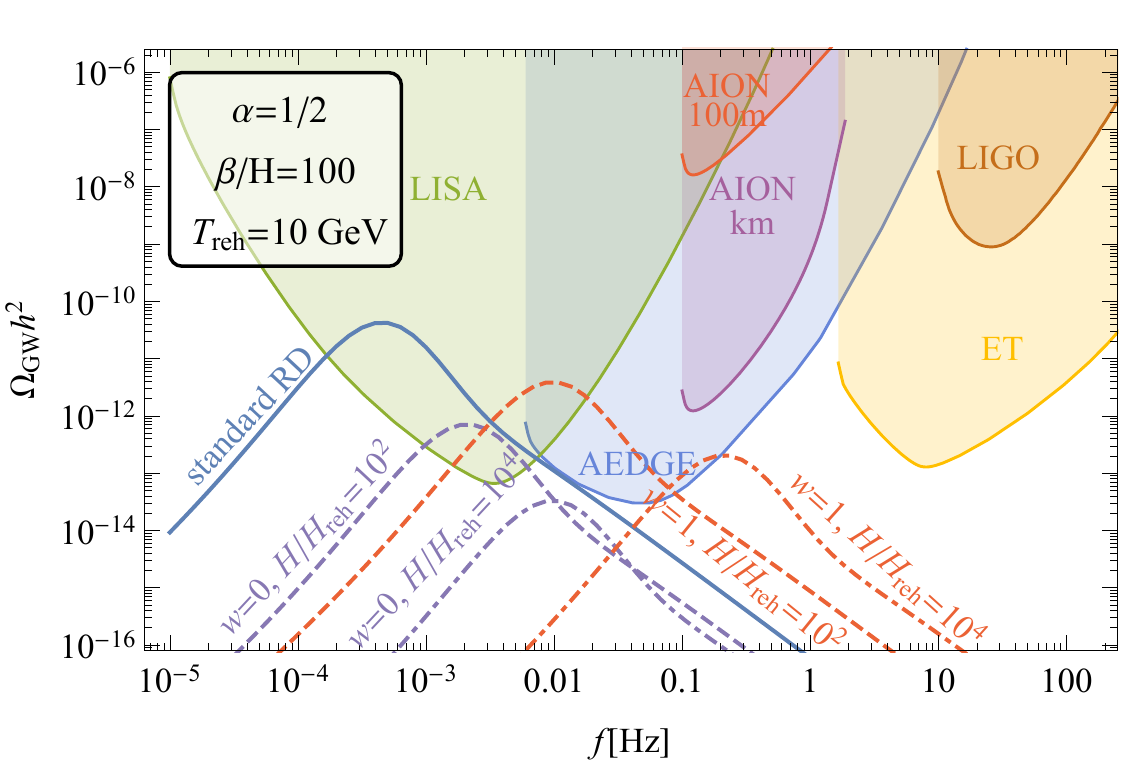}
\caption{Left panel: Example of a GW spectrum from a network of cosmic strings in standard cosmology (solid black curve) together with spectra from strings evolving in cosmologies featuring an EMD epoch (dot dashed curves) and a period of kination (dashed curves) doth ending at $T_\Delta=5 $\,GeV (blue lines) and $T_\Delta=200 $\,GeV (red curves). Right panel: Example of a GW signal from a first order phase transition with $\alpha=1/2,$ $\beta/H=100$ at the temperature $T_*=1$\,TeV. The solid blue curve, also below shows the signal in standard RD expansion while dashed and dot-dashed curves show the cases of $H/H_{\rm reh}=10^2$ and $H/H_{\rm reh}=10^4$ of nonstandard expansion. Red curves correspond to modification of standard expansion with a period of kination with $w=1$ while the purple ones correspond to modification of standard expansion with MD ($w=0$).}
\label{fig:modGWs}
\end{figure*}

\pagebreak
\subsubsection{First order phase transitions}

While study of production of GWs in such a transition is a daunting task usually approached through lattice simulations~\cite{Huber:2008hg,Hindmarsh:2015qta,Hindmarsh:2017gnf,Cutting:2018tjt,Cutting:2019zws,Pol:2019yex} or through simplified analytical modelling~\cite{Caprini:2009yp,Kahniashvili:2009mf,Kisslinger:2015hua,Hindmarsh:2016lnk,Jinno:2016vai,Jinno:2017fby,Ellis:2018mja}, typically the results are parametrised by just a few crucial parameters~\cite{Grojean:2006bp,Caprini:2015zlo,Caprini:2019egz}. These are, firstly, strength of the phase transition,
\begin{equation}
\alpha(T) = \frac{\Delta V(T) - \frac{T}{4} \frac{\partial \Delta V(T)}{\partial T}}{\rho_{\mathrm{R}}} \, ,
\label{eq:alpha}
\end{equation}
which is the ratio of released vacuum energy to the radiation background. Secondly the time scale $\beta$ which assuming exponential nucleation can be defined as
\begin{equation}
\label{eq:beta_Taylor}
\Gamma \propto  e^\frac{-S_3(T)}{T} = e^{\beta(t_*-t) + ...} \,  \rightarrow \beta=-\frac{d}{dt}\frac{S_3(T)}{T}=H(T)\,  T\frac{d}{dT}\frac{S_3(T)}{T} \, .
\end{equation}
with the action of the nucleating bubble $S_3(T)$ defined in Eq.~\eqref{eq:PTdecaywidth}. All the quantities marked with a $*$ are evaluated at the transition temperature. There are three main sources of GWs in a phase transition: collisions of bubble walls~\cite{Huber:2008hg,Cutting:2018tjt}, sound waves produced by bubbles in the plasma~\cite{Hindmarsh:2015qta,Hindmarsh:2016lnk,Hindmarsh:2017gnf} and turbulence ensuing after the transition~\cite{Caprini:2009yp,Kahniashvili:2009mf,Kisslinger:2015hua,Pol:2019yex}. All the expressions necessary to calculate the signal can be found for example in~\cite{Ellis:2019oqb}. 

In the case of modified cosmology, the first obvious modification we have to make is to include any nonstandard components contributing to the expansion at the time of the transition in the abundance of GWs. If the extra components dominate the expansion during the transition this comes down to only a trivial modification of the standard formulas for $\Omega_{\rm GW*}$ through
\begin{equation}\label{eq:PTWGabundanediminish}
\frac{\alpha}{\alpha+1}=\frac{\rho_V}{\rho_V+\rho_R}\rightarrow  \frac{\rho_V}{\rho_{\rm tot}} = \alpha \left(\frac{H_{R*}}{H_*} \right)^2 \,,
\end{equation}
where $\rho_{\rm tot}$ includes the new component dominating the expansion rate at the time of the transition and $H_{R*}$ includes just the subdominant radiation component.

The last key parameter is the temperature at which the transition completes. This leads to another important modification with respect to the standard expressions which  comes in through modified redshifting. GWs redshift in the same way as radiation:\footnote{We approximate that $g_*(T_{\rm reh}) = g_{*,s}(T_{\rm reh})$.}
\begin{equation} \label{eq:PTOmegaredshift}
\Omega_{{\rm GW},0} = \left(\frac{a_*}{a_0}\right)^4  \left(\frac{H_*}{H_0}\right)^2 \Omega_{{\rm GW},*}
=  1.67\times 10^{-5} h^{-2} \left(\frac{100}{g_*(T_{\rm reh})}\right)^\frac13 \left(\frac{H_*}{H_{\rm reh}}\right)^{2\frac{3w-1}{3w+3}} \Omega_{{\rm GW},*} \,, 
\end{equation}
while the frequency scales as $f~ a^{-1}$:
\begin{equation} \label{eq:PTfredshift}
f_0 = \frac{a_*}{a_0} f_* =\frac{a_{\rm reh} H_{\rm reh}}{a_0} \frac{a_* H_*}{a_{\rm reh} H_{\rm reh}} \frac{f_*}{ H_*} 
= 1.65\times 10^{-5} \,{\rm Hz}\, \left( \frac{T_{\rm reh}}{100\,{\rm GeV}} \right) \left( \frac{g_*(T_{\rm reh})}{100} \right)^\frac{1}{6}  \left(\frac{f_*}{H_*}\right)\left(\frac{H_*}{H_{\rm reh}}\right)^{\frac{3w+1}{3w+3}} \, ,
\end{equation}
where again the "reh" index denotes quantities at reheating. This prescription effectively means using the standard formulas at $T_{\rm reh}$ and including the extra redshift  through the last bracket in equations above. There are several examples of interest:
\begin{enumerate}
\item For the case of extra energy constituent redshifting faster than radiation the reheating time and $w$ are enough to calculate $\left(H_*/H_{R*}\right)^2=(H/H_{\rm reh})^{2\frac{3w-1}{3w+3}}$. Remembering that Eq.~\eqref{eq:PTWGabundanediminish} enters the final GW abundance squared we see that the net result including redshift is still a suppression of the amplitude of the signal and increase of the frequency compared to the standard radiation dominated case. The upshot of this scenario is that the large Hubble rate at early times can help facilitate electroweak baryogenesis (see Sec.~\ref{sec:PTdynamics}).
\item The second possibility is that the PT occurs in standard circumstances, however, its strong enough for the energy stored in the field itself to dominate the expansion as the field oscillates around its minimum effectively causing a short MD period before the field decays and reheats the Universe~\cite{Ellis:2019oqb}. Thus, in this case Eq.~\eqref{eq:PTWGabundanediminish} should not be used, yet the redshift is still modified according to Eq.~\eqref{eq:PTOmegaredshift} and Eq.~\eqref{eq:PTfredshift}.
\item The case of PTs in matter domination (or other constituent redshifting slower than radiation)  is much less attractive as now both the redshifting and initial smaller abundance diminishes the signal. Thus the dominating MD period has to decay very soon after the PT to leave an observable abundance.
\end{enumerate}
We show an example of the resulting modification in Fig.~\ref{fig:modGWs} for the first two of the above cases. Furthermore the spectrum will be modified for the modes beyond the horizon size at the time of the transition. The typical $\propto f^3$ result expected for the spectrum in RD~\cite{Caprini:2009fx} at these scales is modified as we apply the transfer function upon re-entry of the modes~\cite{Barenboim:2016mjm,Cai:2019cdl,Domenech:2019quo,Ellis:2020nnr} and as an example we find $\propto f$ for modes re-entering during EMD era. While in principle this modification could allow us to distinguish between RD and modified expansion, the feature in the spectra will only appear at frequencies much below the peak $f\lesssim f_{\rm peak} (\beta/H)^{-1}$ so it could only be visible in exceptionally strong transitions~\cite{Ellis:2020nnr}. 

\section{Summary}
\label{sec:summary}
We have reviewed several causes and consequences of deviations from radiation domination in the early Universe -- taking place either before or after Big Bang Nucleosynthesis -- and the constraints on them, as they have been discussed in the literature during the recent years. In Sec.~\ref{sec:causes}, we discussed the causes of nonstandard expansion and how they can broadly be grouped as modifications to the properties of the standard \LCDM~components, and the introduction of new components altogether. These two categories can be further refined into modifications to the background and/or perturbative evolution. As specific examples, we have discussed post-inflationary reheating, multiple stages of inflation, heavy particles, dark sectors, and moduli fields, as well as nonstandard post-BBN expansion history caused by e.g. decaying dark matter, nonstandard neutrino interactions, or early dark energy. In Sec.~\ref{sec:consequences}, we reviewed various consequences of nonstandard expansion eras including the effects such eras can have on models of dark matter, early Universe phase transitions and baryogenesis, models of inflation, and generation of the primordial curvature perturbation, as well as microhalos and primordial black holes. In Sec.~\ref{sec:constraints}, we summarized the known constraints on nonstandard expansion eras, as well as some future ways to put such scenarios into test. More specifically, we discussed constraints coming from BBN and the CMB, non-detection of microhalos and PBHs, and stochastic gravitational wave backgrounds.

The standard radiation domination that started at the end of inflation and lasted until the usual matter-radiation equality remains as a paradigm. However, several interesting proposals regarding various topics such as the generation of dark matter or matter-antimatter asymmetry during a nonstandard expansion phase have been recently made. We  hope that this review on the possible causes and consequences of such eras and the known constraints on them helps in guiding further research on these topics.

\section*{Acknowledgments}
\small{

\noindent R. Allahverdi is supported by NSF Grant No. PHY-1720174. 

\noindent M. A. Amin is supported by the NASA ATP-theory grant NASA-ATP Grant No. 80NSSC20K0518. 

\noindent A. Berlin is supported by the James Arthur Fellowship.

\noindent N. Bernal received funding from Universidad Antonio Nari\~no grants 2018204, 2019101, and 2019248, the Spanish MINECO under grant FPA2017-84543-P, and the Patrimonio Aut\'{o}nomo - Fondo Nacional de Financiamiento para la Ciencia, la Tecnolog\'{i}a y la Innovaci\'{o}n Francisco Jos\'{e} de Caldas (MinCiencias - Colombia) grant 80740-465-2020. This project has received funding from the European Union's Horizon 2020 research and innovation programme under the Marie Sk\l{}odowska-Curie grant agreement No 860881-HIDDeN.

\noindent C. Byrnes was supported by a Royal Society University Research Fellowship. 

\noindent M. S. Delos is partially supported by NSF CAREER Grant No. PHY-1752752 and by a Dissertation Completion Fellowship from the University of North Carolina at Chapel Hill.

\noindent A. L. Erickcek is partially supported by NSF CAREER Grant No. PHY-1752752. 

\noindent M. Escudero is supported by the European Research Council under the European Union's Horizon 2020 program (ERC Grant Agreement No 648680 DARKHORIZONS).

\noindent D. G. Figueroa is supported by a Ram\'on y Cajal contract by Spanish Ministry MINECO, with Ref.~RYC-2017-23493. 

\noindent K. Freese is supported by the Jeff and Gail Kodosky Endowed Chair in Physics at the University of Texas, DOE grant DE-SC007859 at the University of Michigan, by the Vetenskapsr\aa det (Swedish Research Council) through contract No. 638-2013-8993, and the Oskar Klein Centre for Cosmoparticle Physics at Stockholm University. 

\noindent T. Harada is supported by JSPS KAKENHI Grants No.~JP19K03876 and JP19H01895.

\noindent D. Hooper is supported by Fermi Research Alliance, LLC under Contract No. DE-AC02-07CH11359 with the U.S. Department of Energy, Office of High Energy Physics.

\noindent D. I. Kaiser is supported in part by the U.S. Department of Energy under Contract No.~DE-SC0012567.

\noindent T. Karwal is supported by funds provided by the Center for Particle Cosmology at the University of Pennsylvania. 

\noindent K. Kohri is supported by JSPS KAKENHI Grant No.~JP17H01131 and MEXT Grant-in-Aid for Scientific Research on Innovative Areas JP15H05889, JP18H04594, JP19H05114, JP20H04750.

\noindent G. Krnjaic is supported by Fermi Research Alliance, LLC under Contract No. DE-AC02-07CH11359 with the U.S. Department of Energy, Office of High Energy Physics.

\noindent M. Lewicki is supported by the UK STFC Grant ST/P000258/1 and by the Polish National Science Center grant 2018/31/D/ST2/02048. 

\noindent K. Sinha is supported by DOE Grant DE-SC0009956. 

\noindent T. L. Smith is supported by the Research Corporation, through a Cottrell Scholar Award, and NASA ATP grant number 80NSSC18K0728.

\noindent T. Takahashi is partially supported by JSPS KAKENHI Grant Number 17H01131, 19K03874 and MEXT KAKENHI Grant Number 15H05888, 19H05110.

\noindent T. Tenkanen was supported by the Simons Foundation. 

\noindent J. Unwin is supported by NSF grant DMS-1440140 while in residence at the MSRI, Berkeley. 

\noindent V. Vaskonen is supported by the UK STFC Grant ST/P000258/1 and by the Estonian Research Council grant PRG803.

\noindent S. Watson is supported in part by DOE grant DE-FG02-85ER40237 and NSF grant AST-1813834.

\noindent M. A. Amin, A. L. Erickcek, D. G. Figueroa, M. Lewicki, K. Sinha, T. Tenkanen and S. Watson acknowledge hospitality and support from KITP-UCSB, where part of this work was carried out, supported in part by the National Science Foundation Grant No. NSF PHY-1748958.
}

\bibliography{refs_journal}

\end{document}